\documentclass[psamsfonts,12pt]{report}

\usepackage{amsmath} % already loaded, but doesn't hurt
\usepackage{amsthm} % already loaded, but doesn't hurt
\usepackage{amsfonts} % already loaded, but doesn't hurt
\usepackage{amssymb} % extra symbols not included in amsfonts
\usepackage{mathrsfs}
\usepackage{bbold} % good for those double-struck numbers
\usepackage{natbib}
\usepackage{hyperref}
\hypersetup{colorlinks,bookmarksopen,bookmarksnumbered,citecolor=blue,
	pdfstartview=FitH}
\usepackage{tocbibind}
\usepackage[all]{xy}
\usepackage{graphicx}
\usepackage[usenames,dvipsnames]{color}
\usepackage{verbatim}
\usepackage{tabularx} % for better tables 
\date{May 2008}
\title{Abelian Chern-Simons theory with toral gauge group, modular tensor categories, and group categories}
\author{Spencer D. Stirling\\
  Department of Mathematics\\
  University of Texas at Austin\\
  stirling@math.utexas.edu}

\begin{document}

\newcommand{\Vcat}{\mathcal{V}}
\newcommand{\id}{\text{id}}
\newcommand{\grp}{\mathcal{D}}
\newcommand{\qmodz}{\mathbb{Q}/\mathbb{Z}}
\newcommand{\modfunct}{\mathscr{F}}
\newcommand{\unitobj}{\mathbb{1}}
\newcommand{\idmat}{\mathbb{1}}
\newcommand{\cplx}{\mathbb{C}}
\newcommand{\trio}{{(\grp,q,c)}}
\newcommand{\triotrunc}{{(\grp,q)}}
\newcommand{\grpcat}{\mathscr{C}_\triotrunc}
\newcommand{\ribv}{\text{Rib}_\Vcat}
\newcommand{\ribi}{\text{Rib}_I}

\setcounter{secnumdepth}{1}
\numberwithin{equation}{chapter}

% set up some amsthm theorem environments
\theoremstyle{plain}
\newtheorem{theorem}[equation]{Theorem}
\newtheorem{proposition}[equation]{Proposition}
\newtheorem{corollary}[equation]{Corollary}
\newtheorem{lemma}[equation]{Lemma}
\newtheorem{fact}[equation]{Fact}
\newtheorem{conjecture}[equation]{Conjecture}
\theoremstyle{definition}
\newtheorem{definition}[equation]{Definition}
\newtheorem{example}[equation]{Example}
\theoremstyle{remark}
\newtheorem{remark}[equation]{Remark}

\maketitle

\pagenumbering{roman}
\begin{abstract}
Classical and quantum Chern-Simons with gauge group $\text{U}(1)^N$ were classified
by Belov and Moore in \cite{belov_moore}.  They studied both ordinary
topological quantum field theories as well as spin theories.  On the other hand
a correspondence is well known between ordinary $(2+1)$-dimensional TQFTs and modular
tensor categories.  We study group categories and extend them slightly to produce modular tensor
categories that correspond to toral Chern-Simons.
Group categories have been widely studied in other contexts
in the literature \cite{frolich_kerler},\cite{quinn},\cite{joyal_street},\cite{eno},\cite{dgno}.
The main result is a proof that the associated projective representation of
the mapping class group is isomorphic to the one from toral Chern-Simons.  
We also remark on an algebraic theorem
of Nikulin that is used in this paper.
\end{abstract}

\tableofcontents
%\listoffigures
%\listoftables

\chapter*{Acknowledgements}
The author warmly thanks his advisor, Dan Freed, for suggesting this problem
and for providing years of patient and persistent guidance during this work.  The author
also wishes to thank his wife, his family (especially his parents), 
and his friends for providing limitless moral
support.

\pagenumbering{arabic}

\chapter{Introduction}
\label{ch:introduction}

The study of topological quantum field theories emerged in the 1980's
in \cite{witten_tqft} where a supersymmetric quantum theory was introduced
that is linked to Floer homology and the Donaldson invariants.  It was shown that this quantum field
theory is metric independent.  A short time later
groundbreaking connections were made in \cite{witten} between Chern-Simons field theory
and low-dimensional topology (knot theory and $3$-manifold invariants).

Contemporarily, algebraists and representation theorists were constructing
\textit{quantum groups}, and equally powerful connections were
made between quantum groups, knot theory, and $3$-manifold invariants
(\cite{reshetikhin_turaev1},\cite{reshetikhin_turaev2},
\cite{kirby_melvin}).

In Chern-Simons theory the basic data that characterizes a theory is a 
compact Lie group $G$ along with an element $k\in H^4(BG,\mathbb{Z})$ 
called the \textit{level}.
Witten considered compact semisimple
Lie groups (e.g. $\text{SU}(2)$) where $k$ is an integer.
On the other hand the basic data that characterizes a quantum group is a compact
semisimple Lie group $G$ along with a deformation parameter $t$.

It was noticed immediately that there is an agreement
between Chern-Simons theory and quantum groups when comparing the
induced link invariants and $3$-manifold invariants.
For example, for $G=\text{SU}(2)$ they agree if the
level and the deformation parameter are related by
\begin{equation}
 t=\exp\left(\frac{\pi i}{2(k+2)}\right)
\end{equation}
In light of this (actually somewhat before) Atiyah proposed an
axiomatic umbrella formulation of TQFTs \cite{atiyah_book} that
unifies both approaches into a common language.

Simultaneously a third line of development based on category theory was
emerging.  The braided and ribbon categories 
described in \cite{joyal_street},\cite{shum},
combined with aspects formulated in \cite{moore_seiberg},\cite{reshetikhin_turaev1},\cite{reshetikhin_turaev2}, 
resulted in
\textit{modular tensor categories} (c.f. \cite{turaev}).  In particular quantum
groups are examples of modular tensor categories, and many crucial aspects
of conformal field theory are also encoded in modular tensor categories.
By the early 1990's a clearer picture had emerged: 
\begin{equation}
  \text{Quantum Groups}\subset \text{MTCs}\Longleftrightarrow (2+1)\text{-dim TQFTs} \supset \text{Chern-Simons}
\end{equation}
The relationship between MTCs and TQFTs is discussed further below (in
particular - to the author's knowledge - the direction $\text{MTC}\Leftarrow \text{TQFT}$ is 
not yet constructed for \textit{all} cases).

Several examples of Chern-Simons theories that have been quantized are listed in
the left column of table~(\ref{tab:zoology}) (more cases that have been quantized
include most simple groups $G$ and direct products).  
In particular Chern-Simons theories with gauge
group $\text{U}(1)$ were studied by Manoliu \cite{manoliu} using a real
polarization technique, and more recently Chern-Simons theories with gauge
group $\text{U}(1)^N$ were classified by Belov and Moore \cite{belov_moore}
using K\"ahler quantization.  It was shown that the data that
determines quantum toral Chern-Simons is a trio $\trio$ where $\grp$ is a finite abelian
group, $q:\grp\rightarrow \qmodz$ is a quadratic form, and $c$ is an integer mod 24 (subject to a
constraint).
It is natural to ask what the corresponding modular
tensor categories are.  This paper answers that question.

\begin{table}[hbt]
\caption{Zoology}
\label{tab:zoology}
\centerline{
\begin{tabularx}{\linewidth}{|X|X|X|} 
%{| p{4cm} | p{4cm} || p{4cm} |}
\hline
 Classical Chern-Simons\newline $G$ compact Lie group\newline $k\in H^4(BG,\mathbb{Z})$ &  
   Modular Tensor Category & Link Invariants in $S^3$\newline (link has canonical framing)\\
\hline\hline
$G=\text{SU}(2)$\newline $k\in\mathbb{Z}$ & Quantum group $U_t(\mathfrak{sl}_2(\cplx))$\newline
   $t=\text{exp}(\frac{\pi i}{2(k+2)})$ & Jones polynomial\\
\hline
$G=\text{SU}(N)$\newline $k\in\mathbb{Z}$ & Quantum group $U_t(\mathfrak{sl}_N(\cplx))$\newline
   $t=\text{exp}(\frac{\pi i}{2(k+2)})$ & HOMFLY polynomial\\
\hline
$G=\text{finite group}$\newline $k=$vacuous \cite{freed_finite}\newline\cite{witten_dijkgraaf} 
  & Quantum double $\mathcal{D}(G)$ & No uniform description\\
\hline
$G=\text{U}(1)$ \cite{manoliu}\newline
$G=\text{U}(1)^N$ \cite{belov_moore}\newline
  $k=$even lattice & this paper & Deloup invariants \cite{deloup1} \\
\hline
\end{tabularx}}
\end{table}

We note that Belov and Moore \cite{belov_moore} classified more general
\textit{spin}
\footnote{The ordinary theories below are constructed on manifolds with
extra structure: 2-framings \cite{atiyah}.  \textit{Spin} theories are
really theories of \textit{framed} manifolds.  See the recent work by Hopkins-Lurie
on the Baez-Dolan hypothesis.}
toral Chern-Simons
theories as well.  Unfortunately there is no well-developed
notion of \textit{spin modular tensor category}, however the work done here
makes an excellent
toy model that we can use to decide what the ``right'' definition for
spin MTC should be.  We plan to expand these ideas in
a forthcoming paper.

Physicists will be mainly interested in the applications to the
fractional quantum Hall effect (FQHE).  The \textit{abelian} states at
filling fraction $\nu=\frac{1}{3}$ remain the only rigorously-established
experimental states to coincide with Chern-Simons, hence the abelian case
remains relevant despite being useless for topological quantum computation.

Before we proceed let us mention the very closely
related work of Deloup \cite{deloup1,deloup2,deloup3}.  Deloup
begins with the data of a finite abelian group $\grp$ and
a quadratic form $q:\grp\rightarrow\qmodz$.  Because of
the abelian nature of the data it is possible to construct
invariants of links and (eventually) a $(2+1)$-dim TQFT ``by hand'' appealing
to reciprocity alone.  

This bypasses modular tensor categories entirely.
However, the price is that no braiding is described (the braiding
is rather more subtle than what one might first expect).
We emphasize this difference
since the modular tensor categories described here allow us
to construct an \textit{extended $(2+1)$-dim TQFT} (see chapter~(\ref{ch:tqft})).
In particular we can describe quasiparticles completely, whereas
Deloup's theories cannot.
We also emphasize that Deloup did not connect his work to 
Chern-Simons.  It is the main result in this paper that the TQFTs constructed
here are the same as those from toral Chern-Simons.

Finally, some simple examples of ribbon categories are considered
in the appendix in \cite{deloup1}.  These examples are briefly considered here
in chapter~\ref{ch:mtc}, and we argue that these do \textit{not}
correspond to toral Chern-Simons since many of them are not
modular tensor categories.  

The organization of this paper is as follows: in chapter~(\ref{ch:tqft}) we
give a brief overview of TQFTs starting with the motivating example of Chern-Simons.
In chapter~(\ref{ch:chernsimons}) we review
toral Chern-Simons as was classified by Belov-Moore.  In chapter~(\ref{ch:mtc}) we
provide the relevant definitions for ribbon categories and modular tensor categories,
and we construct $(2+1)$-dim TQFTs from them.  This chapter differs from 
\cite{turaev} and \cite{bakalov_kirillov} in that we emphasize \textit{non-strict}
categories.
In chapter~(\ref{ch:groupcategories}) we study group categories and build
modular tensor categories out of them.
The main result is proven in chapter~(\ref{ch:correspondence}) - the projective
actions of the mapping class group induced from toral Chern-Simons and separately
from group categories are isomorphic.

\chapter{$(2+1)$-dim Topological Quantum Field Theories}
\label{ch:tqft}

\section{Introduction}
In this chapter we give a quick account of $(2+1)$-dim
topological quantum field theories (TQFTs).  A $(2+1)$-dim 
Chern-Simons TQFT is essentially determined by the
$(1+1)$-dim conformal field theory (CFT) 
on the boundary (the Knizhnik-Zamolodchikov equations determine the
braidings and the twists that appear in the theory).
The language of modular tensor categories (MTCs) is rather different,
but underneath the details MTCs axiomatically encode the relevant
structures that appear in CFTs.  Hence it is no surprise that the
Chern-Simons TQFTs form a subset
of the TQFTs constructed from modular tensor categories (it is
in debate whether the opposite inclusion is true \cite{rowell}).

The axiomatic approach toward the end of the chapter is 
taken from chapter 3 in Turaev's book \cite{turaev}
as well as the book of Bakalov and Kirillov \cite{bakalov_kirillov}.
The original axioms were formulated by Atiyah \cite{atiyah_book}
long ago.

Witten's work relies on the earlier work of
Segal \cite{segal} and Moore and Seiberg \cite{moore_seiberg} in conformal field
theory.  Briefly, the conformal field theory that appears on the boundary is the
Wess-Zumino-Witten (WZW) model (actually the chiral/holomorphic part).  For
a geometric perspective on CFTs and Chern-Simons we recommend \cite{kohno}.

Although Atiyah's axioms apply in any dimension, we wish to restrict
ourselves to $(2+1)$-dimensions.  In this case all of the known examples are 
considerably richer than Atiyah's axioms might suggest.  Framed links (ribbons) appear
that physically are meant to encode
the worldlines of exotic \textit{anyonic
quasiparticles} \cite{wilczek} undergoing creation, annihilation, twisting, and braiding.
\footnote{the ribbons must be ``colored'' with the particle species.}
Mathematically more general colored ribbon \text{graphs} are studied,
and \textit{surgery} provides a route from the ribbon graph
construction to Atiyah's axioms.
\footnote{we note that all manifolds must be \textit{oriented}
throughout this paper.}

\section{Chern-Simons}
In \cite{witten} Witten studied the
Chern-Simons quantum field theory defined by the action
\begin{equation}
  \exp(2\pi i k S_{CS})=
  \exp\left(2\pi i k\frac{1}{8\pi^2} \int_{X_3} \text{Tr}(A\wedge dA+\frac{2}{3}A\wedge
    A \wedge A)\right)
  \label{eq:purechernsimons}
\end{equation}  
We will discuss such actions more coherently in 
chapter~(\ref{ch:chernsimons}), but for now $X_3$ is a compact oriented 3-manifold
equipped with a vector bundle $E\rightarrow X_3$ with structure group $G$.
Witten only considered the case where $G$ is a compact 
simply-connected simple Lie group (e.g. $SU(2)$).
The operation $\frac{1}{8\pi^2}\text{Tr}$ is meant to denote the (normalized) Killing
form on the Lie algebra $\mathfrak{g}$.
\footnote{The $\text{Tr}$ notation is somewhat misleading.  We actually require a
symmetric bilinear form $<>:\mathfrak{g}\times\mathfrak{g}\rightarrow\mathbb{R}$.  However,
for simple Lie groups any bilinear form is a scalar
multiple of the Killing form.}

The action as written is not always well-defined since the vector potential
$A$ is not always globally well-defined.
However, obstruction theory tells us that $E$ is trivializable for a connected
simply-connected compact Lie group $G$ 
(this is \textit{not} true in general, 
nor even in the remainder of this paper).
Once we choose a trivialization
\footnote{This choice of trivialization is unimportant.
Chern-Simons is defined to be a gauge theory where the gauge group is
$\mathcal{G}=\text{Map}(X_3,G)$, i.e.
configurations of $A$ that are related by gauge transformations
are physically indistinguishable and must be identified.
However, any two trivializations are related by a gauge transformation.
Hence we only require (for now) that the bundle be \textit{trivializable}.}
then this determines a standard
flat covariant derivative
$D^0$ (the trivialization determines parallel transport).  
Then given any other covariant derivative $D$
we can define the vector potential $A$ via $D=D^0+A$.
\footnote{$E$ has structure group $G$, and $D$ must respect this (e.g. parallel transport
takes orthonormal frames to orthonormal frames for $G=\text{SO}(3)$).
Thus, $A$ is valued in the Lie algebra $\mathfrak{g}$.}

The term $\frac{2}{3}A\wedge A\wedge A$ is written abusively.  It should be
interpreted to mean
\begin{equation}
  \frac{2}{3}A\wedge \frac{1}{2}[A\wedge A]
\end{equation}
where
\begin{equation}
 \frac{1}{2}[A\wedge A](v_1,v_2):=[A(v_1),A(v_2)]
\end{equation}
The bracket on the RHS is the bracket in $\mathfrak{g}$.

It is well known (see e.g. \cite{freed_chernsimons1}) 
that the integral in equation~(\ref{eq:purechernsimons}) is
\textit{not} gauge invariant.  However under gauge transformations (if
$X_3$ is closed)
the integral changes by integer values $M$ only:
\begin{equation}
  \frac{1}{8\pi^2}\int_{X_3} \text{Tr}(A\wedge dA+\frac{2}{3}A\wedge
    A \wedge A)\rightarrow
\frac{1}{8\pi^2}\int_{X_3} \text{Tr}(A\wedge dA+\frac{2}{3}A\wedge
    A \wedge A) + M
\end{equation}
Hence $\exp(2\pi i k S_{CS})$ \textit{is invariant} as long as the \textit{level} $k$ 
is any arbitrary integer.

More generally if $X_3$ has boundary $\Sigma_2$ then under gauge transformations the integral
instead picks up a chiral Wess-Zumino-Witten term:
\begin{equation}
  \frac{1}{8\pi^2}\int_{X_3} \text{Tr}(A\wedge dA+\frac{2}{3}A\wedge
    A \wedge A)\rightarrow
\frac{1}{8\pi^2}\int_{X_3} \text{Tr}(A\wedge dA+\frac{2}{3}A\wedge
    A \wedge A) + S_{cWZW}
\end{equation}
\textit{It is a fact that $S_{cWZW}$ depends only on the configuration
on the boundary $\Sigma_2$}.  Clearly 
the action is \textit{not} gauge invariant if interpreted
in the usual sense.  However the WZW term satisfies
crucial \textit{cocycle conditions}, 
and a more formal construction yields a gauge
invariant theory (see pgs. 16-21 in \cite{freed_chernsimons1}).
%Briefly, 
%if $X_3$ has nonempty boundary $\Sigma_2$ then the number $\exp(2\pi i k S_{CS})$
%\textit{changes} under gauge transformations.  However, it changes in a controlled
%manner.  Denote the restriction of $A$ to $\Sigma_2$ by $\partial A$.  
%We associate to $\partial A$ a complex line
%$L_{\partial A}$.  
%$L_{\partial A}$ is formally defined to be an 
%equivalence class of
%complex lines $\cplx_g$ for different gauge choices $g$.  The complex
%lines $\cplx_g$ are identified using the WZW term that
%appears in
%the integral $\exp(2\pi i k S_{CS})$.  We reinterpret the expression $\exp(2\pi i k S_{CS})$
%to not be a complex number, but rather an equivalence class of complex numbers:
%\begin{equation}
% \exp(2\pi i k S_{CS})\in L_{\partial A} 
%\end{equation}
%In this manner the Chern-Simons action is gauge invariant.

It is instructive to consider briefly a physical system that Chern-Simons
is thought to describe.  In the fractional quantum Hall effect (FQHE)
a $2$-dimensional gas of electrons (trapped between semiconductor layers)
is cooled to a few milliKelvin and placed under a magnetic field pointing in
the $z$ direction (if the 2-d gas lies in the $xy$-plane).  Schematically
the action is
\begin{equation}
  S := S_{\text{cyclotron}}+S_{\text{e-e interaction}}
\end{equation}
where the cyclotron term describes the electrons orbiting in circular
paths due to the magnetic field, and the interaction term describes
Coulomb repulsion between electrons.  The magnetic field breaks the
parity reversal symmetry of the system - hence the system is \textit{chiral}.
Consider the $2$-d electron gas propagating in time.  Then this is a
$(2+1)$-dimensional classical field theory.

Ignoring the e-e interaction term momentarily the quantum description is given in
terms of (degenerate) Landau levels where the energy of the $N$th level goes as
$E_N=\hbar\omega( N+\frac{1}{2})$ where $\omega$ is the cyclotron frequency.  
Hence the system is gapped, and sufficiently 
lowering the temperature restricts the system to the degenerate ground state $N=0$.

The ground Landau level obtains interesting structure
when e-e interactions are again considered.  It is shown (for a special case) in 
\cite{halperin_lee_read} that the
action (through a change of variables) can be written as the effective action
\begin{equation}
  S := S_{\text{cyclotron}}+S_{\text{e-e interaction}}\overset{N=0}{\longrightarrow}
    S_{CS}+S_{\text{quasiparticles}}
\end{equation}
where $S_{CS}$ is the Chern-Simons action introduced in equation~(\ref{eq:purechernsimons}), 
and $S_{\text{quasiparticles}}$
is a term encoding the dynamics of exotic \textit{anyonic quasiparticles} \cite{wilczek}.  The 
quasiparticles can be viewed as
quantum excitations of cooperating electrons and magnetic flux quanta \cite{khare}.
However, we always treat them semiclassically in the sense that their
\textit{trajectories} are treated as \textit{classical} paths.
\footnote{Note that the Chern-Simons vector potential $A$ is usually
\textit{not} the vector potential associated to the magnetic field.}

The quasiparticles are coupled to $A$, hence
they can be viewed as \textit{detectors} that measure
the properties of $A$.  We can imagine quasiparticle/antiquasiparticle
pairs being created, possibly braiding around each other, and annihilating.
Then their worldlines form links in $(2+1)$-dimensions.  Furthermore each 
quasiparticle species has a $(2+1)$-dimensional analogue of spin - the
\textit{twist} - which is a phase factor that a quasiparticle picks up
when it is spun one full counterclockwise turn (viewing the $xy$-plane
from above).  Hence the worldlines should be thought of as \textit{framed}
links, or ribbons, to encode the twists.

As a first attempt to understand the role of the quasiparticles
let us alter the classical setup slightly.  Instead of a Hamiltonian
scenario where the $2$-d electron gas propagates forward in time (i.e.
a $3$-manifold of the form $\Sigma\times I$),
suppose we have a
\textit{closed} compact oriented $3$-manifold $X_3$ with a fixed
vector potential $A$.
Although we are in a classical
setting we put in \textit{by hand} quasiparticles (which are quantum-mechanical).
However, as already mentioned we only allow classical trajectories, 
and we treat them only as detectors to measure aspects of $A$.  We also
ignore the possible twisting of the quasiparticles (this will be remedied
later).

Then the creation and annihilation of a quasiparticle/antiquasiparticle
pair forms a simple closed curve $\gamma$ in $X_3$.  The quasiparticle is labeled
by a representation $R$ of $G$, and the antiquasiparticle is labeled by the
dual representation $R^*$
(the appearance of representations is consistent since the quasiparticles 
are quantum mechanical objects put in by hand).  The measured observable
is defined to be the \textit{Wilson loop}
\begin{equation}
  W_R(\gamma,A):= \text{Tr}_R \text{Hol}_\gamma(A)
\end{equation}
the trace of the holonomy of $A$ around $\gamma$ in the representation $R$.

Now we wish to treat $A$ quantum mechanically (leaving the quasiparticles in
their semiclassical \textit{detector} roles).  It is clearer if we use
the path integral perspective.  Then the quantum observable
associated to a simple closed curve $\gamma$ colored with the species $R$
is a \textit{weighted average} of $W_R(\gamma,A)$ over all configurations of $A$:
\begin{equation}
  <W_R(\gamma,A)>:=\int_\mathscr{A} \mathscr{DA} \exp(2\pi i k S_{CS}) W_R(\gamma,A) 
\end{equation}
It is easy to generalize this to multiple link components with different colorings
$R_i$.  In the absence of link components we obtain a $3$-manifold invariant of $X_3$:
\begin{equation}
  Z(X_3):=\int_\mathscr{A} \mathscr{DA} \exp(2\pi i k S_{CS})
\end{equation}

Unfortunately the path integral quantization procedure is not typically calculable, is not
rigorously defined, and the quasiparticles have rather limited properties in this formulation
(i.e. no twisting and no fusing into composite quasiparticles).
Furthermore we have restricted ourselves to closed $3$-manifolds.
The ultimate remedy is a Hamiltonian quantization procedure involving
K\"ahler quantization (no quasiparticles) and conformal field theory (includes quasiparticles)
which we briefly discuss now.

\subsection*{Phase space}          
Now let us briefly recall some aspects of the phase space 
described in \cite{witten}.  It is simplest to first consider a theory
on $\Sigma\times I$ where $\Sigma$ is a closed oriented $2$-surface
\textit{without marked arcs}.
\footnote{Arcs become ribbons when propagated
in time - these are the worldlines of quasiparticles.  The marking (coloring) is
the particle species.}
As usual the canonical formalism
begins by describing the space of configurations on the initial time surface
$\Sigma\times\{0\}$.  For Chern-Simons the initial configuration is
a smooth Lie algebra-valued vector potential $A_2$ 
(a field configuration) on $\Sigma\times \lbrace 0\rbrace$.
\footnote{We note that a given $A_2$ configuration on the initial 
time surface
cannot be completely arbitrary because for some vector potentials
we would have no hope of solving forward to produce a solution
of the Euler-Lagrange equations on the whole 3-manifold.  
Hence we can only consider vector potentials on the
2-surface
that are subject to the \textit{Gauss law constraint}.}

Given an arbitrary field configuration $A_2$
on the initial time
slice $\Sigma\times \lbrace 0\rbrace$ (subject to the
appropriate constraints) we can use the
equations of motion to propagate it forward in time (producing
a vector potential configuration $A$ on the whole 3-manifold).
\footnote{Usually it is necessary to specify the initial field configuration
\textit{and} time derivative(s) in order to solve forward using
the equations of motion (since typically Euler-Lagrange equations are
second-order differential equations).  However (as we shall see)
the Euler-Lagrange equations are first-order for Chern-Simons, hence
the time derivatives are not necessary.}
In this way the different ``configuration spaces'' at different
time slices $\Sigma\times \lbrace t_1\rbrace$ and
$\Sigma\times \lbrace t_2\rbrace$ can be identified and we need only think of 
\textit{the} configuration space.  On the other hand the resulting
connection $A$ on the entire 3-manifold $\Sigma\times I$
(by construction) is a solution to the Euler-Lagrange equations, 
hence alternatively
we can view the ``configuration space'' as the space of solutions
to the Euler-Lagrange equations on the 3-manifold.

For Chern-Simons the Euler-Lagrange
equation
\begin{equation}
  F=0
\end{equation}
says that classically the allowed connections on
the 3-manifold  
$\Sigma\times I$ must be flat.  Hence naively the configuration
space should be the space $\mathscr{A}^{F=0}$ of flat vector potentials $A$ 
on $\Sigma\times I$.

However Chern-Simons has in addition the assumed mathematical
redundancy that defines it as a gauge theory, so instead
the configuration space is the space of flat vector potentials on
$\Sigma\times I$ modulo gauge transformations, the \textit{moduli space
of flat connections}
\begin{equation}
 \mathscr{M}:=\mathscr{A}^{F=0}/\sim
\end{equation}

Alternatively, we can work over the initial time slice
$\Sigma\times \lbrace 0\rbrace$ and consider the space
$\mathscr{A}_2$ of vector potentials over the 2-manifold that satisfy
the Gauss law constraint.
For Chern-Simons the Gauss constraint is easy - 
the curvature of an allowed configuration $A_2$
over $\Sigma\times \lbrace 0\rbrace$ must
vanish, i.e. $F_2=0$.  
When restricting a vector potential $A$ on the $3$-manifold
to the initial time slice $\Sigma\times \lbrace 0\rbrace$ we must
use up part of the gauge freedom in order to kill the
time component of the 1-form $A$.  This is \textit{temporal gauge}.

Even in temporal gauge
there is still gauge freedom left.  Modding out by this
residual gauge freedom we obtain the configuration space,
again called the \textit{moduli space of flat connections}
\begin{equation}
  \mathscr{M}:= \mathscr{A}^{F_2=0}_2/\sim_2
\end{equation}
We will freely switch back and forth between the two definitions
of configuration space.

\begin{example}
\label{ex:yangmills}
$\mathscr{M}$ was studied in \cite{atiyah_bott}
and \cite{jeffrey},
however there it arises from Yang-Mills theory on a 2-dimensional oriented
surface $\Sigma$ \textit{with Riemannian metric}.
\footnote{$\Sigma$ must have a metric because the Hodge star $(*)$ operator
is used in the Yang-Mills action.}

Since it will be useful later let us remind ourselves of
some elementary facts about Riemann surfaces (see e.g. \cite{schlichenmaier}).
For an oriented 2-surface $\Sigma$ the metric
induces a unique complex structure.
\footnote{One can define an \textit{almost} complex 
structure $J$ via the following map: for a tangent vector $\xi$, $J(\xi)$ is the unique vector
that is
\begin{enumerate}
  \item the same length as $\xi$,
  \item orthogonal to $\xi$,
  \item the pair $(\xi,J(\xi))$ has positive orientation.
\end{enumerate}
Any ``almost'' complex structure on a surface is integrable, so this
is actually a complex structure.
}
Conversely, the uniformization theorem says that a complex
structure on a 2-surface $\Sigma$ induces an
orientation and a \textit{class} of metrics that are all equivalent
up to local conformal transformations (angles are preserved, but not
necessarily lengths).  One of those has normalized constant scalar
curvature.
\footnote{normalized to $-1$, $0$, or $1$}

Hence for an orientable 2-surface we have a one-to-one
correspondence
\begin{equation}
  \text{complex structures}\leftrightarrow\text{conformal classes of metrics and orientations}
\end{equation}
Since in Yang-Mills $\Sigma$ is endowed with a Riemannian metric 
we might as well give $\Sigma$ the induced complex structure.

Let $E_2\rightarrow \Sigma$ be a
vector bundle with structure group $G$ on which Yang-Mills lives.
It is straightforward to show
that \textit{if the vector
bundle $E_2$ is trivial} then the 2-dimensional Yang-Mills
equations of motion are
\begin{equation}
 F_2=0 
\end{equation}
Modding out by gauge transformations we 
recover the moduli space of
flat connections $\mathscr{M}$.

However a \textit{flat} $G$-connection corresponds to
a homomorphism
\begin{equation}
  \pi_1(\Sigma)\rightarrow G
\end{equation}
since a connection can be encoded as monodromies along
paths ($E_2$ is a \textit{trivial bundle} so that the monodromy
along a non-closed path makes sense).  Two paths that
start at a point $x_1$ and end at a point $x_2$ form a loop,
and the difference in monodromies is just the holonomy around
the loop.
However, 
the holonomy of a \textit{flat} connection around a contractible 
loop is always the identity.  Using this it is easy to show
that a homotopy of a non-closed path (leaving the endpoints fixed) 
leaves the monodromy invariant.
Hence the space of flat $G$-connections (even before
modding out
by gauge transformations) is determined by the holonomies around
generators of $\pi_1(\Sigma)$.

As a very easy example consider $\Sigma=S^2$.  Then $\pi_1(S^2)=0$
hence there is only the trivial homormorphism
$\pi_1(\Sigma)\rightarrow G$.  Thus there is only
a single flat connection, so $\mathscr{M}$ is just a point.  In particular
we see that $\mathscr{M}$ is \textit{compact} and even-dimensional;
these are 
features that persist for general $\Sigma$.
\footnote{Here we can see that the assumed triviality of the 
vector bundle $E_2$ is essential.  If not then
we could consider the example $\Sigma=S^2$ and take as the
vector bundle the tangent bundle $TS^2$.  Give $S^2$ a metric (say
a metric of constant curvature 1 by thinking of $S^2$ as
standardly embedded in $\mathbb{R}^3$).  Then the tangent bundle
is an $SO(2)$-bundle.  Since $\pi_1(S^2)=0$ we might conclude
by the argument above that the tangent bundle admits a unique
flat connection.  However, the Gauss-Bonnet theorem implies that
no flat connection exists on $TS^2$ since the Euler characterstic
is $\chi(S^2)=2$ whereas the integral of a flat connection is
just $0$.  The resolution is that $TS^2$ is not trivial.
}
\end{example}

In other theories the Euler-Lagrange equations are typically second order differential
equations.  In the canonical formalism it is customary 
to formally pass to a first-order theory at the cost
of adding extra \textit{momentum} variables.  At the initial time
slice $\Sigma\times\lbrace 0 \rbrace$ the \textit{phase space} is
the space of allowed positions \textit{and} momenta, and
we propagate this phase space forward to any other time slice
using Hamilton's equations.

In Chern-Simons, however, the Euler-Lagrange equations are already first-order differential
equations.  Thus it is inappropriate to introduce auxiliary canonical
momenta (any attempt to do so will yield a constrained mechanical system where the
momenta $\Pi$ can be written in terms of the configuration variables $A$).  Hence the moduli space
of flat connections $\mathscr{M}$ (in addition to being the
configuration space) also plays the role of phase space
equipped with a symplectic structure and a Hamiltonian.

Let us remark briefly about the origin of the symplectic structure on $\mathscr{M}$.  
We refer the reader to \cite{jeffrey} for
more details.
First, in order
to be a symplectic manifold we need that $\mathscr{M}$ is
even dimensional.  Given the identification above of
a flat connection with a homomorphism
\begin{equation}
  \pi_1(\Sigma)\rightarrow G
\end{equation}
the dimension of $\mathscr{M}$ is $2g\cdot\text{dim }G$ where $g$ is the
genus of $\Sigma$, hence manifestly the dimension is even.

Second, consider the space of all $G$-connections 
$\mathscr{A}=\Omega^1(\Sigma,\mathfrak{g})$
over the 2-manifold $\Sigma$.
Since $\mathscr{A}$ is
an affine space (actually here it is a vector space because there is
a distinguished $A=0$ corresponding to the chosen standard flat
connection $D^0$), each tangent space $T\mathscr{A}_A$ can be 
identified with $\mathscr{A}$ itself.  Hence a symplectic form
on the \textit{manifold} $\mathscr{A}$ is determined by a symplectic form
on the \textit{vector space} $\mathscr{A}$.
A natural symplectic form is given by (up to normalization)
\begin{equation}
  \omega(A_1,A_2)=\int_\Sigma\text{Tr } A_1\wedge A_2  
\end{equation}
We leave it to the references
for proof that these statements descend to $\mathscr{M}$.

\subsection*{Prequantization}
We turn our attention towards quantization of the \textit{compact} symplectic
phase space $\mathscr{M}$.
However, we should expect difficulties since in other theories
typically phase space is non-compact.

Since $\mathscr{M}$ also plays the role of configuration
space we might try to make sense of $L^2(\mathscr{M},\mathbb{C})$.
\footnote{We feel that this would be an interesting problem to
compare in this context using \textit{spin networks}.  See for example
\cite{baez_spinnetworks} and \cite{baez_spinfoams}.}  
Indeed if $\text{dim}(\mathscr{M})=2n$ then we have
the usual Liouville volume form
\begin{equation}
  \text{vol}=(-1)^{n(n-1)/2}\frac{1}{n!}\omega\wedge\omega\wedge\ldots\wedge\omega
\end{equation}
where the wedge product is over $n$ copies of the symplectic
form $\omega$.

Hence we know how to integrate functions on $\mathscr{M}$, so
$L^2(\mathscr{M},\mathbb{C})$ is well-defined.  
Intuitively the number of
quantum basis wavefunctions should be proportional to the volume
(a quantum basis state corresponds to a box of side $\hbar$ in phase space).
Since $\mathscr{M}$ is compact the 
total volume of $\mathscr{M}$ is finite, hence we expect finitely-many
quantum basis wavefunctions.
Unfortunately, even though 
$\mathscr{M}$
is compact, $L^2(\mathscr{M},\mathbb{C})$ is
infinite dimensional.
\footnote{For example Fourier series 
provides a countably-infinite basis for functions on the compact manifold $S^1$.}
Therefore we assert that $L^2(\mathscr{M},\mathbb{C})$ is too large
to describe the quantum states.

The technique of geometric quantization \cite{woodhouse} provides 
a more rigorous quantization that agrees with our intuition.
We briefly describe the main ideas.

Instead of $L^2(\mathscr{M},\mathbb{C})$ we can consider $L^2$
sections of a hermitian line bundle $\mathscr{L}$ (a $U(1)$ bundle equipped with
a $U(1)$ covariant derivative $\nabla$) over $\mathscr{M}$.  Denote the
space of these $L^2$ sections $L^2(\mathscr{L})$.  We refer the
reader to pgs 16-18 of \cite{freed_chernsimons1} for the construction of 
$\mathscr{L}$ from the Wess-Zumino-Witten model.
%\footnote{There is a hermitian line bundle 
%$\mathscr{P}$ uniquely determined by the Weil construction:
%if $\frac{\omega}{2\pi}$ is \textit{integral}
%(lives in the image of 
%$H^2(\mathscr{M},\mathbb{Z})\rightarrow H^2(\mathscr{M},\mathbb{R})$) then
%$\mathscr{P}$ is determined with the constraint that
%the first chern class $F=-i\omega$.  Although we did not prove this, the 
%symplectic form $\omega$ defined above is integral.  For Chern-Simons at
%level $k$ we have that $\mathscr{L}\cong\mathscr{P}^{\otimes k}$.}

$L^2(\mathscr{L})$ is the \textit{prequantum Hilbert
space}.  Unfortunately (exactly as is the case for 
$L^2(\mathscr{M},\mathbb{C})$) $L^2(\mathscr{L})$ is infinite dimensional.
In order to shrink to a finite-dimensional physical Hilbert space 
it is instructive to recall that $\mathscr{M}$ \textit{also} 
plays the role of phase space.  In this light $L^2(\mathscr{L})$ is too large 
since it is analogous to ``$L^2(p,q)$'', i.e. $L^2$ functions 
on both the position and momentum variables.

\subsection*{K\"ahler quantization}
\textit{Choosing a polarization} is the process of
picking a foliation of $\mathscr{M}$ by leaves that are precisely
half the dimension of $\mathscr{M}$.  At a point $x\in\mathscr{M}$ 
the leaf $P$ that passes through $x$ determines locally a
``momentum'' submanifold  of $\mathscr{M}$.  The physical
Hilbert space is defined to be the subspace of $L^2(\mathscr{L})$ 
of sections that are \textit{constant} in the momentum
direction.

More precisely at $x\in P\subset\mathscr{M}$ the tangent space 
$TP_x\subset T\mathscr{M}_x$ must be a Lagrangian
subspace (maximal isotropic) with respect to the symplectic
form $\omega$, i.e. $TP_x$ is an $n$-dimensional subspace 
($\mathscr{M}$ is $2n$ dimensional) such
that if $A_1,A_2\in TP_x$ then
$\omega(A_1,A_2)=0$.
The physical Hilbert space is comprised of sections $s$ such that
$\nabla_{A} s=0$ for every $A\in \Gamma(TP)$.
There are several methods for choosing a polarization, however
each requires that we impose \textit{extra structure} on $\mathscr{M}$.

We now describe a similar method for reducing the phase space.
The idea is to equip
$\mathscr{M}$ with a complex structure $J$ and restrict
to \textit{holomorphic sections}.  
For technical reasons it is useful if $\mathscr{M}$ can be made K\"ahler.
We already have a symplectic form (possibly not normalized properly) 
\begin{equation}
  \omega(A_1,A_2)=\int_\Sigma\text{Tr } A_1\wedge A_2  
\end{equation}
and a choice of complex structure $J$.
Then $\mathscr{M}$ is K\"ahler if we define the 
Riemannian metric $g(A_1,A_2)$ to be
\begin{equation}
  g(A_1,A_2)=\omega(A_1,J\cdot A_2)
\end{equation}

Now shrink the prequantum Hilbert space using standard complex analysis:  
an almost complex structure is a (fiberwise) linear
map $J:T\mathscr{M}\rightarrow T\mathscr{M}$
that satisfies $J^2=-1$.
$T\mathscr{M}$ is a real vector bundle, but over the reals $J$ has no eigenvalues.
However, if we complexify $T\mathscr{M}$ (which doubles the real dimension) then
$T\mathscr{M}_\mathbb{C}$ splits into $\pm i$ eigenspaces of $J$,
i.e.
\begin{equation}
  T\mathscr{M}_\mathbb{C}=T\mathscr{M}^{(1,0)}\oplus T\mathscr{M}^{(0,1)}
\end{equation}
(we should also complexify the symplectic form 
$\omega_\mathbb{C}$ and the covariant derivative $\nabla^{\mathbb{C}}$ in
the line bundle $\mathscr{L}$).
Then the \textit{holomorphic sections} of $\mathscr{L}$ are sections $s$ such that 
$\nabla^{\mathbb{C}}_{A} s=0$ for every 
$A\in \Gamma (T\mathscr{M}^{(0,1)})$.  Define the physical Hilbert
space $\mathscr{H}$ to be the space of holomorphic sections of $\mathscr{L}$.

\subsection*{Extra assumption: complex structure on $\Sigma$}
The only issue left to resolve is the \textit{choice} of
complex structure $J$ on $\mathscr{M}$.  However, recall
that $\mathscr{M}$ is the moduli space of flat connections on
$\Sigma$.

Let us equip $\Sigma$ with a Riemannian metric.  Then 
there is an induced natural complex structure $J$
on the manifold $\mathscr{A}=\Omega^1(\Sigma,\mathfrak{g})$ that
can be seen as follows.  Since
$\mathscr{A}$ is an affine space (actually a vector space because
of the distinguished $A=0$ due to a choice of standard flat connection
$D^0$)
the tangent space $T_A\mathscr{A}$ at a point $A\in\mathscr{A}$ can be 
identified 
with the vector space
$\mathscr{A}$ itself.  Hence a complex structure on the manifold
$\mathscr{A}$ is determined by a linear operator $J$ acting
on the vector space $\mathscr{A}$ such that $J^2=-1$.  Such
a map is given by
\begin{equation}
  J(A)=*A
\end{equation}
where $*$ is the Hodge dual.  Because $\Sigma$ is 2-dimensional
it is trivial to verify that $J^2=(*)^2=-1$ on 1-forms - so this
defines a complex structure on $\mathscr{A}$ (which
descends to a complex structure on the moduli space $\mathscr{M}$).

The symplectic form is
\begin{equation}
  \omega(A_1,A_2)=\int_\Sigma\text{Tr } A_1\wedge A_2  
\end{equation}
and the complex structure
\footnote{Again we ignore integrability of this almost complex
structure.}
is defined by
\begin{equation}
  J(A):=*A
\end{equation}
Hence a K\"ahler structure on $\mathscr{A}$ is achieved by using
the Riemannian metric
\begin{equation}
  g(A_1,A_2)=\omega(A_1,J\cdot A_2)=\int_\Sigma\text{Tr } A_1\wedge *A_2  
\end{equation}
Passing to moduli space we obtain a K\"ahler structure
on $\mathscr{M}$.

\begin{example}
\label{example:complexstructure}
We note that the full strength of a Riemannian metric on $\Sigma$ 
is not required to produce the complex structure $J$\ on
$\mathscr{M}$.

Recall from example~(\ref{ex:yangmills}) that a given orientation and
Riemannian metric on a 2-surface $\Sigma$ induces a complex structure $j$ on $\Sigma$
(see below in local coordinates).  However, let us forget the Riemannian metric on $\Sigma$
and \textit{start} with
a complex structure $j$ on $\Sigma$.  Then $j$ induces a complex structure $J^\prime$ on the
affine manifold $\mathscr{A}=\Omega^1(\Sigma,\mathfrak{g})$
(since each tangent space $T_A\mathscr{A}$ is identified with
the vector space $\mathscr{A}$ itself).  Passing to the
moduli space we obtain a complex structure $J^\prime$ on $\mathscr{M}$.  

In local coordinates it is straightforward to see that $J^\prime$
is actually the \textit{opposite} complex structure to the $J$ defined
using a Riemannian metric on $\Sigma$ and the Hodge star operator 
(see \cite{griffiths_harris} for the relevant complex geometry).

For example consider the 2-dimensional plane $\mathbb{R}^2$ equipped with
the standard inner product and standard orientation.
Let us ignore the fact that the forms in $\mathscr{A}$ are $\mathfrak{g}$-valued.
Take the oriented
orthonormal basis
\begin{equation}
  \left(\frac{\partial}{\partial x},\frac{\partial}{\partial y}\right)
\end{equation}
The volume form for this orientation and metric is just $dx\wedge dy$, hence
the Hodge dual gives us 
\begin{equation}
  J(dx):=*dx=dy\quad\text{and}\quad J(dy)=*dy=-dx
\end{equation}

On the other hand the standard inner product on $\mathbb{R}^2$ induces a complex
structure map (a counterclockwise quarter turn)
\begin{equation}
  j\left(\frac{\partial}{\partial x}\right)=\frac{\partial}{\partial y}\text{    and    }
  j\left(\frac{\partial}{\partial y}\right)=-\frac{\partial}{\partial x}
\end{equation}
The dual of $j$ defines a linear operator $J^\prime$ on the space of 1-forms
$A\in\mathscr{A}$
\begin{equation}
  (J^\prime(A))\left(a\frac{\partial}{\partial x}+b\frac{\partial}{\partial y}\right):=
  A\left(j\left(a\frac{\partial}{\partial x}+b\frac{\partial}{\partial y}\right)\right)
\end{equation}
where $a$ and $b$ are real coefficients.
Using the above action of $j$ a quick calculation shows
\begin{equation}
  J^\prime(dx)=-dy\text{    and    }J^\prime(dy)=dx   
\end{equation}
which is clearly the opposite of $J$.

Hence if we complexify $\mathbb{R}^2$ then
the holomorphic differential $dz_{J^\prime}$ associated to $J^\prime$ is
equal to the antiholomorphic differential $d\bar{z}_J$ associated to $J$.
Let us complexify explicitly and produce the formulas for $J^\prime$
(then the reader can check that the corresponding formulas for
$J$ are the conjugates).  We have
\begin{equation}
  \mathbb{R}^2=\mathbb{R}\left\lbrace \frac{\partial}{\partial x},\frac{\partial}{\partial y} \right\rbrace
\end{equation}
Allowing complex coefficients gives
\begin{equation}
  \mathbb{C}^2=\mathbb{C}\left\lbrace \frac{\partial}{\partial x},\frac{\partial}{\partial y} \right\rbrace
\end{equation}
Define
\begin{align}
  \frac{\partial}{\partial z}&:=\frac{1}{2}\left(\frac{\partial}{\partial x}-i\frac{\partial}{\partial y}\right)\\
  \frac{\partial}{\partial \bar{z}}&:=\frac{1}{2}\left(\frac{\partial}{\partial x}+i\frac{\partial}{\partial y}\right)
\end{align}
Then it is easy to check (using the above formulas for $J^\prime$) that
\begin{align}
  J^\prime\left(\frac{\partial}{\partial z}\right)&=i\frac{\partial}{\partial z}\\
  J^\prime\left(\frac{\partial}{\partial \bar{z}}\right)&=-i\frac{\partial}{\partial \bar{z}}\\
\end{align}
So the holomorphic tangent space (relative to $J^\prime$) is just
\begin{equation}
  (T\mathbb{C}^2)^{(1,0)}:=\mathbb{C}\left\lbrace\frac{\partial}{\partial z}\right\rbrace
\end{equation}
and the antiholomorphic tangent space (relative to $J^\prime$) is just
\begin{equation}
  (T\mathbb{C}^2)^{(0,1)}:=\mathbb{C}\left\lbrace\frac{\partial}{\partial \bar{z}}\right\rbrace
\end{equation}
The same calculations end up conjugated when we use the complex structure $J$
instead.
\end{example}

In view of this example we do not need a
Riemannian structure on $\Sigma$ in order to K\"ahler
quantize, but merely a complex structure
$j$.  In the next section (using instead the conformal field theory approach) we 
dispense even with the complex structure.

\section{Conformal field theory}
\label{sec:CFT}
In the last section we outlined K\"ahler quantization
and described how to construct a finite-dimensional quantum Hilbert
space $\mathscr{H}$ associated to the initial time slice $\Sigma\times\{0\}$.
In the Hamiltonian formalism (on the manifold $\Sigma\times I$) 
$\mathscr{H}$ is evolved forward using the Hamiltonian $H$.
However it is easy to verify that for Chern-Simons $H=0$.  There are no dynamics
on $\Sigma\times I$ where $(\Sigma,j)$ is a \textit{closed} Riemann surface, 
hence we conclude that K\"ahler quantization is rather mundane.
Furthermore the chiral WZW action appears on the boundary in Chern-Simons, but
this was not used in K\"ahler quantization.  Motivated by
this we turn to the richer structure provided by conformal field theory (which
agrees with K\"ahler quantization on closed Riemann surfaces $\Sigma$ \cite{beauville_laszlo}).

A detailed analysis of the Wess-Zumino-Witten model is provided in
(for example) \cite{kohno}.
\footnote{The strategy for the WZW model is to first \textit{avoid} closed
surfaces and instead study the WZW action on Riemann surfaces $(\Sigma,j)$ with
\textit{at least} one boundary circle.  The WZW action is not a priori 
well-defined on Riemann surfaces with boundary, however a study of the
unit disk $(\Sigma,j)=D$ yields a construction based on a central extension of
the loop group.  Gluing laws can then be defined.  In particular this defines the
theory on \textit{closed} Riemann surfaces since any such surface can be decomposed into 
two surfaces glued along nonemtpy boundary.} 
However here we restrict ourselves to the axiomatic framework described in \cite{segal}.
The most primitive notion introduced by Segal is a \textbf{modular functor}.
\footnote{A modular functor is part of the underlying structure of
a \textit{chiral} conformal field theory
(a \textit{weak} conformal field theory in the language of \cite{segal}). 
Given two opposite-chirality weak conformal field theories based on the
same \textit{unitary} modular functor it is possible to combine them
to form an honest conformal field theory.  Since it is a chiral theory that appears in
Chern-Simons we restrict our attention to the modular functor.}
We mention that in the following we consider Riemann surfaces with labeled (colored) 
boundary circles.
A boundary circle should be interpreted as the boundary of an excised disk containing a quasiparticle,
and the color
specifies the particle species.  In addition we require that the boundary circles 
be \textit{parameterized}.
To make contact with our previous
characterization of quasiparticles (and remain consistent with other treatments
(see chapter 5 in \cite{turaev} and chapter 5 in \cite{bakalov_kirillov})
it is not necessary to parameterize boundary circles, but rather
merely select a \textit{basepoint} on each boundary circle.
\footnote{It is clear that a circle $S^1$ parameterized by a
diffeomorphism $S^1\rightarrow U(1)$ 
has a distinguished basepoint
(e.g. the preimage of $\lbrace 1\rbrace$ for example).  
However the space $\text{Diff}^+_\text{pt}(S^1)$ of all (orientation preserving)
diffeomorphisms that 
share the same basepoint is contractible.  Below we shall only be
concerned with $\pi_1$ of the various spaces that appear, hence
only the parameterization up to homotopy is important.}
A third
alternative is to shrink each circle
to a \textit{marked point with distinguished tangent vector} on a \textit{closed} surface $\Sigma$.
These are \textbf{marked arcs}.  However in CFT the boundary circles play a
richer role - on the one hand they are quasiparticles, but on the other hand
Riemann surfaces can be glued together along parameterized boundary
circles (which cannot be done with marked arcs).

\begin{definition}
Let $\phi$ be a finite set of labels (particle species).  Define a
category $\mathscr{G}_\phi$ as follows:
\begin{enumerate}
 \item An object is a compact 
Riemann surface $(\Sigma,j)$ of arbitrary topological type, and possibly with
many connected components and parameterized boundary circles.  The
boundary circles are labeled (\textbf{colored}) with elements from $\phi$. 
If the orientation induced by the parameterization agrees with the boundary orientation then
the circle is \textit{outgoing}.  If they disagree then the circle is \textit{incoming}.
 \item A morphism
$(\Sigma,j)\rightarrow(\overline{\Sigma},\overline{j})$ takes a Riemann surface $(\Sigma,j)$ with an
outgoing and an incoming boundary circle labeled by the same color $i\in\phi$ and glues them along
the parameterizations to form a new Riemann surface $(\overline{\Sigma},\overline{j})$ with
two fewer boundary circles.
\end{enumerate}
\end{definition}

\begin{definition}
A \textbf{Segal modular functor} is a functor
\footnote{We note that a Segal modular functor is stronger
than the modular functor defined
later in this treatment.  A Segal modular functor is defined in
terms of Riemann surfaces, boundary circles can
be glued, and is valid in
2 dimensions only.  

However, the dependence on the complex structure of a Riemann surface $\Sigma$ can
be relaxed.  Presumably then a Segal modular functor is equivalent to
an \textit{extended $2$-d modular functor} as discussed in 
chapter 5 of \cite{turaev} and chapter 5 of \cite{bakalov_kirillov}.  Because of
the gluing property an extended $2$-d modular functor is stronger than
a modular functor defined below (and in chapter 3 of \cite{turaev}).}
\begin{equation}
  \modfunct:\mathscr{G}_\phi\rightarrow\text{finite dimensional complex
    vector spaces}
\end{equation} 
that assigns to a Riemann surface $(\Sigma,j)$ with colored parameterized boundary
a complex vector space $\modfunct((\Sigma,j))$ (not a Hilbert space in general).  This
functor must satisfy
\begin{enumerate}
  \item $\modfunct$ is a \textit{holomorphic functor} (see below)
  \item $\modfunct((\Sigma,j) \coprod (\Sigma^\prime,j^\prime))=\modfunct((\Sigma,j))\otimes
    \modfunct((\Sigma^\prime,j^\prime))$
  \item For the Riemann sphere $\text{dim}(\modfunct(S^2))=1$
  \item Consider cutting a Riemann surface $(\Sigma,j)$ along a parameterized 
    simple closed curve to
    produce a new surface with two more boundary circles (one incoming and
    one outgoing).  Let us color
    both circles with a color $i$ from the finite set of colors $\phi$.  Denote
    this new Riemann surface by $(\Sigma_i,j)$.  We could think about sewing this
    back together, which by definition is just a morphism $f_i:(\Sigma_i,j)\rightarrow (\Sigma,j)$
    (a gluing).
    The functor then gives a linear map 
    $\modfunct(f_{i}):\modfunct((\Sigma_i,j))\rightarrow\modfunct((\Sigma,j))$.
    Summing over all colors we require that the map
\begin{equation}
   \bigoplus_{i\in \phi} \modfunct((\Sigma_i,j))\rightarrow
     \modfunct((\Sigma,j))
\end{equation}
    be a natural isomorphism. 
\end{enumerate}
\end{definition}
In order to define holomorphic functor we mention some more
standard results from complex geometry.
Consider the space $\mathscr{J}(\Sigma)$ of \textit{all} 
complex structures on $\Sigma$ ($\Sigma$ is
a smooth manifold possibly with colored parameterized boundary).  
In other words
$\mathscr{J}(\Sigma)$ is the space of all Riemann surfaces
that are topologically diffeomorphic to $\Sigma$.
$\mathscr{J}(\Sigma)$ is a contractible topological space (consider
the space of smoothly-varying matrices $j(x)$ for $x\in\Sigma$ such
that $j^2=-1$.  This space is contractible in 2 dimensions).

Two Riemann surfaces $(\Sigma,j_1)$ and $(\Sigma,j_2)$ of the
same topological type are \textit{equivalent} if there is an orientation-preserving
diffeomorphism $\phi:\Sigma\rightarrow\Sigma$ that maps
$j_1$ to $j_2$ (i.e. a biholomorphic map).  If $\Sigma$ has boundary then we assume that
$\phi$ maps circles to circles respecting the parameterizations.
The resulting space $\mathscr{J}(\Sigma)/\sim$ is the \textit{moduli space}
$\mathscr{C}_\Sigma$ (see \cite{schlichenmaier} - except note that in
contrast to other treatments here any boundary components are parameterized).

A functor $\modfunct$ is \textbf{holomorphic} if the complex vector spaces 
$\modfunct((\Sigma,j))$ assigned to
Riemann surfaces $(\Sigma,j)$ of a given topological type $\Sigma$
smoothly vary as the complex structure $j$ varies.  More precisely,
$\modfunct$ is holomorphic if we obtain a holomorphic vector bundle 
$\modfunct(\mathscr{C}_\Sigma)\rightarrow\mathscr{C}_\Sigma$.

Consider the following (proposition 5.4 in \cite{segal}):
\footnote{Unfortunately Segal avoids
proving this for \textit{closed} oriented surfaces $\Sigma$ since
then the moduli space $\mathscr{C}_\Sigma$
may have singularities.  We ignore this
source of complication.}
\begin{proposition}[Segal] Associated to any arbitrary modular functor $\modfunct$ is
 a canonical flat connection on the \textbf{projective} bundle
 $\mathbb{P}\modfunct(\mathscr{C}_\Sigma)\rightarrow\mathscr{C}_\Sigma$
\end{proposition}

This implies that we can identify
the projective vector spaces $\mathbb{P}\modfunct((\Sigma,j_1))$ and 
$\mathbb{P}\modfunct((\Sigma,j_2))$ 
\textit{once a path has been specified} in
$\mathscr{C}_\Sigma$ from $(\Sigma,j_1)$ to $(\Sigma,j_2)$.
Since the connection is \textit{flat} only the homotopy type of
the path is relevant.  

Choose a complex structure $(\Sigma,j)$ and associate to $\Sigma$
the vector space
\begin{equation}
  \mathscr{H}:=\modfunct((\Sigma,j))
\end{equation}
From the comments above $\mathscr{H}$ is a projective representation
of $\pi_1(\mathscr{C}_\Sigma)$, and the choice of $j$
is equivalent to the choice of basepoint for $\pi_1(\mathscr{C}_\Sigma)$.  
Let us now study $\pi_1(\mathscr{C}_\Sigma)$.

\begin{example}
\label{ex:modulispaceclosed}
It is a standard result that when $\Sigma$ is a
\textit{closed} oriented surface then $\mathscr{C}_\Sigma$ is a finite-dimensional
complex variety but perhaps \textit{with singularities}.

For the Riemann sphere the moduli space $\mathscr{C}_{S^2}$ is a point (there is
a unique Riemann sphere $\mathbb{C}\cup\lbrace\infty\rbrace$ up to 
automorphisms of the complex structure via the action of $\text{PSL}(2,\mathbb{C})$).

A closed genus 1 surface is obtained from the complex plane $\mathbb{C}$ 
in the usual way by identifying points related by translations using a rank 2 lattice
$\mathbb{Z}\oplus\mathbb{Z}$.  Explicitly we identify 
$z\mapsto z+1$ and $z\mapsto z+\tau$ where $\text{Im}(\tau)>0$.
\footnote{
We see that the universal covering space of a torus
is just $\mathbb{C}$.
The automorphisms (transformations that preserve
the complex structure) of $\mathbb{C}$ are
just the affine transformations $z\mapsto az+b$ where $a,b\in\mathbb{C}$ and
$a\neq 0$.  Using these automorphisms we can transform a given lattice
generated by arbitrary vectors $\alpha$ and $\beta$ into a unique lattice generated by
vectors of the form $1$ and $\tau$ with $\text{Im}(\tau)>0$.  The resulting
complex structure on the torus is unaffected.
}
Hence the complex tori are determined by a choice of $\tau\in U$
in the upper half plane.

However given a fixed lattice in $\mathbb{C}$ 
even a basis of the form $(\tau,1)$ is not unique.
We can apply a unimodular matrix
$\bigl( \begin{smallmatrix} a&b\\c&d\end{smallmatrix}\bigr)
\in \text{SL}(2,\mathbb{Z})$ 
\footnote{Explicitly $a, b, c, d\in\mathbb{Z}$ and $ad-bc=1$}
to the basis $(\tau,1)$ to give
a new basis $(a\tau+b,c\tau+d)$ for the same lattice.
Next let us again use the automorphisms of the complex plane 
(affine transformations)
to put this new basis back into the form $(\tau^\prime,1)$.  A
small calculation shows that
\begin{equation}
  \tau^\prime=\frac{a\tau_1+b}{c\tau_1+d} 
\end{equation}
We note that we can multiply both numerator and denominator
in the above equation by $-1$ and still get the same $\tau^\prime$,
hence we need only consider projective unimodular matrices
$\text{PSL}(2,\mathbb{Z})$.
Summarizing, two complex tori are equivalent if related by a transformation in
$\text{PSL}(2,\mathbb{Z})$ acting
on the upper half plane $U$.
Since $\text{PSL}(2,\mathbb{Z})$ is discrete we have that the action on
$U$ is \textit{discontinuous} in the sense of \cite{farkas_kra} pg. 203.
It can be shown that \textit{as a naive set}
\begin{equation}
  \mathscr{C}_T\cong U/\text{PSL}(2,\mathbb{Z})=\mathbb{C} 
\end{equation}

However, we note that the above action is \textit{not free}.
Hence $\mathscr{C}_T$ is not a smooth manifold, but in fact
has 2 singular points with extra internal structure.
In other words the moduli space $\mathscr{C}_T$ is
a \textit{stack} and it is \textit{not true} that
$\pi_1(\mathscr{C}_T)=\pi_1(\mathbb{C})=1$.  In fact
it turns out (for a suitably-defined definition of
the fundamental group) that
$\pi^{\text{stack}}_1(\mathscr{C}_T)\cong \text{MCG}(T)\cong\text{PSp}(2,\mathbb{Z})
\cong \text{PSL}(2,\mathbb{Z})$
where $\text{MCG}(T)$ is the mapping class group of the
torus.

For closed higher genus ($\geq 2$) surfaces a similar result holds (technically
the construction is easier because a \textit{fine} moduli space
can be extracted from the coarse moduli space).  It happens
that $\text{dim}_\mathbb{C}\mathscr{C}_{\Sigma}=3g-3$, but
again there are singularities which force us to treat
$\mathscr{C}_\Sigma$ as a stack.  
It turns out that again
\begin{equation}
  \pi^{\text{stack}}_1(\mathscr{C}_{\Sigma})\cong \text{MCG}(\Sigma)
\end{equation}
We refer the reader to chapter 6 of \cite{bakalov_kirillov}.
\end{example}

\begin{example}
Now let us consider surfaces with \textit{parameterized} holes.
\footnote{from now
on by ``hole'' we mean a removed open disk, i.e. $\Sigma$ has parameterized
boundary circles.  This is in contrast to a puncture, i.e. a removed point - see \cite{farb_margalit}
pg. 64}
In this case there are no singularities
in $\mathscr{C}_\Sigma$.
\footnote{we emphasize that the
boundary here is parameterized.  For a constrasting example
suppose $\Sigma$ is an annulus with \textit{unparameterized boundary}.  
Then the moduli space is the real 
interval $(0,1)$, which disagrees with the result stated here.  See \cite{farkas_kra} page 211}
The $k$-holed sphere requires 
special treatment and must be dealt with separately in the three regimes
$k=1$, $k=2$, and $k\geq 3$.  The $k$-holed torus also must be analyzed
by hand in the regimes $k=1$ and $k\geq 2$.  Higher genus ($g\geq 2$) surfaces can
be dealt with uniformly, although much is still unknown.
\footnote{see \cite{birman}, although here we have the additional complication
of parameterized holes rather than simple punctures}
We start with the sphere.

First, let us consider the sphere with one parameterized hole, i.e. the unit 
disk $\Delta=\lbrace z\colon |z|\leq 1 \rbrace$.
\footnote{By the classification of \textit{exceptional} Riemann surfaces the only
simply connected Riemann surfaces are $\mathbb{C}\cup\lbrace\infty\rbrace$,
$\mathbb{C}$, and $\Delta=\lbrace z\colon |z|\leq 1 \rbrace$.  
Hence there is only one ``disk'' to consider
here.  See \cite{farkas_kra} pg. 207}
The unit disk conformally maps
to the upper half plane $U$ via the map $z\mapsto i\frac{1-z}{1+z}$, and the
upper half plane has a unique complex structure, hence there is a unique
complex structure on the unit disk $\Delta$.  So we expect that
$\mathscr{C}_\Delta\cong\lbrace\text{pt}\rbrace$.  However, we have forgotten
about the parameterization of the boundary $S^1$ so we must take into
account the group $\text{Diff}^+(S^1)$.  To make the analysis easier
for our purposes it suffices
to think about the boundary with a distinguished
basepoint (rather than a full parameterization).  Hence let us
consider the upper half plane $U$ with a distinguished
basepoint on the real axis.

The automorphism group (the group that preserves the complex structure)
of the upper half plane $U$ is just
$PSL(2,\mathbb{R})$.  In particular we can think about the affine transformation
$z\mapsto z+a$ for any real number $a$.  But this maps any choice
of basepoint on the real axis to any other choice of basepoint, 
so we conclude that the choice of basepoint is irrelevant.
\footnote{More trivially instead we could just think of rigid rotations
acting on the unit disk $\Delta$ (these preserve the
complex structure).  Any arbitrary basepoint on the boundary circle can
be rotated to the point $z=1$.}
Hence
even with a parameterized boundary we have
$\mathscr{C}_\Delta\cong\lbrace\text{pt}\rbrace$.
\end{example}

\begin{example}
Now consider a sphere with two parameterized holes (an annulus).
Again by the classification for \textit{exceptional} Riemann 
surfaces the only Riemann surfaces with $\pi_1(\Sigma)\cong\mathbb{Z}$
are $\mathbb{C}\setminus\lbrace 0\rbrace$, $\Delta\setminus\lbrace 0\rbrace$,
and the family of standard annuli $\Delta_r=\lbrace z\in\mathbb{C}\colon r\leq|z|\leq 1\rbrace$ 
where $r\in(0,1)$ (i.e. \textit{all} annuli
are just standard annuli).  Hence (as we have already mentioned) the 
moduli space of complex
annuli $\mathscr{C}_{\Delta_r}$ is just the interval $(0,1)$.  Here again,
however, we
have forgotten the boundary parameterizations.  Like before (and from now
on) we do not consider the full parameterizations, but rather a distinguished basepoint 
on each boundary circle.  It
is clear that we can perform a rigid rotation (which preserves the
complex structure on the annulus) to rotate any arbitrary basepoint on 
the outer circle $\lbrace z\in\mathbb{C}\colon |z|=1\rbrace$ to
the point $z=1$, hence the choice of basepoint on the outer circle is 
irrelevant.

Now we have used up the rigid rotation automorphism (which is the only
automorphism of an annulus) hence we cannot dispense with
the choice of basepoint on the inner circle (we have a whole $S^1$
worth of choices).
In view of this we see that the moduli space of annuli (with parameterized
boundary) is just $\mathscr{C}_{\Delta_r}\cong(0,1)\times S^1$.
\footnote{This is merely a homotopy equivalence because we are
considering basepoints rather than parameterizations.}
There are no singularities nor stack structure, hence
we directly calculate 
$\pi_1(\mathscr{C}_{\Delta_r})\cong\mathbb{Z}\cong \text{MCG}_\partial(\Delta_r)$.
\footnote{This is an enlarged mapping class group for surfaces with \textit{basepointed}
boundary circles.  In this case a Dehn twist in a collar neighborhood of a
boundary circle is a non-trivial element
of the mapping class group.  If the boundary circles were not parameterized/basepointed then
such a Dehn twist could be smoothly deformed (untwisted) back to the identity.}
See figure~(\ref{fig:ccwtwist})
\begin{figure}
  \centering
  \input{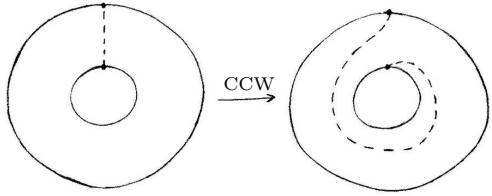}
  \caption{A counterclockwise twist of a boundary circle with respect
    to a second boundary circle.  Instead of using parameterizations we
    depict distinguished basepoints.  We provide visual markings to
    show the diffeomorphism.}
  \label{fig:ccwtwist}
\end{figure} 
\end{example}

\begin{example}
More generally recall that we saw for closed surfaces 
$\pi^{\text{stack}}_1(\mathscr{C}_\Sigma)\cong \text{MCG}(\Sigma)$.
For compact oriented surfaces with
$k\geq 1$ holes we now sketch that the same result is true 
although the presence of parameterized boundary circles enlarges the mapping
class group considerably (for a detailed account see \cite{birman} and
\cite{farb_margalit}).

In order to understand $\pi_1(\mathscr{C}_\Sigma)$ 
let us remind ourselves that previously we obtained the moduli
space $\mathscr{C}_\Sigma$ from $\mathscr{J}(\Sigma)$ by identifying
any two Riemann surfaces $(\Sigma,j_1)$ and $(\Sigma,j_2)$ if there is a 
biholomorphic diffeomorphism $\Sigma\rightarrow\Sigma$
mapping one complex structure to the other.  Now we have boundary circles (equipped
with basepoints)
hence we further require that any diffeomorphism maps basepoints to
basepoints.  Denote this space of biholomorphic basepoint-preserving
diffeomorphisms $\text{Diff}_\partial^+(\Sigma)$.
\footnote{obviously the diffeomorphism must preserve orientation as well.}

Rather than mod out by \textit{all} such biholomorphic diffeomorphisms
let us consider a weaker notion of equivalence by defining 
\textbf{Teichm\"uller space} $\mathscr{T}_\Sigma$ where
we identify any two Riemann surfaces if there is a biholomorphic 
diffeomorphism in $\text{Diff}_\partial^+(\Sigma)$ 
\textit{that can be smoothly deformed to the identity}
(clearly such diffeomorphisms
must be the identity on each boundary circle separately).  Denote this
restricted subset $\text{Diff}^+_{\partial,0}(\Sigma)\subset\text{Diff}^+_{\partial}(\Sigma)$. In
symbols we have
\begin{equation}
  \mathscr{C}_\Sigma:=\mathscr{J}(\Sigma)/\text{Diff}_\partial^+(\Sigma) 
\end{equation}
and
\begin{equation}
  \mathscr{T}_\Sigma:=\mathscr{J}(\Sigma)/\text{Diff}^+_{\partial,0}(\Sigma) 
\end{equation}

On the other hand by definition $\text{Diff}^+_\partial(\Sigma)/\text{Diff}^+_{\partial,0}(\Sigma)$ 
is the
mapping class group $\text{MCG}_\partial(\Sigma)$,
so we see that
\begin{equation}
  \mathscr{C}_\Sigma=\mathscr{T}_\Sigma/\text{MCG}_\partial(\Sigma) 
\end{equation}
In this context the mapping class 
group is often called the \textbf{Teichm\"uller group} $\text{Teich}(\Sigma)$.

In the language of covering space theory we can view 
Teichm\"uller space as a covering of moduli space
\begin{equation}
  \mathscr{T}_\Sigma\rightarrow \mathscr{C}_\Sigma 
\end{equation}
where the deck transformations are just given by elements of
$\text{MCG}_\partial(\Sigma)$.
The usual covering space results tell us that
\footnote{We can use this result if the deck group action is
free, which is evidenced by the fact that the resulting quotient manifold
$\mathscr{C}_\Sigma$ has no singularities.}
\begin{align}
   \text{Deck Transformations}&\cong\pi_1(\mathscr{C}_\Sigma)/\pi_1(\mathscr{T}_\Sigma)\\
      \text{MCG}_\partial(\Sigma)&\cong\pi_1(\mathscr{C}_\Sigma)/1
\end{align}
In the above equation we have used the fact that $\mathscr{J}(\Sigma)$ is
actually a contractible space, and since modding out by diffeomorphisms that
can be deformed to the identity does not change the homotopy type,
we see that the Teichm\"uller space $\mathscr{T}_\Sigma$ is
also contractible.  So $\pi_1(\mathscr{T}_\Sigma)=1$.
\footnote{
This explains the somewhat interchangeable roles that $\pi_1(\mathscr{C}_\Sigma)$,
$\text{MCG}_\partial(\Sigma)$, and $\text{Teich}(\Sigma)$ play in the literature.}
\end{example}

\begin{example}
We have shown that for arbitrary compact oriented surfaces with/without
parameterized holes that
\begin{equation}
 \pi_1(\mathscr{C}_\Sigma)\cong\text{MCG}_\partial(\Sigma)
\end{equation}
for suitably defined fundamental group and mapping class group.  Hence
it is worthwhile to study $\text{MCG}_\partial(\Sigma)$ a bit further.
We already mentioned the explicit results for the sphere with $k=0$,
$k=1$, and $k=2$ punctures.  We also mentioned that for the
closed torus $\text{MCG}_\partial(T)\cong \text{PSL}(2,\mathbb{Z})$.
 
Now consider a special family of examples - the \textit{unit disk} 
with $k\geq 2$ parameterized holes in the interior.  This is \textit{not}
the sphere with $k+1$ holes because here the
outer ($k+1$)st boundary circle is considered \textit{distinguished and fixed}.
These disks can be used as building blocks to analyze certain aspects of
all surfaces.

For concreteness consider the two-holed disk ($k=2$)
$\Delta_2$
embedded in $\mathbb{R}^2$ using whatever standard
embedding that the reader prefers (see the left disk in 
figure~(\ref{fig:ccwbraid}) for our convention).
\begin{figure}
  \centering
  \input{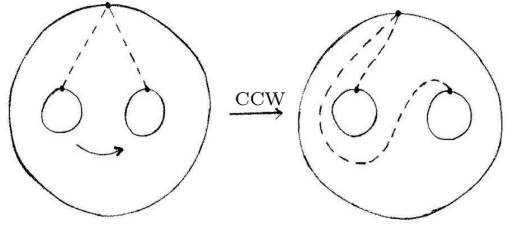}
  \caption{A counterclockwise braiding of two boundary circles with respect
    to the distinguished outer boundary circle.  Instead of using parameterizations we
    depict distinguished basepoints.  We provide visual markings to
    show the diffeomorphism.}
  \label{fig:ccwbraid}
\end{figure}

Now consider the counterclockwise \textit{braiding} diffeomorphism 
$c:\Delta_2\rightarrow\Delta_2$ depicted in figure~(\ref{fig:ccwbraid}).
This is a diffeomorphism of $\Delta_2$ that cannot be smoothly deformed to the
identity, hence is a nontrivial element of the mapping class group.
More generally for a disk with $k$
parameterized holes we expect that the \textit{braid group} $B_k$ on $k$ strands 
is a subgroup of $\text{MCG}_\partial(\Sigma)$.

Likewise each of the interior holes can be (separately) \textit{twisted} 
via a full counterclockwise Dehn twist $\theta_i:\Delta_2\rightarrow\Delta_2$
for $i=1,...,k$ (see figure~(\ref{fig:ccwtwist}) for the case $k=1$).
Hence we convince ourselves that $\mathbb{Z}^k$ is a subgroup of
$\text{MCG}_\partial(\Sigma)$.

It is not difficult to see that a braiding operation, followed by any
twist operation, followed by the inverse braiding operation, can be
written as a different twist operation.  In other words 
$B_k$ is in the normalizer for $\mathbb{Z}^k$.

In light of this
it is not surprising that $\text{MCG}_\partial(\Sigma)$
is the semidirect product of $\mathbb{Z}^k$ with $B_k$:
\begin{equation}
  \text{MCG}_\partial(\Sigma)\cong \mathbb{Z}^k\rtimes B_k 
\end{equation}
 
We have been incomplete in our analysis, however, since we have
forgotten that in conformal field theory each boundary component
must be labelled by a \textit{color} from a finite set.  It only
makes sense to swap holes that have the same coloring, so
we are forced to consider instead of the full braid group
$B_k$ the \textit{colored braid group} $CB_k$.  So we have
\begin{equation}
  \text{MCG}_\partial(\Sigma)\cong \mathbb{Z}^k\rtimes CB_k 
\end{equation}
\end{example}

\begin{example}
Now consider the special case of the sphere with $k$ parameterized holes.  
It is fairly trivial to
analyze this case by
excising a \textit{special disk} (from the last example) that contains 
all of the holes.
The result is two pieces - a disk $\Delta$ and a disk 
$\Delta_k$ with $k$ holes.  
Then the mapping class group is
\begin{equation}
  \text{MCG}_\partial(\Sigma)\cong (\mathbb{Z}^k\rtimes CB_k)/\mathbb{Z}_{\text{everything}}
\end{equation}
$\mathbb{Z}_{\text{everything}}$ is the subgroup of $\mathbb{Z}^k\rtimes CB_k$ 
generated by the central element that takes 
a full Dehn twist of the entire interior of $\Delta_k$ (leaving the outer
circle fixed, of course).  When the disks are glued together this Dehn twist can be
pushed onto $\Delta$ instead, and any Dehn twist of $\Delta$
can be smoothly deformed to the identity.  So we conclude that $\mathbb{Z}_{\text{everything}}$ is
trivial for the sphere with holes.

For example, for $k=2$ holes (with the same coloring) the braid group
$B_2$ becomes the symmetric group $S_2$ when modding out by $\mathbb{Z}_{\text{everything}}$.
\end{example}

$\text{MCG}(\Sigma)$ in genus $g\geq 1$ is significantly more complicated and much
is not known.  We refer
the reader to \cite{farb_margalit}.

\section{Axiomatic definition of an $(n+1)$-dimensional TQFT}
The axioms for an $(n+1)$-dimensional TQFT were originally proposed by Atiyah (see, e.g., \cite{atiyah_book}).  They appear
in various incarnations throughout the literature, but we follow
chapter 3 of \cite{turaev}.

\subsection*{Modular functor}
Consider the category $\mathscr{U}$ defined by
\begin{enumerate}
 \item The objects are (possibly \textit{extended}) $n$-dimensional closed oriented
manifolds $\Sigma$.
We are interested in the case $n=2$, and for us the extended structure on
a closed genus $g$ surface $\Sigma$ is
a \textit{parameterization} diffeomorphism
\begin{equation}
  \phi:\Sigma^\text{standard}_g\rightarrow \Sigma
\end{equation}
where $\Sigma^\text{standard}_g$ is a fixed genus $g$ surface.
\footnote{The parameterization can be relaxed to a much weaker extended structure.
    See \cite{atiyah},\cite{walker},\cite{freed_gompf}.}
 \item The morphisms are orientation-preserving
diffeomorphisms $\Sigma\rightarrow\Sigma^\prime$.
\end{enumerate}
$\mathscr{U}$ has a canonical commutative strict monoidal structure (see chapter~(\ref{ch:mtc})):
\begin{enumerate}
 \item The tensor product is given by disjoint union:
\begin{equation}
  \Sigma\boxtimes\Sigma^\prime:=\Sigma\sqcup\Sigma^\prime
\end{equation}
 \item The unit object $\unitobj$ is the empty set $\varnothing$ (since
$\Sigma\sqcup\varnothing=\Sigma$).
 \item $\mathscr{U}$ is commutative, i.e. $\Sigma\sqcup\Sigma^\prime=\Sigma^\prime\sqcup\Sigma$.
\end{enumerate}

Now consider the category $\text{Vect}^{\text{fin}}_{\cplx}$ of
finite-dimensional complex vector spaces.  This
is also a commutative strict monoidal category (using the ordinary vector space
tensor product $\otimes$).  The unit object here is $\cplx$.

\begin{definition}
A \textbf{modular functor} $\modfunct$ is a covariant strict monoidal functor
(see chapter~(\ref{ch:groupcategories}))
\begin{equation}
  \modfunct:\mathscr{U}\rightarrow\text{Vect}^{\text{fin}}_{\cplx}
\end{equation}
\end{definition}

In other words, 
to each $n$-dimensional extended closed oriented manifold
$\Sigma$ we assign a vector space $\modfunct(\Sigma)$:
\begin{equation}
  \xymatrix {
  \Sigma\ar@{|->}[r]^-{\modfunct} & \modfunct(\Sigma) \\
  }
\end{equation}
To each orientation-preserving diffeomorphism $f:\Sigma\rightarrow\Sigma^\prime$
we assign a vector space isomorphism 
$\modfunct(f):\modfunct(\Sigma)\rightarrow\modfunct(\Sigma^\prime)$ (which we denote
$f_\sharp$):
\begin{equation}
  \xymatrix {
  f\ar@{|->}[r]^-{\modfunct} & f_\sharp \\
  } 
\end{equation}
Functoriality means $(fg)_\sharp = f_\sharp g_\sharp$ and 
$\id_\Sigma \mapsto \id_\sharp = \id_{\modfunct(\Sigma)}$.

Being a strict monoidal functor means that in addition
\begin{equation}
   \modfunct(\Sigma\sqcup\Sigma^\prime)=\modfunct(\Sigma)\otimes\modfunct(\Sigma^\prime)
\end{equation}
There are extra associativity and naturality axioms for strict monoidal functors
that can be found in chapter~(\ref{ch:groupcategories}).
Most notably we have the identity assignment
\begin{equation}
  \modfunct(\varnothing) = \cplx
\end{equation}

It is interesting to contrast with the Segal modular functor in section~\ref{sec:CFT}.  
Most conspicuous is the lack of gluing in this version.  
An $n=2$ modular functor as defined here is weaker than a Segal modular functor.
\footnote{The nomenclature is confusing.  In chapter 5 of \cite{turaev} is described
a so-called \textit{$2$-d modular functor}.  The construction has much more structure
than a modular functor in $2$ dimensions (as defined here and in chapter 3 of \cite{turaev}).
Following \cite{bakalov_kirillov} we prefer to 
call the stronger version an \textbf{extended $2$-d
modular functor}.  Presumably extended $2$-d modular functors are in
one-to-one correspondence with the Segal modular functors defined above.}
We mention that the extended structure on $\Sigma$ for the case $n=2$ 
can be weakened to a choice of distinguished 
Lagrangian subspace of $H_1(\Sigma)$.

\subsection*{$(n+1)$-dimensional TQFT}
We require 2 more categories.  First consider the
bordism category $\text{Bord}_{n+1}$ defined by
\begin{enumerate}
 \item The objects are the same as the objects in $\mathscr{U}$ (extended
closed oriented $n$-manifolds).
 \item The morphisms are $(n+1)$-dimensional compact oriented bordisms,
i.e. for objects $\Sigma$ and $\Sigma^\prime$ a morphism $\Sigma\rightarrow\Sigma^\prime$
is an $(n+1)$-dimensional oriented manifold $X$ such that
$\partial X=-\Sigma\sqcup\Sigma^\prime$.  The bordisms may also have
extended structure.
\footnote{For a $(2+1)$-dimensional theory there is no need to endow bordisms
with extended structure in order to define a theory with anomaly (see below). 
However an anomaly-free theory requires an extended structure on
$X$ (in the language of \cite{turaev} these are \textit{weighted extended bordisms}).
See \cite{atiyah},\cite{walker},\cite{freed_gompf}.}
\end{enumerate}

Consider a different category of bordisms $\mathscr{B}$ defined by
\begin{enumerate}
 \item The objects $X$ in $\mathscr{B}$ are the morphisms in $\text{Bord}_{n+1}$, i.e.
(extended) compact oriented $(n+1)$-dimensional \textit{bordisms} between
extended oriented closed $n$-manifolds.
 \item The morphisms are orientation-preserving diffeomorphisms between
    bordisms $f:X\rightarrow X^\prime$.
\end{enumerate}
$\mathscr{B}$ has a canonical commutative strict monoidal structure:
\begin{enumerate}
 \item The tensor product is given by disjoint union:
\begin{equation}
  X\boxtimes X^\prime:=X\sqcup X^\prime
\end{equation}
 \item The unit object $\unitobj$ is the empty set $\varnothing$ (since
$X\sqcup\varnothing=X$).
 \item $\mathscr{B}$ is commutative, i.e. $X \sqcup X^\prime=X^\prime\sqcup X$.
\end{enumerate}

\begin{definition}
An \textbf{$(n+1)$-dimensional topological quantum field theory} $\tau$ based
on $(\modfunct,\mathscr{U},\text{Bord}_{n+1},\mathscr{B})$ is a rule:
\begin{enumerate}
 \item Given a bordism $X\in\text{Mor}(\Sigma,\Sigma^\prime)$
between $\Sigma\in\text{Ob}(\text{Bord}_{n+1})$ and $\Sigma^\prime\in\text{Ob}(\text{Bord}_{n+1})$ 
assign a linear map
\begin{equation}
  \tau(X):\modfunct(\Sigma)\rightarrow\modfunct(\Sigma^\prime)
\end{equation}
\item This rule must be \textit{projectively functorial} with respect
to the category $\text{Bord}_{n+1}$ (i.e. satisfy a gluing property).  Consider a bordism $X$ between
$\Sigma$ and $\Sigma^\prime$ and another bordism $X^\prime$ between
$\Sigma^\prime$ and $\Sigma^{\prime\prime}$.  Then glue the bordisms
together along $\Sigma^\prime$ to form a bordism 
$X\cup_{\text{glue}}X^\prime:\Sigma\rightarrow\Sigma^{\prime\prime}$.  We require
that:
\footnote{The anomaly $k$ measures how far
$\tau$ is from being a functor $\text{Bord}_{n+1}\rightarrow\text{Vect}^{\text{fin}}_{\cplx}$.}
\begin{equation}
  \tau(X\cup_{\text{glue}}X^\prime)=k\tau(X^\prime)\circ\tau(X) 
\end{equation}
where $k\in\cplx^\times$ is an invertible number called the \textbf{gluing anomaly} (if
$k=1$ then the theory is said to be \textit{anomaly-free}).  

Since the cylinder $\Sigma\times I$
is the identity morphism $\Sigma\rightarrow\Sigma$ in the category $\text{Bord}_{n+1}$,
\textit{projective functoriality} also requires that
\begin{equation}
 \tau(\Sigma\times I)=\id_{\modfunct(\Sigma)}
\end{equation}
 \item In terms of the category $\mathscr{B}$ we have an assignment
\begin{equation}
  \tau:\mathscr{B}\rightarrow \text{finite-dim linear maps} 
\end{equation}
We require this map be a strict monoidal functor.
This means (among other things) that
\begin{equation}
  \tau(X_1\sqcup X_2)=\tau(X_1)\otimes\tau(X_2)
\end{equation}
 \item Finally we require a compatibility on the categories $\mathscr{U}$,
$\mathscr{B}$, and $\text{Bord}_{n+1}$: if 
$f:X\rightarrow X^\prime$ is a morphism in $\mathscr{B}$ ($f:X\rightarrow X^\prime$ is
an orientation-preserving diffeomorphism of bordisms)
then the following diagram must commute:
\begin{equation}
 \xymatrix {
   \modfunct(\partial_-X)\ar[r]^-{\tau(X)}\ar[d]^-{(f|_{\partial_-X})_\sharp} & 
    \modfunct(\partial_+X)\ar[d]^-{(f|_{\partial_+X})_\sharp} \\
   \modfunct(\partial_-X^\prime)\ar[r]^-{\tau(X^\prime)} & 
    \modfunct(\partial_+X^\prime)
 }
\end{equation}
\end{enumerate}
\end{definition}

\subsection*{Extended $(2+1)$-dim TQFTs and extended $2$-d modular functors}
The definition of TQFT provided above applies in any dimension.  However
in $(2+1)$-dimensions most known theories satisfy stronger properties and
can be interpreted as \textbf{extended $(2+1)$-dim TQFT}
(or \textit{TQFT with corners}).  We refer the reader to chapter 4 of
\cite{bakalov_kirillov} for the relevant extended axioms,
\footnote{In particular the theories of Deloup described in \cite{deloup1},\cite{deloup2},\cite{deloup3}
are \textit{not} extended.
Links are intrinsic in the construction, however
ribbon graphs do not appear.  Furthermore the boundary surfaces are always \textit{closed}
manifolds.}
but briefly this means that the objects in $\mathscr{U}$
are not \textit{closed} $2$-surfaces, but instead are
compact surfaces with marked arcs (or parameterized boundary circles).
The bordisms are also extended to include
colored ribbon graphs with ends that terminate on the marked arcs. 
The construction provided here in chapter~(\ref{ch:mtc})
is manifestly \textit{extended}.

Likewise, the notion of modular functor can be strengthened to an
\textbf{extended $2$-d modular functor} (see chapter 5 of \cite{bakalov_kirillov}).  
The main additional feature is that colored boundary circles are allowed,
and they can be glued
(compare with the Segal modular functor). 

The known causality relationships between these notions are depicted in the
following diagram (as described in section 5.8 of \cite{bakalov_kirillov}):
\begin{equation}
  \xymatrix{
    *\txt{Modular\\Tensor\\Category}\ar@{=>}[r]\ar@{->}[dr] &  
      *\txt{Extended\\ $(2+1)$-dim TQFT} \ar@{=>}[d]\ar@{=>}[r] & 
      *\txt{$(2+1)$-dim TQFT} \ar@{=>}[d]\\
      & *\txt{Extended $2$-d Modular Functor} \ar@/^/@{.>}[ul]\ar@{=>}[ur]\ar@{=>}[r] &
      *\txt{$2$-d Modular Functor} \\
    & *\txt{Segal Modular Functor}\ar@{<:>}[u]^-{?} &
    }
\end{equation}
The broken line indicates that under certain circumstances an extended $2$-d modular 
functor reproduces a modular tensor category (see theorem 5.7.10 in ~\cite{bakalov_kirillov}).

\chapter{Toral Chern-Simons Theories}
\label{ch:chernsimons}

In this chapter we aim to give a brief summary of toral
Chern-Simons theories as described by Belov and Moore in \cite{belov_moore}.
Belov and Moore give a much more general description that includes
spin TQFTs, but in the context of modular tensor categories we
are confined to ordinary TQFTs.  Hence in this paper we shall mostly
limit ourselves to the ordinary (non-spin) 
Chern-Simons 
theories.

We will strive to keep the notation found in \cite{belov_moore} to
avoid confusion.

\section{Classical toral Chern-Simons theories}
Classical Chern-Simons theories for connected simply-connected compact
Lie groups were studied by Freed in \cite{freed_chernsimons1}.  The
theory for arbitrary compact Lie groups was developed in
\cite{freed_chernsimons2,witten_dijkgraaf},
and the $U(1)$ theory in particular was studied later by
Manoliu \cite{manoliu}.

To begin we consider Chern-Simons theory for a connected
simply-connected compact Lie group $G$.  Let $X_3$ be a closed
\footnote{We leave it to the references to define a theory on
manifolds with boundary.}
oriented 3-manifold.
Let $\pi:P\rightarrow X_3$ be a principal $G$-bundle.  A \textit{connection}
$\Theta$
on $P$ is a $\mathfrak{g}$-valued 1-form
\footnote{$\Theta$ is a 1-form on $P$, not on $X_3$}
that is $G$-equivariant
\begin{equation}
  \text{Ad}_g (R^*_g\Theta) = \Theta
\end{equation}
and in addition is just the Maurer-Cartan form $\theta$ when restricted to each fiber:
\begin{equation}
  i^*_x \Theta = \theta_x
\end{equation}
(here $i_x:P_x\rightarrow P$ is the inclusion of the fiber $P_x$ for any point
$x\in X_3$).

The curvature $\Omega\in\Omega^2_P(\mathfrak{g})$ is defined by
\begin{equation}
  \Omega = d\Theta + \frac{1}{2}[\Theta\wedge\Theta]
\end{equation}
where
\begin{equation}
  \frac{1}{2}[\Theta\wedge\Theta](v_1,v_2):=[\Theta(v_1),\Theta(v_2)] 
\end{equation}
The bracket on the RHS is the bracket in $\mathfrak{g}$.  The curvature
restricted to any fiber vanishes by the Maurer-Cartan equation
\begin{equation}
  i^*_x\Omega = d\theta + \frac{1}{2}[\theta\wedge\theta]=0
\end{equation}
In other words the curvature form $\Omega$ vanishes on vectors that
are tangent to each fiber, i.e. $\Omega$ is \textit{horizontal}.  
It is easy to verify that $\Omega$ is $G$-equivariant. Collecting
these results a standard argument shows that there is a 2-form $\omega$ on the base $X_3$ 
such that
\begin{equation}
  \Omega = \pi^*\omega
\end{equation}
$\omega$ is said to be a \textit{transgression} of $\Omega$.

Let $<>:\mathfrak{g}\times\mathfrak{g}\rightarrow R$ be an
$\text{Ad}$-invariant symmetric bilinear form.
\footnote{We note that $<>$ is often denoted by $\frac{1}{8\pi^2}\text{Tr}$
for compact simply-connected simple Lie groups $G$.  The trace
denotes the Killing form (for such groups any $\text{Ad}$-invariant symmetric 
bilinear form is a scalar multiple of the Killing form).}
Alternatively, 
$<>\in\text{Sym}^2_G(\mathfrak{g^*})$ can be viewed as
an $\text{Ad}$-invariant rank 2 homogeneous polynomial on $\mathfrak{g}$.
Define the Chern-Simons 3-form $\alpha(\Theta)\in\Omega^3_P(\mathbb{R})$
via the formula
\begin{equation}
  \alpha(\Theta):=<\Theta\wedge\Omega>-\frac{1}{6}<\Theta\wedge[\Theta\wedge\Theta]> 
\end{equation}
This is an antiderivative of $<\Omega\wedge\Omega>$.

In the case that $G$ is connected and simply-connected we know from obstruction theory 
that any $G$-bundle over
a manifold of dimension $\leq 3$ is trivializable.
Pick a trivialization for $P$, i.e. a global section $p:X_3\rightarrow P$.
\footnote{For a straightforward account of Chern-Simons actions for trivializable
bundles see \cite{baez_muniain}.}  
Define the Chern-Simons action (on $X_3$) by
\begin{equation}
  S_{X_3}(p,\Theta):=\int_{X_3}p^*\alpha(\Theta)
\end{equation}

Different trivializations $p$ and $p^\prime$ are related by a gauge transformation.
It is a 
basic physical axiom of gauge theory that if two
configurations are related by a gauge transformation then they
are physically indistinguishable, i.e. the mathematical
description of a gauge theory is redundant.  
Unfortunately, a calculation shows that the actions $S_{X_3}(p,\Theta)$ and $S_{X_3}(p^\prime,\Theta)$
are not the same (i.e. the action is \textit{not} gauge invariant).  However, for
certain choices of the bilinear form $<>$ the difference is an integer, i.e.
$S_{X_3}(p,\Theta)-S_{X_3}(p^\prime,\Theta)\in\mathbb{Z}$.  Hence we see that
\begin{equation}
  \label{eq:expchernsimons}
  \exp\left(2\pi i S_{X_3}(p,\Theta)\right)
\end{equation}
is well-defined independent of the choice of trivialization $p$.
\footnote{We note that picking a choice $p$ is \textit{not} the same
as gauge fixing.}
The correct choices for $<>$ comprise a lattice in $\text{Sym}^2_G(\mathfrak{g^*})$.  This
lattice is characterized by the following: 
the Chern-Weil construction (see \cite{freed_chernsimons2}) provides
a natural isomorphism
\begin{equation}
 \text{Sym}^2_G(\mathfrak{g^*})\cong H^4(BG;\mathbb{R})
\end{equation}
The appropriate lattice is just $H^4(BG;\mathbb{Z})\subset H^4(BG;\mathbb{R})$.
So we see that a classical Chern-Simons theory is determined (in this case)
by a connected simply-connected compact Lie group $G$ and an integral
bilinear form chosen from $H^4(BG;\mathbb{Z})$ (the \textit{level}).

On the other hand, $U(1)^N$ is not simply-connected
and it is \textit{not} true that any principal $U(1)^N$-bundle over a 3-manifold
is trivializable.
A different
technique must be used to define the Chern-Simons action \cite{witten_dijkgraaf}.
Choose a compact oriented 4-manifold
$Z_4$ such that $X_3$ is the boundary of $Z_4$ (such a manifold always
exists by Rokhlin's theorem \cite[pg. 87]{prasolov_sossinsky}).
\footnote{In fact a well-defined Chern-Simons theory exists for
\textit{arbitrary} compact gauge groups without appealing to 4-manifold 
extensions.  If $H_3(BG)=0$ (which it does for any torus) then a
Chern-Simons theory can be constructed directly using results in
\cite{freed_chernsimons2}.  Even more generally it is shown there that 
$H_3(BG)$ is at most a finite group,
and even then a classical Chern-Simons theory can be constructed
by studying $H^4(BG)$.}
In some cases (depending on the gauge group $G$) the bundle $P$ can
be extended to a principal $G$-bundle $\tilde{P}\rightarrow Z_4$.  For $G$ a
torus this is always possible.
\footnote{It is
pointed out in \cite{belov_moore} that any obstruction to such
an extension lives in the oriented bordism group $\Omega_3(BG)$
of the classifying space $BG$.  It is also mentioned in \cite{belov_moore} 
that for $G$ abelian $\Omega_3(BG)=0$, hence we will always be able
to extend the bundle in this paper.}

Given the extension $\tilde{P}\rightarrow Z_4$ we
can arbitrarily extend the connection $\Theta$ on $P$ to a connection $\tilde{\Theta}$
on $\tilde{P}$ (using a partition of unity).  
If $\tilde{\Omega}$ denotes the curvature of $\tilde{\Theta}$ then we can
define the Chern-Simons action to be the integral of the 
second Chern class
\begin{equation}
  \exp\left(2 \pi i \int_{Z_4} <\tilde{\Omega}\wedge \tilde{\Omega}>\right)  
\end{equation}
%We have used similar arguments as those above to absorb the representation
%information into the even-valued integer coefficient $B$.
It is not difficult to check using Stokes' theorem that if $P$ 
is trivializable then this action reduces to our first naive action.

A standard argument shows that this expression does not depend on the
choice of 4-manifold $Z_4$.  Given two such
manifolds $Z_4$ and $Z_4^\prime$ we can glue them together along
their common boundary $X_3$ to produce a closed oriented 4-manifold
$(-Z_4)\cup Z_4^\prime$ (here $-Z_4$ denotes reversed orientation).  
Now the integral of a Chern class over a closed oriented
manifold is an integer $N$, i.e.
\begin{equation}
  \exp\left(2\pi i \int_{(-Z_4)\cup Z_4^\prime}<\tilde{\Omega}\wedge \tilde{\Omega}>\right) =
  \exp\left(2\pi i N\right) = 1
\end{equation}
Furthermore this integer is independent of the extending connection
$\tilde{\Theta}$.
On the other hand the LHS is just
\begin{equation}
  \exp\left(2\pi i \left(-\int_{Z_4} <\tilde{\Omega}\wedge \tilde{\Omega}>+
    \int_{Z_4^\prime} <\tilde{\Omega}\wedge \tilde{\Omega}>\right)\right)
\end{equation}
Hence
\begin{equation}
  \exp\left(2\pi i \int_{Z_4} <\tilde{\Omega}\wedge \tilde{\Omega}>\right) =
  \exp\left(2\pi i \int_{Z_4^\prime} <\tilde{\Omega}\wedge \tilde{\Omega}>\right)
\end{equation}
So we see that in general a classical Chern-Simons theory is determined
by a compact gauge group $G$ and a choice of \textit{integral} bilinear form (the \textit{level})
in $H^4(BG;\mathbb{Z})$ (the bundle $P$ is \textit{not} part of the data since
we want to consider \textit{all} bundles.)

In particular consider the case $G=U(1)$.  Then $\mathfrak{g}\cong i\mathbb{R}$
and hence the Chern-Simons action becomes
\begin{equation}
  \exp\left(2 \pi i \frac{k}{4\pi^2} \int_{Z_4} \tilde{\Omega}\wedge \tilde{\Omega}\right)  
\end{equation}
where the level $<>$ is encoded in $\frac{k}{4\pi^2}$ where $k$ is
any integer.  It is customary to redefine the action in terms of an even integer
$B=2k$.  The action is (for $B$ an \textit{even} integer)
\begin{equation}
  \exp\left( \pi i \frac{B}{4\pi^2} \int_{Z_4} \tilde{\Omega}\wedge \tilde{\Omega}\right)  
\end{equation}

For $U(1)^N$ the analogue of the even integer $B$ is an integer-valued 
symmetric matrix $B_{\alpha\beta}$
with \textit{even} integers along the diagonal.  
We will call such a symmetric bilinear form \textit{even}.  Following
\cite{belov_moore} in the
remainder of this paper we restrict our attention to \textit{nondegenerate}
integer-valued symmetric bilinear forms.

%Expanding this becomes       
%\begin{equation}
%  \exp\left(\pi i\left( 
%    B_{\alpha\alpha} \int_{Z_4} 
%    \tilde{F}^\alpha\wedge\tilde{F}^\alpha+
%    \sum_{\alpha<\beta}2B_{\alpha\beta} \int_{Z_4} 
%    \tilde{F}^\alpha\wedge\tilde{F}^\beta\right)\right)  
%\end{equation}

%In light of what we saw for the $U(1)$ case it is necessary
%that each diagonal entry in $B$ be an \textit{even integer}.
%Since an extra factor of $2$ already appears in the second term from
%double-counting the off-diagonal entries are arbitrary integers.

It is worth noting that 
we equip $X_3$ with a spin structure then there
exists a compatible extending \textit{spin} 4-manifold $Z_4$
\cite{belov_moore}.
In that case the integral of the second Chern 
class is already an \textit{even} integer.  Hence in that case the action
is well defined if we allow \textit{arbitrary} integers along the
diagonal of $B$.

Every nondegenerate integer-valued symmetric bilinear form $B$ (not necessarily even)
can be thought of as the inner product on a lattice $\Lambda$.
We summarize these results is the following proposition:

\begin{proposition}
Classification of classical toral Chern-Simons
\begin{enumerate}
  \item The set of ordinary classical toral Chern-Simons theories
        is in one-to-one correspondence with \textit{even lattices} $(\Lambda,B)$.  
  \item The set of spin classical toral Chern-Simons theories
        is in one-to-one correspondence with \textit{arbitrary lattices} $(\Lambda,B)$.
\end{enumerate}
\end{proposition}

\section{Quantization of lattices}
\label{sec:latticequantization}
In the previous section we have seen that an 
abelian classical Chern-Simons theory (including a spin theory) is
determined by an integer lattice $\Lambda$ equipped with a
symmetric bilinear form 
$B:\Lambda\times\Lambda\rightarrow\mathbb{Z}$.

Since we are not interested in the general spin case for now we mostly
limit our discussion to \textit{even} symmetric bilinear
forms.  In basis-independent language we mean symmetric bilinear 
forms $B$ such that
$B(X,X)\in 2\mathbb{Z}$ for every $X\in\Lambda$.

It will happen that the canonical quantization program 
described in section~(\ref{sec:canonicalquantization}) will
rely heavily on the aspects of lattices described here.  We
abusively call this ``quantization of lattices''.
The easiest piece of data that can be harvested from a lattice
$(\Lambda, B)$
(even or not) is the signature $C\in\mathbb{Z}$ of the bilinear form.

For the remaining data we require the following definition:
\begin{definition}
Let $R$ be a ring.
A nondegenerate $R$-valued \textbf{quadratic form} on an abelian
group (e.g. a lattice) is a function $Q:\Lambda\rightarrow R$ such that:
\begin{itemize}
  \item $Q(X+Y)-Q(X)-Q(Y)+Q(0)$ defines a bilinear and nondegenerate symmetric
     form
  \item We say that $Q$ is a \textbf{pure quadratic form} if 
     $Q(nX)=n^2Q(X)$ for every integer $n$ (in particular $Q(0)=0$).
\end{itemize}
\end{definition}  
In this paper if we accidentally drop the ``pure'' modifier than
we still mean pure - we will explicitly say ``generalized''
otherwise.

Any even lattice $(\Lambda, B)$ induces a
\textit{pure quadratic form} $Q:\Lambda\rightarrow\mathbb{Z}$
given by the formula (division by 2 makes sense because
$B$ is even)
\begin{equation}
  Q(x)=\frac{1}{2}B(X,X)
\end{equation}
We note that (for even lattices) the 
pure quadratic form and the bilinear form
determine each other: given a pure quadratic form $Q$ a bilinear
form can be recovered with the formula
\begin{equation}
  B(X,Y)=Q(X+Y)-Q(X)-Q(Y)
\end{equation}
$Q$ is a \textit{pure quadratic refinement} of $B$.

\subsection*{Discriminant group}
From an arbitrary lattice (which determines a classical theory) 
we construct
a finite abelian group $\grp$ (the \textit{discriminant} group).
The bilinear form $B$ descends to a bilinear form 
$b:\grp\times\grp\rightarrow\qmodz$, and
if the lattice is even then the pure quadratic form $Q$ on $\Lambda$ 
descends to a pure quadratic form $q:\grp\rightarrow\qmodz$ as well
\cite{nikulin}.

The content of the work of Belov and Moore is that
quantum toral Chern-Simons theory is (almost) completely determined by
$\triotrunc$, i.e. we have a quantization map
\begin{equation}
  \textit{Ordinary classical Chern-Simons}\rightarrow\textit{Ordinary quantum Chern-Simons}
\end{equation}
that is encoded in the map
\begin{equation}
  \label{eq:evenquantization}
  \textit{Even lattice } (\Lambda, B)\rightarrow\textit{Discriminant Group }
  (\grp,q,c)  
\end{equation}
where $c\equiv C \text{ mod } 24$ ($C$ is the signature of the
bilinear form $B$).
The above map is surjective, however it is not injective.
\footnote{There is a slight error in the main theorem of \cite{belov_moore}.  See appendix~(\ref{appendix1}).}
\footnote{It is important to note that, in contrast to a lattice,
a quadratic form on a finite group supplies more information than a bilinear form.}

The construction of the group is as follows: consider the dual lattice $\Lambda^*$.
Since we have a nondegenerate symmetric bilinear form 
$B$ we have an embedding of the lattice into its dual 
$\Lambda\overset{f}{\rightarrow}\Lambda^*$ given by 
$X\overset{f}{\mapsto} B(X,\cdot)$.  In general this map is not
invertible over the integers (e.g. it is not possible 
to invert the $1\times 1$ matrix $B=(2)$ over
the integers) but it can be inverted over the rationals.
So let $V=\Lambda\otimes\mathbb{Q}$ and $V^*=\Lambda^*\otimes\mathbb{Q}$
be vectors spaces that contain $\Lambda$ and $\Lambda^*$,
respectively. 

In this case $f$ is invertible and hence we have the (restricted) map
$f^{-1}:\Lambda^*\subset V^*\rightarrow V$.  It is
easy to see that $\Lambda$ is in the image of $\Lambda^*$,    
so we can think of $\Lambda$ as
a sublattice of $\Lambda^*$ (all embedded in $V$).
From now on we will think
of both $\Lambda$ and $\Lambda^*$ as being embedded in $V$.  
The finite abelian group is just the
quotient $\grp=\Lambda^*/\Lambda$.  

It is straightforward to check that the bilinear form
$B:V\times V\rightarrow \mathbb{Q}$ descends to a (nondegenerate,
symmetric) bilinear
form $b:\grp\times\grp\rightarrow\qmodz$ and that, if the
lattice is even, the pure quadratic form $Q:V\rightarrow \mathbb{Q}$
also descends to a pure quadratic form $q:\grp\rightarrow\qmodz$.

\begin{example}
\label{ex:z4_1}
As an example, consider the rank $1$ lattice $\Lambda=\mathbb{Z}$
equipped with the bilinear form $B=(4)$.  So $B(1,1)=4$ and, since
this is an even lattice, $Q(1)=\frac{4}{2}=2$.
Tensoring over $\mathbb{Q}$
we see that $\Lambda$ consists of the numbers
\begin{equation}
  \Lambda=\lbrace\dotsc,-1,0,1,2,3,\dotsc\rbrace
\end{equation}
and $\Lambda^*$ (through the map $f^{-1}$) consists of the fractions
\begin{equation}
  \Lambda^*=\lbrace\dotsc,-1/4,0,1/4,1/2,3/4,1,5/4,\dotsc\rbrace
\end{equation}
The discriminant group $\grp$ is just 
\begin{equation}
  \grp=\lbrace 0,1/4,1/2,3/4\rbrace\cong\mathbb{Z}_4
\end{equation}
The induced bilinear form is just
\begin{equation}
  b(1/4,1/4) = B(1/4,1/4)\pmod 1 = 1/4*4*1/4\pmod 1 = 
    1/4\pmod 1
\end{equation}
and the induced quadratic form is
\begin{equation}
  q(1/4) = \frac{1}{2}B(1/4,1/4)\pmod 1 = 1/8\pmod 1 
\end{equation}
The value of $b$ and $q$ on the generator $1/4$ determines all
of the values completely.
\footnote{True since $b$ is bilinear and $q$ is pure.  For a choice
of generator $x$ any two arbitrary elements can be written as $nx$
and $mx$ for integers $n$ and $m$.  Hence
\begin{equation}
  b(nx,mx)=mnb(x,x)
\end{equation}
and
\begin{equation}
  q(nx)=n^2q(x)
\end{equation}
}

Since the rank of the lattice (here rank $N=1$) 
is just the rank of the original gauge group
$U(1)^N$ we say that the above example is ``$U(1)$ Chern-Simons
at level $B=4$''.  Obviously the ``level'' becomes a matrix in
higher rank.
\end{example}

\begin{example}
\label{ex:z4_2}
Let us consider another example.  For this let us forget
the lattice and just consider the same finite abelian
group
\begin{equation}
  \grp=\lbrace 0,1/4,1/2,3/4\rbrace\cong\mathbb{Z}_4
\end{equation}
We keep the same bilinear form
\begin{equation}
  b(1/4,1/4) = 1/4\text{ mod }1
\end{equation}
but use a \textit{different} pure quadratic refinement
\begin{equation}
  q(1/4) = 5/8\text{ mod }1
\end{equation}
(we obtained this quadratic form by taking the value of the
previous quadratic form on the generator and adding $1/2$).
It is easy to verify that this pure quadratic form is a refinement of $b$.
This is clearly \textit{not} $U(1)$ at level $4$.  It is also not
clear that this data lifts to a lattice.
\footnote{However, we will see below that it does.  All pure quadratic
forms on finite abelian groups will be realized by \textit{even} lattices.}  

So for $\grp=\mathbb{Z}_4$ and the same bilinear
form we have found two distinct pure quadratic refinements.
\end{example}

\begin{example}
\label{ex:z3}
Consider a rank $1$ lattice $\Lambda=\mathbb{Z}$ with bilinear form 
$B=(3)$.  This lattice is not even, so
it does not induce a pure quadratic refinement.  The discriminant group is
\begin{equation}
  \grp=\lbrace 0, 1/3, 2/3\rbrace\cong\mathbb{Z}_3
\end{equation}
and the induced bilinear form is
\begin{equation}
  b(1/3,1/3)=B(1/3,1/3)\text{ mod } 1=1/3*3*1/3=1/3\text{ mod }1
\end{equation}
As stated, a pure quadratic form is \textit{not} induced
by this lattice.

However, if we disregard the classical lattice and simply consider
the group $\grp=\mathbb{Z}_3$ equipped with the bilinear
form $b$ as above then we can produce a pure quadratic refinement of $b$:
\begin{align}
  q(1/3)&=2/3\text{ mod }1\\
  q(2/3)&=1/3\text{ mod }1\\
  q(0)&=0\text{ mod }1
\end{align}
Again, it is enough to specify $q$ on the generator, but we list
all of the values explicitly for clarity.  It is routine to verify that this
pure quadratic form is a refinement of $b$.  It is also straightforward
to check that this is the \textit{unique} pure quadratic form that is
compatible with $b$ (see lemma~(\ref{lemma:purecyclic})).

However, this theory is \textit{not} $U(1)$ at level $3$ (the first part of
this example) since
that lattice did not induce a pure quadratic form (it is not an even
theory).  The theory described here, however, can be lifted (as we
shall see) to
a different (greater rank) even lattice since $q$ is pure.
\end{example}

By studying these two examples and considering the possible
bilinear forms and corresponding pure quadratic refinements on an
arbitrary cyclic group we have the following proposition
(which is clearer if the readers prove it for themselves)

\begin{lemma}\label{lemma:purecyclic}
Let $\grp$ be a cyclic group $\mathbb{Z}_N$ equipped with
a symmetric bilinear form $b:\grp\times\grp\rightarrow\qmodz$
(possibly degenerate).
Then:
\begin{enumerate}
  \item If $b=0$ then $q=0$ identically.
  \item If $b\neq 0$ and $N$ is even then there are exactly 
    two pure quadratic refinements
    of $b$ (on a generator $x$ we have either $q_0(x)$ or
    $q_1=q_0(x)+\frac{1}{2}$).
  \item If $b\neq 0$ and $N$ is odd then there is a unique pure 
    quadratic refinement of $b$.
\end{enumerate}
\end{lemma}
\begin{proof}
Pick a generator $x$ for $\grp$.  Since $Nx\equiv 0$ we have that
\begin{equation}
  b(x,x)=\frac{m}{N}
\end{equation} 
for some integer $m<N$.  Since $q$ is pure we have that
\begin{equation}
  b(x,x)=q(x+x)-q(x)-q(x)=q(2x)-2q(x)=4q(x)-2q(x)=2q(x)
\end{equation}
so $q(x)=\frac{1}{2}b(x,x)$.  Hence we are left to consider
the ambiguity when dividing by $2$ in $\qmodz$. 

If $N$ is even then we obtain
two possibilities for $q$ on a generator $x$:
\begin{equation}
  q(x)=\frac{m}{2N}\text{ or }\frac{m}{2N}+\frac{1}{2}
\end{equation}
It is easy to verify that both of these options are well defined
(i.e. $q(Nx)=0$).  The value on an arbitrary element $nx$ is
defined by asserting purity $q(nx)=n^2q(x)$.

If $N$ is odd then having a $2N$ in the
denominator does not produce a well-defined pure quadratic
form.  Since $N$ is odd there exists instead a unique integer $m^\prime\text{ mod }N$
such that
\begin{equation}
  2m^\prime \equiv m\text{ mod }N
\end{equation}
So we define $q(x)=m^\prime/N$.

It is also straightforward to check that a different choice of generator
$x$ gives back one of these examples (hint: write the new generator
in terms of the old).
\end{proof}

In particular, since an arbitrary finite abelian group can be 
decomposed (not uniquely!) as a direct sum of cyclic groups of
prime power order we have

\begin{lemma}\label{lemma:purequadraticexistence}
Any arbitrary finite abelian group $\grp$ equipped with a
symmetric bilinear form $b$ (perhaps degenerate) admits a pure 
quadratic refinement.
\end{lemma} 
\begin{proof}
Choose a decomposition of $\grp$ into cyclic groups.  Each
cyclic factor $\mathbb{Z}_{N_i}$ considered by itself has a 
(possibly degenerate) symmetric bilinear form $b_i$ which
is just the restriction of $b$ to $\mathbb{Z}_{N_i}$.  By
lemma~(\ref{lemma:purecyclic}) choose a pure quadratic refinement
$q_i$.  

Now we must combine the $q_i$'s into a pure quadratic
refinement $q$ defined on the whole group.  Given an element of the
form $x+y\in\grp$ where
$x$ is in one factor $i$ and $y$ is in another $j$ define
\begin{equation}
  q(x+y):= b(x,y)+q_i(x)+q_j(y)
\end{equation}
It is easy to see that this is the only possibility (and that
$q$ is pure).  
\end{proof}

The existence of a pure quadratic refinement will be
useful in the sequel.

\subsection*{Gauss sums (reciprocity)}
We hinted above in equation~(\ref{eq:evenquantization}) that we
must manually keep around information about the signature $C$ of $B$ when
we quantize since passing to the discriminant group ``loses memory''
of the signature (for our purposes we actually only need to keep the 
value of $c=C\text{ mod }24$).

However some information about $C$ is maintained in $(\grp,q)$ alone.  Gauss
proved a relation (a \textit{Gauss sum} or \textit{reciprocity}) on
rank $1$ even lattices that has since been extended to arbitrary even lattices.
For reference see Milnor and Husemoller
\cite{milnor_husemoller} (especially the appendix.  We note that the
majority of the book applies to unimodular lattices only, i.e. 
$\det B=\pm 1$).  Other references include Nikulin \cite{nikulin}

In fact the induced quadratic refinement $q$ on $\grp$ can reproduce
information about the signature (but only mod $8$) according to the formula
\begin{equation}
  \frac{1}{\sqrt{\lvert\grp\rvert}}\sum_{x\in\grp}{\exp{(2\pi i
    q(x))}}=\exp{(2\pi i C/8)}
\end{equation}

\begin{example}
Consider again example~(\ref{ex:z4_1}) which is $U(1)$ at level 4.
Computing the Gauss sum gives $C\equiv 1\text{ mod }8$ which agrees
with expectation since this theory arises from a rank $1$ lattice
equipped with a bilinear form with signature $1$.

Now consider example~(\ref{ex:z4_2}) which was a theory \textit{different}
from $U(1)$ at level 4.  From Gauss sum considerations we see that,
if the theory is realized by an even lattice (which it is), then
the signature of the lattice mod 8 is $C\equiv 5\text{ mod }8$ (clearly
not a rank $1$ lattice).
\end{example}

\subsection*{Generalized quadratic forms and spin theories}
Let us return momentarily to arbitrary (not necessarily
even) lattices $(\Lambda, B)$.  Although spin theories are
not the subject of this paper, we wish to clarify for
ourselves some
of the constructions that are discussed in \cite{belov_moore}.  In
addition we make explicit some observations that are not
mentioned there. 

We have seen that we have a quantization map encoded in
the map
\begin{equation}
  \textit{Even lattice } (\Lambda, B)\rightarrow\textit{Discriminant Group }
  (\grp,q,c)  
\end{equation}
However, \cite{belov_moore} specifies a
 quantization for \textit{arbitrary}
lattices, so we should have a more general map
\begin{equation}
  \textit{Lattice } (\Lambda, \text{?}_1)\rightarrow\textit{Discriminant Group }
  (\grp,\text{?}_2,c)  
\end{equation}
It is not immediately clear what should play the role of $\text{?}_1$ and
$\text{?}_2$.  Let us describe the construction. 

It is easy to see that for any 
symmetric nondegenerate bilinear form
$B$ (even or not) on $\Lambda$ there exists an element
$W\in\Lambda^*$ such that
$B(X,X)=B(X,W)\text{ mod }2$ for every $X\in\Lambda$.
In fact, if $W$ satisfies this then it is trivial to
show that $W+2\lambda$ does as well for any
$\lambda\in\Lambda^*$.  Conversely, since $B$ is
nondegenerate it is also trivial to see that if $W$
and $W^\prime$ satisfy the condition then $W^\prime=W+2\lambda$
for some $\lambda\in\Lambda^*$.

In other words there exists a \textit{unique class} 
$[W]\in\Lambda^*/2\Lambda^*$ 
such that   
$B(X,X)=B(X,[W])\text{ mod }2$ for every $X\in\Lambda$.
Such a class is called the \textit{characteristic class} 
\cite{belov_moore}
or the \textit{Wu class} \cite{deloup1} for the lattice $(\Lambda, B)$.
We call a specific choice of $W$ in $\Lambda^*$ a \textit{Wu representative}.

As a special case 
if the lattice is even then (by definition) $B(X,X)=0\text{ mod }2$
for every $X\in\Lambda$, hence $[W]=[0]\in\Lambda^*/2\Lambda^*$.
Conversely if $[W]=[0]$ then the lattice is even.  Since one
of the representatives of 
$[0]$ is just the identity element $W=0\in\Lambda^*$ 
we
have - in the case of even lattices - a canonical choice $W=0$
picked out.  For odd lattices there is no such distinguished representative. 

So for even lattices (that we have already considered) the construction
that follows momentarily
reduces to a single pure quadratic form by setting $W=0$.
For the general theory there will
be no preferred representative, hence no preferred \textit{generalized}
quadratic form; we will be forced to be content with an equivalence
class of (generalized) quadratic forms on $\grp$.

Let us start with a definition:  

\begin{definition}
Let $q$ and $q^\prime$ be two $\qmodz$-valued generalized quadratic 
forms on a finite
abelian group $\grp$.  Then we say that $q$ is equivalent to $q^\prime$
if there exists a fixed $\delta\in\grp$ such that 
$q^\prime(x)=q(x-\delta)$ for every $x\in\grp$. 
\end{definition}

Now finally we are ready to construct a set of generalized quadratic forms.
Consider a lattice $(\Lambda, B)$ where generically $B$ is odd.  Consider
the induced discriminant group $\grp$ and the induced bilinear form
$b$.  Since $B$ is generically odd we do not have an induced
pure quadratic form.

From the lattice (which defines a Wu class $[W]$)
we need an algorithm to construct a
generalized quadratic form $Q:V\rightarrow \mathbb{Q}$
that descends to a well-defined generalized quadratic form
$q:\grp\rightarrow\qmodz$.  Let all of the Wu representatives of
$[W]$ be denoted by $\{W_i\}_{i\in\mathbb{Z}}$.  Since we have infinitely-many representatives
$W_i$ we will not be able to construct a single quadratic form,
but rather a family of quadratic forms (we shall see momentarily why this constant
term is used):
\begin{equation}
  Q_{W_i}(X):= \frac{1}{2}B(X,X-W_i)+
    \frac{1}{8}B(W_i,W_i)
\end{equation}

Each $Q_{W_i}$ descends to a well-defined generalized quadratic
form on $\grp$ 
\begin{equation}
  q_i(x):= \frac{1}{2}B(X,X-W_i)+\frac{1}{8}B(W_i,W_i)
    \text{ mod }1
\label{eq:quadraticfromwu}
\end{equation}
where $X\in\Lambda^*$ is an arbitrary lift of $x\in\grp$ (the choice
of lift does not affect the value of the form because of the defining
property for $W_i$).

It is routine to verify that each $q_i$ is a generalized quadratic
\textit{refinement} of $b$ (i.e. $q_i(x+y)-q_i(x)-q_i(y)+q_i(0)=b(x,y)$).

Perhaps more interesting, if $W_i,W_j\in\Lambda^*$ are two 
Wu representatives of $[W]$ then it is easy to show
(using that fact that $W_i=W_j+2\lambda$ for some $\lambda\in\Lambda^*$) that
the generalized quadratic refinements $q_i$ and $q_j$ are
\textit{equivalent} in the sense defined above.  

Even further, it
is a simple calculation to show that an entire equivalence class
of generalized quadratic forms is realized by the set of
all Wu representatives $\{W_i\}_{i\in\mathbb{Z}}$.
So $[W]$ determines completely
an equivalence class of generalized quadratic refinements which we
denote by
\begin{equation}
  [q_W]
\end{equation}      

Now we know exactly what to substitute for $\text{?}_1$ and $\text{?}_2$ in
the more general quantization map above:
\begin{equation}
\label{eq:quantization}
  \textit{Lattice } (\Lambda, \lbrace Q_{W_i}\rbrace)
    \rightarrow\textit{Discriminant Group }(\grp,[q_W],c)  
\end{equation}

It is easy to see that this map reduces to the old quantization map
defined only on even lattices (where $[W]=[0]$) by 
picking the special pure quadratic refinement defined by $W=0$ out
of the equivalence class.

The reason for choosing the constant term as in equation~(\ref{eq:quadraticfromwu}) is
that then the Gauss reciprocity formula generalizes
to arbitrary generalized quadratic forms (see pg 70 in Hopkins and
Singer \cite{hopkins_singer}).  Hence partial information
(mod $8$) about the signature of $B$ is retained in the same formula 
\begin{equation}
  \frac{1}{\sqrt{\lvert\grp\rvert}}\sum_{x\in\grp}{\exp{(2\pi i
    q_i(x))}}=\exp{(2\pi i C/8)}
\label{eq:gausssum}
\end{equation}
Obviously different $q_i$'s in the same equivalence class give the
same number on the LHS, hence define the same $C\text{ mod }8$.

\subsection*{The quantization map is surjective}
The ``lattice quantization'' map in equation~(\ref{eq:quantization})
is surjective.  However the map is
\textit{not} injective (in fact infinitely-many classical theories
will map onto a given quantum theory).

Consider an arbitrary finite abelian group $\grp$ equipped with
an equivalence class of nondegenerate generalized quadratic forms
$[q]$.  Use the Gauss sum formula (equation~(\ref{eq:gausssum})) 
to define a ``signature'' integer $C\text{ mod }8$.  The term
``signature'' doesn't technically make sense because there is
no classical lattice here, but we use it anyway.
$C\text{ mod }8$ is determined by $\grp$ and $[q]$, so it is not extra information.

However, we require not just an integer mod $8$, but rather
an integer mod $24$.  So suppose that, in addition, we are given an integer $c\text{ mod }24$ such that
$c\equiv C\text{ mod }8$.  Obviously for a given $C$ 
there are only 3 possibilities for such a $c$.

Then we can ask the following question: does the trio of data
$(\grp,[q],c)$ lift to a classical lattice?
\footnote{We note that $[q]$
determines a bilinear form $b$, hence we could write the data
as a quartet $(\grp,b,[q],c)$).}
The answer is yes.  We shall start with the simpler case (which is
the only one relevant for the remainder of this paper).

We know that an even lattice $(\Lambda, B)$ maps under 
equation~(\ref{eq:evenquantization}) to a trio $(\grp,q,c)$ where $q$ is
a \textit{pure} nondegenerate quadratic form and $c$ is an integer
mod $24$ that satisfies the Gauss sum in equation~(\ref{eq:gausssum}).    

On the other hand, given such a trio $\lbrace$ $(\grp,q,c)$ where $\grp$ is
a finite abelian group, $q$ is a nondegenerate pure quadratic form,
and $c$ is an integer mod $24$ that satisfies the Gauss formula $\rbrace$
can this be lifted to an even
lattice $(\Lambda, B)$?  The following result answers this positively 
(corollary 1.10.2 pg 117 in \cite{nikulin}):
\begin{corollary}{(V.V. Nikulin, 1979)}
\label{cor:nikulinpure}
Let $r_+\geq 0$ and $r_-\geq 0$ be integers. 
Consider a finite abelian group $\grp$ equipped with a $\qmodz$-valued 
nondegenerate \textit{pure} quadratic form $q$.  
Define the ``signature'' mod $8$ of $q$ by the Gauss sum formula
in equation~(\ref{eq:gausssum}).
Then if the quantity $r_++r_-$ is sufficiently
large and if $r_+-r_-\equiv\text{sign }q\text{ mod }8$
then there exists an \textit{even} lattice $(\Lambda, B)$
such that
\begin{enumerate}
 \item $(\grp, q)$ is the discriminant group and quadratic form from $(\Lambda, B)$  
 \item $(\Lambda, B)$ has $r_+$ positive eigenvalues and
    $r_-$ negative eigenvalues
\end{enumerate}
\end{corollary}
Nikulin's original statement provides estimates on ``sufficiently large'', but
we do not need them.  Note that the modifier ``pure'' is left out of Nikulin's
version because in \cite{nikulin} all quadratic forms are \textit{defined} to
be pure.

As can be seen, a given trio lifts to infinitely-many even lattices.
We conclude that the even quantization map in equation~(\ref{eq:evenquantization})
is surjective but not injective. 

Now consider a trio $\lbrace$ $(\grp,[q],c)$ where $\grp$ is
a finite abelian group, $[q]$ is an equivalence class of nondegenerate 
generalized quadratic forms,
and $c$ is an integer mod $24$ that satisfies the Gauss formula $\rbrace$.
Can this be lifted to a (generically odd) lattice?  Consider
Nikulin's results about odd lattices (Corollary
1.16.6 \cite{nikulin}):
\begin{corollary}{(V.V. Nikulin, 1979)}
\label{cor:nikulingeneralized}
Let $r_+\geq 0$ and $r_-\geq 0$ be arbitrary positive integers. 
Consider a finite abelian group $\grp$ equipped with a $\qmodz$-valued 
nondegenerate symmetric bilinear form $b$.  Then if the quantity $r_++r_-$ is sufficiently
large then there exists a (possibly odd) lattice $(\Lambda, B)$
such that
\begin{enumerate}
 \item $(\grp, b)$ is the discriminant group and bilinear form from $(\Lambda, B)$  
 \item $(\Lambda, B)$ has $r_+$ positive eigenvalues and
    $r_-$ negative eigenvalues
\end{enumerate}
\end{corollary}
Again what we present here is weaker than the corollary presented
in the original work.

This corollary shows that the data $(\grp, b, c)$ lifts to a
(possibly odd) lattice $(\Lambda, B)$ where $\text{signature
}B=C=r_+-r_-\equiv c\text{ mod }24$ for
\textit{arbitrary} integer $c\text{ mod }24$.  
Note the appearance of $b$ rather than
$[q]$ in the trio here.  This indicates that the bilinear
form lifts, but we have still not seen that $[q]$ lifts ($[q]$ lifts
means that it is derived from the Wu class $[W]$ on the lift lattice).
We have not seen the following extension of Nikulin's theorem
explicitly stated and proven in the literature, hence we 
prove it here for completeness:

\begin{proposition}
The trio $(\grp,[q], c)$ lifts to a (possibly odd) lattice.
\end{proposition}
\begin{proof}
To see that $[q]$ lifts as well let us compare it to $[q_W]$ where
$[W]$ is the Wu class of the lifted lattice $(\Lambda, B)$.  We need to
show that $[q]=[q_W]$ so pick a Wu representative $W$
and consider the induced generalized quadratic form
\begin{equation}
  q_W(x)\equiv \frac{1}{2}B(X,X-W)+\frac{1}{8}B(W,W)
    \text{ mod }1
\end{equation}
where $X\in\Lambda^*$ is an arbitrary lift of $x\in\grp$.
Pick one of the quadratic forms $q$ out of the equivalence
class $[q]$ as well.  We want to compare $q$ and $q_W$ (their induced bilinear 
forms $b$ are at least the same because $q_W$ is constructed from
a lift of $b$.  Also we have already seen that $C\equiv c\text{ mod }24$ by
construction of the lift so
$q$ and $q_W$ satisfy the Gauss sum formula for the same value of
$C\text{ mod }8$).

It is easier to compare them if we strip off the constants,
so define $\tilde{q}(x)=q(x)-q(0)$ and 
$\tilde{q_W}(x)=q_W(x)-q_W(0)=\frac{1}{2}B(X,X-W)$.
Clearly
\begin{align}
  &\tilde{q}(x+y)-\tilde{q}(x)-\tilde{q}(y)=\\
  &[q(x+y)-q(0)]-[q(x)-q(0)]-[q(y)-q(0)]=\\
  &q(x+y)-q(x)-q(y)+q(0)=b(x,y)
\end{align}
so the bilinear form is not changed when passing from
$q$ to $\tilde{q}$.  A similar statement holds for
$q_W$ to $\tilde{q_W}$.

Since $\tilde{q}$ and $\tilde{q_W}$ refine the same bilinear
form $b$ they differ by a linear term.  This can be seen from
\begin{equation}
  [\tilde{q}-\tilde{q_W}](x+y)-[\tilde{q}-\tilde{q_W}](x)-
  [\tilde{q}-\tilde{q_W}](y)=[b-b](x,y)=0
\end{equation}
which shows that $[\tilde{q}-\tilde{q_W}]$ is linear.  But
$b$ is nondegenerate so any linear function is of the form
$b(x,\delta)$ for some fixed $\delta\in\grp$.  So
\begin{equation}
  [\tilde{q}-\tilde{q_W}](x)=b(x,\delta)=B(X,\Delta)\text{ mod }1
\end{equation}
for some fixed $\delta\in\grp$ ($\Delta\in\Lambda^*$ is an arbitrary lift of
$\delta\in\grp$).  Therefore
\begin{align} 
  \tilde{q}(x)&=\tilde{q_W}+B(X,\Delta)\text{ mod }1\\
    &=\frac{1}{2}B(X,X-W)+B(X,\Delta)\text{ mod }1\\
    &=\frac{1}{2}B(X,X-(W-2\Delta))\text{ mod }1
\end{align}
The last line is of the form $\tilde{q_{W^\prime}}$
where $W^\prime=W-2\Delta$ is just another choice of representative
for the same Wu class $[W]$.

So we see that $\tilde{q}=\tilde{q_{W^\prime}}$.  Now all that
we need to do is put the constants back in.  We need to check if
\begin{equation}
  q(x)=\tilde{q}(x)+q(0)
\end{equation}
equals
\begin{align}
  q_{W^\prime}(x)&=
    \tilde{q_{W^\prime}}(x)+\frac{1}{8}B(W^\prime,W^\prime)\text{ mod }1\\
    &=\frac{1}{2}B(X,X-W^\prime)+\frac{1}{8}B(W^\prime,W^\prime)\text{ mod }1
\end{align}

Now it is clear that since $q_{W^\prime}$ is in the same equivalence class as $q_W$ (since
$W$ and $W^\prime$ are just different representatives for the
same Wu class) they both satisfy the Gauss sum 
(equation~(\ref{eq:gausssum})) for the same value of $C\text{ mod }8$.

On the other hand we already mentioned that $q$ and $q_W$ also satisfy
the Gauss sum for the same value of $C\text{ mod }8$ (by the lift
construction).  Hence they all
satisfy the Gauss sum for the same value of $C\text{ mod }8$.  
Now the Gauss sum can be viewed as a constraint that determines the 
constants (because when we stripped off the constants we showed that
 $\tilde{q}$ equals $\tilde{q_{W^\prime}}$).  
In this case we have no choice but to conclude
$q(0)=\frac{1}{8}B(W^\prime,W^\prime)\text{ mod }1$.

Summarizing, $q=q_{W^\prime}$ for some Wu representative
$W^\prime$, hence the equivalence class of quadratic refinements
$[q]$ actually lifts through the Nikulin construction (to $[q_W]$).  
We conclude that the trio $(\grp,[q],c)$ lifts.
\end{proof}

\section{Canonical quantization of Belov and Moore}
\label{sec:canonicalquantization}
In the last section we discussed the quantization of lattices.
We use the term quantization since the resulting trio of data $\trio$ encodes the 
quantization of toral (spin or non-spin) Chern-Simons
gauge theory.  In this section we transcribe the relevant Hilbert space structure
that arises from the wavefunctions constructed in \cite{belov_moore} and recall that
this provides a (non-extended) $2$-d modular functor (see chapter~(\ref{ch:tqft})).

\subsection*{Hilbert space preliminaries}
First it is useful to mention some preliminaries before reproducing the action of the mapping
class group for \textit{closed} surfaces
\footnote{Note that Belov and Moore study only \textit{fixed} vortices (marked arcs, or
colored boundary circles).
The braiding and twisting of such quasiparticles must also be described to specify
an extended $2$-d modular functor (see chapter~(\ref{ch:tqft})).  
Hence we restrict our attention to closed surfaces.}
on the Hilbert space of wavefunctions as described in section 5.6 of
\cite{belov_moore}.

Following Belov and Moore we avoid the special 
considerations that must be taken into account when the surface 
$\Sigma$ is the Riemann sphere (see chapter~(\ref{ch:tqft})) and skip to the case where
$\Sigma$ is a \textit{closed} oriented Riemann surface with genus $g\geq 1$.

Let us pick a \textit{canonical basis} for the first homology group $H_1(\Sigma,\mathbb{Z})$,
i.e. an \textit{ordered} set of loops $\lbrace a_i, b_i \rbrace_{i=1,\ldots,g}$ in $\Sigma$ such that
the oriented intersection numbers are given by
\begin{align}
  I(a_i,b_j) &= -I(b_j,a_i)= \delta_{ij} \\
  I(a_i,a_j) &= 0 \\
  I(b_i,b_j) &= 0 
\end{align}
Such a basis always exists (but is not unique) for
any closed Riemann surface $\Sigma$.
\footnote{This \textit{choice} of canonical basis is
a variant of the \textit{extra structure} that is required
on $\Sigma$ in order to define an anomaly-free TQFT.  
See chapter~(\ref{ch:tqft}).  Also we shall not bother to
distinguish between homology \textit{classes} and representative loops.}
Clearly this intersection matrix defines a
symplectic inner product on $H_1(\Sigma,\mathbb{Z})$.

Orientation-preserving diffeomorphisms map loops to loops \textit{and preserve intersection
numbers}, 
hence on the canonical basis $\lbrace a_i,b_i \rbrace$ the mapping class group
$\text{MCG}(\Sigma)$ acts via invertible integer-valued matrices 
\textit{that leave the symplectic inner product matrix unchanged}.
Such matrices are elements
of the (integral) symplectic group $\text{Sp}(2g,\mathbb{Z})$.  So we have a map
\begin{equation}
  \text{MCG}(\Sigma)\rightarrow \text{Sp}(2g,\mathbb{Z})
\end{equation}
In general this map is surjective and the kernel is the Torelli group.  It
is claimed in \cite{belov_moore} that for the abelian theories considered
there the Torelli group acts trivially.
In other words the mapping class group action on the wavefunctions is
encoded entirely in $\text{Sp}(2g,\mathbb{Z})$ for abelian theories.

Since $H_1(\Sigma,\mathbb{Z})$ is $2g$-dimensional let us write the choice of
canonical basis using the convention
\begin{equation}
  a_1 =
  \begin{pmatrix}
    1\\
    0\\
    \vdots\\
    0\\
    0\\
    \vdots\\
    0\\
  \end{pmatrix}
  \quad
  \cdots
  \quad
  a_g =
  \begin{pmatrix}
    0\\
    0\\
    \vdots\\
    1\\
    0\\
    \vdots\\
    0\\
  \end{pmatrix}
  \quad
  \quad
  b_1 =
  \begin{pmatrix}
    0\\
    0\\
    \vdots\\
    0\\
    1\\
    \vdots\\
    0\\
  \end{pmatrix}
  \quad
  \cdots
  \quad
  b_g =
  \begin{pmatrix}
    0\\
    0\\
    \vdots\\
    0\\
    0\\
    \vdots\\
    1\\
  \end{pmatrix}
\end{equation}
The symplectic group is then generated by matrices of the form
\begin{align}
  &\begin{pmatrix}
     A & 0 \\
     0 & A^{-1,t}
  \end{pmatrix}
  ,\;A\in \text{GL}(g,\mathbb{Z}),\text{ i.e. det}(A)=\pm 1 \label{eq:symplecticgenerators}\\
  &\begin{pmatrix}
     \idmat_g & B \\
     0 & \idmat_g
  \end{pmatrix}
  ,\;B\text{ is any symmetric integral }g\times g\text{ matrix}\notag\\
  &\begin{pmatrix}
     0 & -\idmat_g \\
     \idmat_g & 0
  \end{pmatrix}\notag
\end{align}
As usual in genus $g=1$ these matrices are $\idmat$, $t$, and $s$ - 
the familiar generators of the modular group 
$\text{SL}(2,\mathbb{Z})\cong\text{Sp}(2,\mathbb{Z})$
\begin{equation}
  \idmat=\begin{pmatrix}
     1 & 0 \\
     0 & 1
  \end{pmatrix}
  \qquad
  t=\begin{pmatrix}
     1 & 1 \\
     0 & 1
  \end{pmatrix}
  \qquad
  s=\begin{pmatrix}
     0 & -1 \\
     1 & 0
  \end{pmatrix}
\end{equation}

The chosen canonical basis $\lbrace a_i,b_i \rbrace_{i\in 1,\ldots,g}$ for
$H_1(\Sigma,\mathbb{Z})$ induces a dual basis $\lbrace \alpha_i,\beta_i\rbrace_{i\in 1,\ldots,g}$
of integral 1-forms $H^1(\Sigma,\mathbb{Z})$.  This is useful since
(chapter~(\ref{ch:tqft})) the K\"ahler quantization procedure
has as classical configuration space the moduli 
space of flat connections $\mathscr{M}$ (which are essentially 1-forms).
The Hilbert space is comprised of wavefunctions of the form $\Psi(\text{1-forms})$.

Using the dual basis $\lbrace \alpha_i,\beta_i\rbrace_{i\in 1,\ldots,g}$
we can decompose any 1-form $\omega$
\footnote{The universal coefficient theorem tells us that $H^1(\Sigma,\mathbb{R})\cong H^1(\Sigma,\mathbb{Z})\otimes\mathbb{R}$.}
into
\footnote{Warning: our notation diverges from that in \cite{belov_moore}.  We use $\omega_1$ and 
$\omega_2$
instead of $a^1$ and $a_2$ to avoid notation collisions.  Our indices are also placed
differently.}
\begin{equation}
  \omega = \omega_1^i \alpha_i + \omega_2^i \beta_i
\end{equation}
for $\omega_1^i,\omega_2^i\in\mathbb{R}$.
The transformations in equation~(\ref{eq:symplecticgenerators}) are transposed when
acting on the dual basis $\lbrace \alpha_i,\beta_i\rbrace_{i\in 1,\ldots,g}$
\begin{align}
  \label{eq:symplecticdualgenerators}
  &\begin{pmatrix}
     A^t & 0 \\
     0 & A^{-1}
  \end{pmatrix}
  ,\;A\in \text{GL}(g,\mathbb{Z}),\text{ i.e. det}(A)=\pm 1 \\
  &\begin{pmatrix}
     \idmat_g & 0 \\
     B^t & \idmat_g
  \end{pmatrix}
  ,\;B\text{ is any symmetric integral }g\times g\text{ matrix}\\
  &\begin{pmatrix}
     0 & \idmat_g \\
     -\idmat_g & 0
  \end{pmatrix}
\end{align}
(obviously $B^t=B$).  The induced action on any wavefunction is given by
\begin{enumerate}
 \item A transform:
  \begin{equation}
    \label{eq:Atransform1}
    (M_A\cdot \Psi)(\omega) := \Psi(M_A\cdot\omega)=\Psi(A^t\cdot \omega_1,A^{-1}\cdot \omega_2)
  \end{equation}
 \item B transform:
  \begin{equation}
    \label{eq:Btransform1}
    (M_B\cdot \Psi)(\omega) := \Psi(M_B\cdot\omega)=\Psi(\omega_1,\omega_2+B\cdot \omega_1)
  \end{equation}
 \item S transform:
  \begin{equation}
    \label{eq:Stransform1}
    (M_S\cdot \Psi)(\omega) := \Psi(M_S\cdot\omega)=\Psi(\omega_2,-\omega_1)
  \end{equation}
\end{enumerate}

Now let us discuss a few further constructions utilized in 
\cite{belov_moore} to write down a \textit{basis} of wavefunctions 
(and to understand the above group action in terms of this basis).

\subsection*{Dependence on spin structure and Wu class}
The basis of wavefunctions depends on the choice of spin structure and choice of
Wu class (see below).
First, it is a fact that any compact oriented 3-manifold $X$ admits at least
one spin structure \cite{stipsicz}.
This is equivalent to saying that the first and second Stiefel-Whitney classes 
(which are valued in $H^1(X,\frac{1}{2}\mathbb{Z}/\mathbb{Z})$)
for the tangent bundle vanish, i.e.
$w_1(TX)=w_2(TX)=0\in H^1(X,\frac{1}{2}\mathbb{Z}/\mathbb{Z})$ (in
fact $TX$ is trivializable).

The group $H^1(X,\frac{1}{2}\mathbb{Z}/\mathbb{Z})$ itself need not be zero, however.
In fact $H^1(\Sigma,\frac{1}{2}\mathbb{Z}/\mathbb{Z})$ enumerates the
different possible spin structures on $X$.
\footnote{The space of spin structures is an 
$H^1(\Sigma,\frac{1}{2}\mathbb{Z}/\mathbb{Z})$-torsor.  However given our
choice of canonical homology basis $\lbrace a_i, b_i \rbrace_{i=1,\ldots,g}$
a preferred spin structure is determined (see pg. 27 of \cite{belov_moore}).
We identify this with $0\in H^1(X,\frac{1}{2}\mathbb{Z}/\mathbb{Z})$ (i.e.
we have fixed a preferred origin for the spin structures, and hence the
space of spin structures can be identified with $H^1(\Sigma,\frac{1}{2}\mathbb{Z}/\mathbb{Z})$
itself).}
Explicitly for a manifold of the form $X=\Sigma\times I$ (as in the current
Hamiltonian formulation) we have that
$H^1(X,\frac{1}{2}\mathbb{Z}/\mathbb{Z})\cong H^1(\Sigma,\frac{1}{2}\mathbb{Z}/\mathbb{Z})$
since $X$ deformation retracts onto $\Sigma$.  But by the universal coefficient
theorem we see that
\begin{equation}
  H^1(\Sigma,\frac{1}{2}\mathbb{Z}/\mathbb{Z})\cong 
 H^1(\Sigma,\mathbb{Z})\otimes\frac{1}{2}\mathbb{Z}/\mathbb{Z}
\end{equation}
Manifestly this has $2^{2g}$ elements that
can be written in terms of the dual basis $\lbrace \alpha_i,\beta_i\rbrace_{i\in 1,\ldots,g}$ 
(but with $\frac{1}{2}\mathbb{Z}/\mathbb{Z}$ coefficients).

In light of this let us encode a fixed spin structure by specifying a
set of coefficients $([\epsilon_1], [\epsilon_2])\in(\frac{1}{2}\mathbb{Z}/\mathbb{Z})^{2g}$
(i.e. a spin structure is given by 
$[\epsilon_1]\cdot\alpha+[\epsilon_2]\cdot\beta\in H^1(\Sigma,\frac{1}{2}\mathbb{Z}/\mathbb{Z})$).
For this fixed spin structure the main idea is to define a
Hilbert space $H_{([\epsilon_1],[\epsilon_2])}$ of wavefunctions using
theta functions.

The spin structure $([\epsilon_1],[\epsilon_2])$ is not the only
piece of data needed to write down a Hilbert space.  Recall from 
section~(\ref{sec:latticequantization}) that the ``quantization''
of a classical lattice is encoded in the data
\begin{equation}
  (\grp,[q_W],c)
\end{equation}
where $\grp$ is a finite abelian group, $[q_W]:\grp\rightarrow\qmodz$ is an equivalence class
of quadratic forms on $\grp$ constructed from the Wu class $[W]\in\Lambda^*/2\Lambda^*$ of
the classical lattice, and 
$c$ is an integer mod $24$ that is essentially a choice of cube root
of the Gauss reciprocity formula.  The content
of the Belov-Moore construction is that the Hilbert space (and action of
the mapping class group) is determined by this data alone.  So we add additional
decoration to the above Hilbert space
\begin{equation}
  H_{([\epsilon_1],[\epsilon_2]),(\grp,[q_W],c)}
\end{equation}
or, more compactly
\begin{equation}
  H_{([\epsilon_1],[\epsilon_2]),[W]}
\end{equation}

$H_{([\epsilon_1],[\epsilon_2]),[W]}$ can only be explicitly written down by
picking a representative 
$(\epsilon_1,\epsilon_2)\in(\frac{1}{2}\mathbb{Z})^{2g}$ 
of $([\epsilon_1],[\epsilon_2])$.
Likewise,
we are forced to pick an explicit representative $W\in\Lambda^*$ from the Wu class $[W]$.
Unfortunately the basis of wavefunctions
\textit{does naively depend} on these representative choices, however 
different bases constructed from 
different representatives
are gauge equivalent by an explicit set of gauge transformations (which we
list below).  Hence there is no loss in generality when picking representatives
$(\epsilon_1,\epsilon_2)$ and $W$:
\begin{equation}
  H_{(\epsilon_1,\epsilon_2),W} 
\end{equation}
As discussed in \cite{belov_moore} there are 
precisely $\vert\grp^{g}\vert$ basis wavefunctions in $H_{(\epsilon_1,\epsilon_2),W}$ 
enumerated by $\gamma\in\grp^{g}$ (i.e. there is a copy of the discriminant group
$\grp$ for each canonical basis loop $b_i$ where $i\in 1,\ldots,g$):
\begin{equation}
  \Psi_{\gamma,(\epsilon_1,\epsilon_2),W}(\text{1-forms})
\end{equation}

The transformation laws that map one basis of wavefunctions determined by a choice
of representative $(\epsilon_1,\epsilon_2),W$ to another choice are derived at the end of
section 5.3 in \cite{belov_moore} (and more succinctly in equation~5.42 in
\cite{belov_moore}).  
Recall that we are not considering
vortices here.
\footnote{In the language of \cite{belov_moore} set $c^1=c_2=0$.}
The dependence on $W$ is shown in \cite{belov_moore}, but we shall 
not need it since
the Wu representative is unaltered by the action of the symplectic group.
The dependence on representative $(\epsilon_1,\epsilon_2)$, however, is necessary
in what follows.  We have
\begin{multline}
  \label{eq:spinstructuretransformation}
  \Psi_{\gamma,(\epsilon_1+n_1,\epsilon_2+n_2),W}=\\ 
  e^{8\pi i q_W(0)[n_1\cdot n_2 + \epsilon_1\cdot n_2 - \epsilon_2 n_1]+
     2\pi i n_2^i[q_W(-\gamma_i)-q_W(\gamma_i)]}
   \Psi_{\gamma+n_1\otimes\overline{W},(\epsilon_1,\epsilon_2),W}
\end{multline}
where $\overline{W}$ is the projection of $W$ into the discriminant group $\grp$.  
The repeated index $i=1,\ldots,g$ is summed over, as usual (manifestly $(n_1,n_2)\in\mathbb{Z}^{2g}$).

The results mentioned in the next subsection show that the action of the mapping class
group on the theta functions (as formally described in
equations~(\ref{eq:Atransform1}), (\ref{eq:Btransform1}), (\ref{eq:Stransform1}))
does \textit{not} preserve the spin structure.  In light of this Belov and
Moore proposed that the full Hilbert space for the theory must be written
as a direct sum over the separate spin structures:
\begin{equation}
  \label{eq:hilbertspacesum}
  H_{[W]}=\bigoplus_{[\epsilon_1], [\epsilon_2]\in(\frac{1}{2}\mathbb{Z}/\mathbb{Z})^{\otimes g}}
    H_{[\epsilon_1], [\epsilon_2],[W]}
\end{equation}

\subsection*{Action of the mapping class group on theta functions}
Using the properties of theta functions (see \cite{belov_moore}) it is possible to cast the
action of the mapping class group (discussed in equations~(\ref{eq:Atransform1}), (\ref{eq:Btransform1}), (\ref{eq:Stransform1}))
into new expressions
(we add the extra decorations to the wavefunctions from here):
\begin{enumerate}
 \item A transform:
  \begin{multline}
    \label{eq:Atransform2}
    (M_A\cdot \Psi_{\gamma,(\epsilon_1,\epsilon_2),W})(\omega) := 
    \Psi_{\gamma,(\epsilon_1,\epsilon_2),W}(M_A\cdot\omega)=\\
    \Psi_{\gamma,(\epsilon_1,\epsilon_2),W}(A^t\cdot \omega_1,A^{-1}\cdot \omega_2)=
    \Psi_{A^t\gamma,(A^t\epsilon_1,A^{-1}\epsilon_2),W}(\omega_1,\omega_2)
  \end{multline}
 \item B transform:
  \begin{multline}
    \label{eq:Btransform2}
    (M_B\cdot \Psi_{\gamma,(\epsilon_1,\epsilon_2),W})(\omega) := 
    \Psi_{\gamma,(\epsilon_1,\epsilon_2),W}(M_B\cdot\omega)=\\
    \Psi_{\gamma,(\epsilon_1,\epsilon_2),W}(\omega_1,\omega_2+B\cdot \omega_1)=
    e^{2\pi i \phi(B)c/24}e^{4\pi i\epsilon_1^i B^{ii}q_W(0) -
    2\pi i B^{ii}[q_W(\gamma_i)-q_W(0)]}\\
    \times e^{-2\pi i\Sigma_{i<j}B^{ij}b(\gamma_i,\gamma_j)}
    \Psi_{\gamma,(\epsilon_1,\epsilon_2-B\epsilon_1-\frac{1}{2}\text{diag}(B)),W}(\omega_1,\omega_2)
  \end{multline}
 \item S transform:
  \begin{multline}
    \label{eq:Stransform2}
    (M_S\cdot \Psi_{\gamma,(\epsilon_1,\epsilon_2),W})(\omega) := 
    \Psi_{\gamma,(\epsilon_1,\epsilon_2),W}(M_S\cdot\omega)=\\
    \Psi_{\gamma,(\epsilon_1,\epsilon_2),W}(\omega_2,-\omega_1)=
    |\grp|^{-g/2}
    \sum_{\gamma^\prime\in\grp^g}e^{2\pi i b(\gamma_i,\gamma_i^\prime)}
    \Psi_{\gamma^\prime,(-\epsilon_2,\epsilon_1),W}(\omega_1,\omega_2)
  \end{multline}
\end{enumerate}
Here $b(\cdot,\cdot):\grp\rightarrow\qmodz$ is the bilinear form determined by
$q_W$ and $i,j\in 1,\ldots,g$ are summed over when the indices are repeated
(except the $i$ in $2\pi i$ means $2\pi\sqrt{-1}$ of course).  The
quantity $\phi(B)$ is an integer determined from the matrix $B$ (see \cite{belov_moore}).

We will always choose the representative
$(\epsilon_1,\epsilon_2)\in(\frac{1}{2}\mathbb{Z})^{2g}$
such that every element $\epsilon_1^i,\epsilon_2^i$ is either $0$ or $\frac{1}{2}$ 
(if the above action on the
basis wavefunctions destroys this
choice then we can use equation~(\ref{eq:spinstructuretransformation}) to put each
element back into this form).

\subsection*{Even (non-spin) theories}
For the case of an even (non-spin) topological quantum field theory 
(see section~(\ref{sec:latticequantization})) we can
always make the special choice for Wu representative $W=0$ 
(the quadratic form $q_W$ is then pure).
In this case the
spin structure $([\epsilon_1],[\epsilon_2])$ is irrelevant. 
The basis wavefunctions are written in terms of the
theta functions \textit{up to non-trivial normalization factors}
(see page 28 in \cite{belov_moore} and the
other references cited there for
greater detail):
\begin{equation}
  \Psi_{\gamma,(\epsilon_1,\epsilon_2),W}(\omega_1,\omega_2)\sim 
  \Theta_{\Lambda+\gamma}^{\epsilon_1\otimes W,\epsilon_2\otimes W}(\omega_1,\omega_2)
\end{equation}
Clearly if we set $W=0$ then different spin structures $([\epsilon_1],[\epsilon_2])$
produce the same wavefunctions.  
The full Hilbert space is \textit{not a direct sum} over spin structures as in
equation~(\ref{eq:hilbertspacesum}).  Instead there are only $|\grp^g|$ basis
wavefunctions, and the action
of the symplectic group reduces to
\begin{enumerate}
 \item A transform (even theory):
  \begin{equation}
    \label{eq:Atransformeven}
    (M_A\cdot \Psi_{\gamma})(\omega)=
    \Psi_{A^t\gamma}(\omega)
  \end{equation}
 \item B transform (even theory):
  \begin{equation}
    \label{eq:Btransformeven}
    (M_B\cdot \Psi_{\gamma})(\omega)=
    e^{2\pi i \phi(B)c/24}e^{-2\pi i B^{ii}q_W(\gamma_i)}
    e^{-2\pi i\Sigma_{i<j}B^{ij}b(\gamma_i,\gamma_j)}
    \Psi_{\gamma}(\omega)
  \end{equation}
 \item S transform (even theory):
  \begin{equation}
    \label{eq:Stransformeven}
    (M_S\cdot \Psi_{\gamma})(\omega)=|\grp|^{-g/2}
    \sum_{\gamma^\prime\in\grp^g}e^{2\pi i b(\gamma_i,\gamma_i^\prime)}
    \Psi_{\gamma^\prime}(\omega)
  \end{equation}
\end{enumerate}

\subsection*{An example in genus 1}
In genus 1 the above symplectic group action on the Hilbert space of wavefunctions
can be made more explicit.  We take this
opportunity to correct some slight calculational errors in subsection 5.6.1 of \cite{belov_moore}
for the benefit of the reader.

Denote the matrix elements of an operator $\mathscr{O}$ acting from 
$H_{\epsilon_1, \epsilon_2,W}$ to $H_{\epsilon_1^\prime, \epsilon_2^\prime,W}$
by the notation $\mathscr{O}_{\gamma^\prime}^\gamma
\bigl [\begin{smallmatrix}2\epsilon_1 && 2\epsilon_2\\
2\epsilon_1^\prime && 2\epsilon_2^\prime\end{smallmatrix}\bigr ]$.
\footnote{Beware: our primed and unprimed indices are exactly opposite to
that in \cite{belov_moore}.  We seek to remain consistent with our previous
notation.}
Then in genus 1 the $t$ and $s$ symplectic matrices induce operators $T$
and $S$ given by the following matrix elements (everything not listed is
zero):
\begin{align}
T_{\gamma^\prime}^\gamma
\bigl [\begin{smallmatrix}0 && 0\\0 && 1\end{smallmatrix}\bigr ]&=
e^{2\pi ic/24-2\pi i[q_W(-\gamma)-q_W(0)]}\delta_{\gamma^\prime}^\gamma\\
T_{\gamma^\prime}^\gamma
\bigl [\begin{smallmatrix}0 && 1\\0 && 0\end{smallmatrix}\bigr ]&=
e^{2\pi ic/24-2\pi i[q_W(\gamma)-q_W(0)]}\delta_{\gamma^\prime}^\gamma\\
T_{\gamma^\prime}^\gamma
\bigl [\begin{smallmatrix}1 && 0\\1 && 0\end{smallmatrix}\bigr ]&=
T_{\gamma^\prime}^\gamma
\bigl [\begin{smallmatrix}1 && 1\\1 && 1\end{smallmatrix}\bigr ]=
e^{2\pi ic/24-2\pi iq_W(-\gamma)}\delta_{\gamma^\prime}^\gamma
\end{align}
The $S$ matrices are
\begin{align}
S_{\gamma^\prime}^\gamma
\bigl [\begin{smallmatrix}0 && 0\\0 && 0\end{smallmatrix}\bigr ]&=
S_{\gamma^\prime}^\gamma
\bigl [\begin{smallmatrix}1 && 0\\0 && 1\end{smallmatrix}\bigr ]=
|\grp|^{-1/2}e^{2\pi ib(\gamma,\gamma^\prime)}\\
S_{\gamma^\prime}^\gamma
\bigl [\begin{smallmatrix}0 && 1\\1 && 0\end{smallmatrix}\bigr ]&=
|\grp|^{-1/2}e^{2\pi ib(\gamma,\gamma^\prime+\overline{W})}\\
S_{\gamma^\prime}^\gamma
\bigl [\begin{smallmatrix}1 && 1\\1 && 1\end{smallmatrix}\bigr ]&=
|\grp|^{-1/2}e^{2\pi ib(\gamma,\gamma^\prime+\overline{W})+4\pi iq_W(0)}
\end{align}

For \textit{even} theories the spin labelling collapses since set $W=0$.
Since $q_W$ is then pure we have $q_W(0)=0$ and $q_W(-\gamma)=q_W(\gamma)$.
The resulting $T$ and $S$ operators are much simpler
\begin{align}
T_{\gamma^\prime}^\gamma&=
e^{2\pi ic/24-2\pi iq_W(\gamma)}\delta_{\gamma^\prime}^\gamma\\
S_{\gamma^\prime}^\gamma&=|\grp|^{-1/2}e^{2\pi ib(\gamma,\gamma^\prime)}
\end{align}

\chapter{Modular Tensor Categories}
\label{ch:mtc}

\section{Introduction}
The goal of this chapter is to provide a brief sketch of modular tensor
categories to lay a foundation for future chapters.  Modular tensor
categories (MTCs) grew somewhat simultaneously out of the study of
conformal field theory by Moore and Seiberg \cite{moore_seiberg}
and quantum groups by Lusztig, Jimbo, 
Reshetikhin and Turaev, and others
(see the references in \cite{reshetikhin_turaev1}, \cite{reshetikhin_turaev2},
and \cite{kirby_melvin} for a more complete listing).

For the majority of this chapter we follow \cite{turaev} and \cite{bakalov_kirillov}
(borrowing conventions and notation from both).  Our arrows will be in exactly
the opposite direction to those in \cite{turaev}.  We also follow the definition
of the $S$-matrix in \cite{bakalov_kirillov}.  We have also found the unpublished
notes of Boyarchenko \cite{boyarchenko} useful.

Both books \cite{turaev} and \cite{bakalov_kirillov} consider in detail \textit{strict}
ribbon categories.  This is not sufficient for our purposes
and hence we shall consider ribbon categories that are not necessarily strict.
However, since strict categories are easier to understand we consider them first
in all of the definitions below.

\section{Monoidal categories}
\subsection*{Strict monoidal categories}
\begin{definition}
A \textbf{strict monoidal category}
is a category $\Vcat$ equipped with
a covariant bifunctor
\footnote{
By \textit{covariant bifunctor} we mean that for any two objects $V,W\in\text{Ob}(\Vcat)$
there is an object $V\otimes W\in\text{Ob}(\Vcat)$, and for any two morphisms
$f:V\rightarrow V^\prime$ and $g:W\rightarrow W^\prime$ there is a morphism
$f\otimes g:V\otimes W\rightarrow V^\prime\otimes W^\prime$.  Functoriality
means that given morphisms
$f^\prime:V^\prime\rightarrow V^{\prime\prime}$,
$g^\prime:W^\prime\rightarrow W^{\prime\prime}$ the following identities are
required to be satisfied:
\begin{equation}
  (f^\prime\circ f)\otimes (g^\prime\circ g)=(f^\prime\otimes g^\prime)\circ(f\otimes g)
\end{equation}
\begin{equation}
  \id_V\otimes\id_W=\id_{V\otimes W}
\end{equation}}
 $\otimes:\Vcat\times\Vcat\rightarrow\Vcat$ 
and a distinguished object $\unitobj$ such
that the following two identities hold:
\begin{enumerate}
 \item Strict identity:
\begin{equation}
  U\otimes\unitobj = \unitobj\otimes U = U
  \label{eq:stricttriangle}
\end{equation}
\item Strict associativity:
\begin{equation}
  (U\otimes V)\otimes W = U\otimes (V\otimes W)
  \label{eq:strictpentagon}
\end{equation}
\end{enumerate}
\end{definition}

\begin{example}
A simple example of a strict monoidal category 
is the category $\text{Vect}_\cplx$ of complex vector spaces
under the usual tensor product.  Here the unit object is $\unitobj=\cplx$.
\end{example}

\begin{example}
\label{ex:ribi}
Now we construct a more complicated strict monoidal category $\ribi$, called
the category
of colored ribbon graphs.  Here $I$ is some auxilliary set of labels (``colors'').

First we require some preliminary definitions.  We will be rather informal
here since the following definition is written carefully in \cite{turaev}:
\begin{definition}
A \textbf{$(k,l)$-ribbon graph} $\Omega$ is an \textit{oriented} surface
in $\mathbb{R}^3$ \textit{up to isotopy}.  
The surface is constructed out of elementary pieces (see figure~(\ref{fig:ribbongraph})):
\begin{enumerate}
 \item oriented \textit{ribbons} (long vertical strips)
 \item \textit{coupons} (horizontal strips)
 \item oriented \textit{annuli}
\end{enumerate}
\begin{figure}
  \centering
  \input{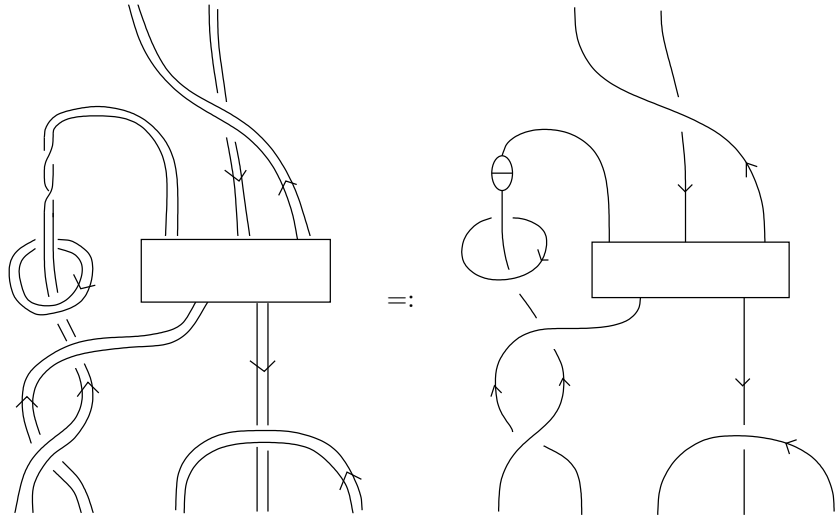}
  \caption{A $(k=5,l=2)$-ribbon graph.  The diagrammatic presentation
    is depicted on the right.}
  \label{fig:ribbongraph}
\end{figure} 
Each coupon has a distinguished bottom side (``in'') and distinguished top side (``out'')
on which ribbon ends can be connected.
\footnote{The graph should be thought of as evolving from the bottom to the top.
Note that the orientations of the ribbons does not have any relationship with being
glued to the ``in'' or ``out'' side.}
Any ribbon end that
terminates on a coupon is \textit{not} allowed to slide from the ``in'' side to the ``out'' side
(or vica versa)
under isotopy.

For a $(k,l)$-ribbon graph there 
are $k\geq 0$ free ribbon ends that are marked as ``inputs'', and 
likewise there are $l\geq 0$ free ribbon ends that are marked
as ``outputs''.  In fact it is always possible to perform an isotopy to put the ribbon graph
$\Omega$ into a \textit{standard drawing position} (see figure~(\ref{fig:ribbongraph})),
i.e.:
\begin{enumerate}
 \item The $k$ ``input'' free ribbon ends are at the bottom.  They are
    ordered from left to right (the ordering can be changed
    by braiding the free ribbon ends over/under each another).
 \item The $l$ ``output'' free ribbon ends are at the top.  They are
    ordered from left to right.
 \item The graph is ``face up'' (determined by the orientation of $\Omega$)
   except in finitely-many localized places where the ribbons
   are \textit{twisted} (see figure~(\ref{fig:twist})). 
 \item The graph sits entirely in the plane of the drawing except
   at a finite number of overcrossings, undercrossings, and twists
   (see figure~(\ref{fig:braid})).
\end{enumerate}
Because of the standard drawing position it is clear that we can represent
any ribbon graph by a \textbf{ribbon diagram}, i.e. a diagram where the oriented ribbons are replaced
by their oriented cores.  The ribbons can be recovered by using the blackboard framing.  See
the right side of figure~(\ref{fig:ribbongraph}).
\begin{figure}
  \centering
  \input{twist.eps_t}
  \caption{On the top is depicted a right twist (a $(1,1)$-ribbon graph).  On the bottom is depicted
    a left twist (a $(1,1)$-ribbon graph).  The diagrammatic presentation
    is depicted on the right for each.}
  \label{fig:twist}
\end{figure} 
\begin{figure}
  \centering
  \input{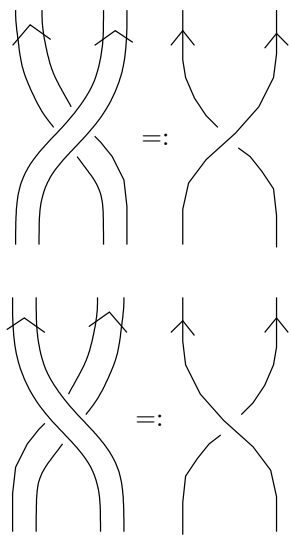}
  \caption{On the top is depicted a right braid (a $(2,2)$-ribbon graph).  On the bottom is depicted
    a left braid (a $(2,2)$-ribbon graph).  The diagrammatic presentation
    is depicted on the right for each.}
  \label{fig:braid}
\end{figure} 

Now let $I$ be a set of labels (colors).  We define a \textbf{colored $(k,l)$-ribbon graph}
as a $(k,l)$-ribbon graph where each ribbon and each annulus is labeled by some element in
$I$ (we do not color the coupons yet).
\end{definition}
\begin{figure}
  \centering
  \input{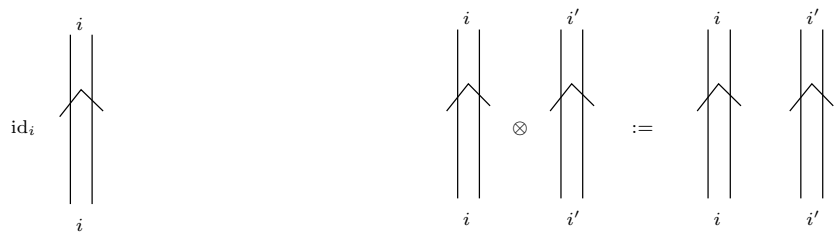}
  \caption{The identity morphism $\id_i:[[i,+1]]\rightarrow[[i,+1]]$ in $\ribi$ is depicted
    on the left.  The tensor product of morphisms in $\ribi$ (in this case two
    identity morphisms) is depicted on the right.}
  \label{fig:ribi}
\end{figure} 

\begin{definition}
Define a strict monoidal category $\ribi$ as follows:
\begin{enumerate}
 \item The objects are ordered lists $[[i_1,\pm 1],[i_2,\pm 1],\ldots]$ where
    $i_1,i_2,\ldots\in I$.  The unit object $\unitobj$ is the empty list $[]$.
 \item Given objects $[[i_1,\pm 1],[i_2,\pm 1],\ldots,[i_k,\pm 1]]$ and
    $[[i^\prime_1,\pm 1],[i^\prime_2,\pm 1],\ldots,[i^\prime_l,\pm 1]]$ a
    morphism between them is a colored $(k,l)$-ribbon graph such that
    the $k$ ``input'' ribbons are labeled (in order) by
    $i_1,\ldots,i_k$ and each ribbon is directed up for $+1$ and
    directed down for $-1$.  Similary the $l$ ``output'' ribbons are
    labeled by $i^\prime_1,\ldots,i^\prime_l$ where they are directed
    up for $+1$ and down for $-1$.  It is obvious that these morphisms
    can be composed by stacking colored ribbon graphs on top of each other.
\end{enumerate}
$\ribi$ is a strict monoidal category since any two ordered lists can be concatenated
\begin{multline}
[[i_1,\pm 1],[i_2,\pm 1],\ldots,[i_k,\pm 1]]\otimes
[[i^\prime_1,\pm 1],[i^\prime_2,\pm 1],\ldots,[i^\prime_l,\pm 1]]=\\
[[i_1,\pm 1],[i_2,\pm 1],\ldots,[i_k,\pm 1],[i^\prime_1,\pm 1],
  [i^\prime_2,\pm 1],\ldots,[i^\prime_l,\pm 1]]
\end{multline}
(this defines $\otimes$ on the objects) and ribbon graphs can be placed adjacent
to each other (this defines $\otimes$ on the morphisms - see e.g. the right side of
figure~(\ref{fig:ribi})).
\end{definition}
\end{example}

\subsection*{(Non-strict) monoidal categories}
We now consider monoidal categories that may not be strict.
\begin{definition}
A \textbf{monoidal category} is a category $\Vcat$ equipped with
a covariant bifunctor $\otimes:\Vcat\times\Vcat\rightarrow\Vcat$ 
and a distinguished object $\unitobj$.  Furthermore we require a family
of natural isomorphisms (for all objects $U$, $V$, $W$, $X$):
\begin{equation}
  \lbrace a_{U,V,W}:(U\otimes V)\otimes W\rightarrow U\otimes (V\otimes W)\rbrace 
\end{equation}
\begin{equation}
  \lbrace r_U:U\otimes\unitobj\rightarrow U\rbrace
  \label{eq:rightidentity}
\end{equation}
\begin{equation}
  \lbrace l_U:\unitobj\otimes U\rightarrow U\rbrace
  \label{eq:leftidentity}
\end{equation}
such that the following diagrams commute:

Pentagon diagram:
\begin{equation}
 \xymatrix @C-30pt @R+30pt {
  & *{(U\otimes V)\otimes (W\otimes X)}\ar[dr]^{a_{U,V,W\otimes X}} & \\
  *{((U\otimes V)\otimes W)\otimes X}\ar[ur]^{a_{U\otimes V,W,X}}
  \ar[d]^{a_{U,V,W}\otimes\id_X}
    && *{U\otimes(V\otimes(W\otimes X))}\\
  *{(U\otimes(V\otimes W))\otimes X}\ar[rr]+<-50pt,0pt>^-{a_{U,V\otimes W,X}}
    && *{U\otimes((V\otimes W)\otimes X)}\ar[u]^{\id_U\otimes a_{V,W,X}}
} 
\label{eq:pentagon}
\end{equation}

Triangle diagram:
\begin{equation}
  \xymatrix @C-20pt @R+30pt {
    *{(U\otimes\unitobj)\otimes V}\ar[rr]+<-30pt,0pt>^-{a_{U,\unitobj,V}}\ar[dr]_{r_U\otimes\id_V} 
    && *{U\otimes(\unitobj\otimes V)}\ar[dl]^{\id_U\otimes l_U} \\
    & *{U\otimes V} &
  } 
\label{eq:triangle}
\end{equation}
The MacLane Coherence Theorem \cite{maclane} states that if these commutative
diagrams are satisfied then \textit{any} diagram involving $a,r,l$
is commutative, i.e.:
\begin{enumerate}
 \item given any ordered list $A$ of objects that are tensored together and grouped
   with parenthesis,
 \item and given the same ordered list $A^\prime$ but with different parenthesis grouping (and possibly
   with unit objects $\unitobj$ appearing/not appearing in different places),
 \item then any two ways of getting from $A$ to $A^\prime$ using any combination of the 
   maps $a,r,l$ are the \textit{same}. 
\end{enumerate}
This implies in particular that any monoidal category is \textit{monoidal equivalent} (see
chapter~(\ref{ch:groupcategories})) to a strict monoidal category.

\begin{example}
There is a straightforward ``non-associative'' generalization of colored $(k,l)$-ribbon graphs
constructed by Bar-Natan in \cite{bar-natan},
and it is not difficult to construct the corresponding (non-strict) monoidal category
$\ribi^{\text{NS}}$.  For example the objects are ordered lists \textit{with parenthesis}
$[([i_1,\pm 1],[i_2,\pm 1]),\ldots]$, and the morphisms
are \textit{non-associative} colored $(k,l)$-ribbon graphs.
\end{example}
\end{definition}

\section{Braided monoidal categories}
%Any monoidal category is monoidal equivalent (chapter~(\ref{ch:groupcategories})) to
%a strict monoidal category.  
In this section we define 
\textit{braided} monoidal categories.
%Unfortunately it is \textit{not} true that any braided monoidal category
%is \textit{braided monoidal equivalent}
%\footnote{see chapter~(\ref{ch:groupcategories})}
%to a braided \textit{strict} monoidal category
%(see pg. 251 of \cite{maclane}).  Hence it is not always appropriate to restrict
%attention to braided strict monoidal categories.
The natural setting for the examples in this paper are braided (non-strict)
monoidal categories.  However, we discuss braided strict monoidal categories
first since they are easier to understand.

\subsection*{Braided strict monoidal categories}
\begin{definition}
A \textbf{braided strict monoidal category}
is a strict monoidal category equipped
with a family of natural \textit{braiding} isomorphisms (for all pairs of objects)
\begin{equation}
  \lbrace c_{U,V}:U\otimes V\rightarrow V\otimes U \rbrace
\end{equation}
The braiding isomorphisms represent a \textit{weak} form of commutativity.
Note that it is \textit{not} usually true that $c_{V,U}\circ c_{U,V}=\id_{U\otimes V}$.
If this condition is satisfied then the category is called \textbf{symmetric} (we
are interested in non-symmetric categories).

The braiding isomorphisms are required to satisfy the following \textit{hexagon relations}:
\begin{equation}
  \xymatrix @R+20pt {
    & *{A\otimes (B\otimes C)}\ar@{=>}[rr]^-{c_{A,B\otimes C}}\ar@{.>}[dl]^-{\id} && 
      *{(B\otimes C)\otimes A} & \\
    *{(A\otimes B)\otimes C}\ar[dr]^-{c_{A,B}\otimes\id_C}
      &&&& *{B\otimes(C\otimes A)}\ar@{.>}[ul]^-{\id} \\
    & *{(B\otimes A)\otimes C}\ar@{.>}[rr]^-{\id} && *{B\otimes (A\otimes C)}
      \ar[ur]^-{\id_B\otimes c_{A,C}}
  } 
\label{eq:stricthexagon1}
\end{equation}
\begin{equation}
  \xymatrix @R+20pt {
    & *{(U\otimes V)\otimes W}\ar@{=>}[rr]^-{c_{U\otimes V,W}}\ar@{.>}[dl]^-{\id} && 
      *{W\otimes (U\otimes V)} & \\
    *{U\otimes (V\otimes W)}\ar[dr]^-{\id_U\otimes c_{V,W}}
      &&&& *{(W\otimes U)\otimes V}\ar@{.>}[ul]^-{\id} \\
    & *{U\otimes (W\otimes V)}\ar@{.>}[rr]^-{\id} && *{(U\otimes W)\otimes V}
      \ar[ur]^-{c_{U,W}\otimes\id_V}
  } 
\label{eq:stricthexagon2}
\end{equation}
\end{definition}
It is easy to check that $\ribi$ is a braided strict monoidal category (use 
the braiding graphs as in figure~(\ref{fig:braid})).  The
hexagon relations have a very simple geometric interpretation in $\ribi$ - it
is instructive for the reader to draw them out for himself/herself.

\subsection*{(Non-strict) braided monoidal categories}
We now consider braided monoidal categories that may not be strict.
\begin{definition}
A \textbf{braided monoidal category} is a monoidal category equipped
with a family of natural \textit{braiding} isomorphisms (for all pairs of objects)
\begin{equation}
  \lbrace c_{U,V}:U\otimes V\rightarrow V\otimes U \rbrace
\end{equation}

In contrast to the strict case the braiding isomorphisms are required to satisfy
more elaborate \textit{hexagon relations}:
\begin{equation}
  \xymatrix @R+20pt {
    & *{A\otimes (B\otimes C)}\ar@{=>}[rr]^-{c_{A,B\otimes C}}\ar[dl]^-{a_{A,B,C}^{-1}} && 
      *{(B\otimes C)\otimes A} & \\
    *{(A\otimes B)\otimes C}\ar[dr]^-{c_{A,B}\otimes\id_C}
      &&&& *{B\otimes(C\otimes A)}\ar[ul]^-{a_{B,C,A}^{-1}} \\
    & *{(B\otimes A)\otimes C}\ar[rr]^-{a_{B,A,C}} && *{B\otimes (A\otimes C)}
      \ar[ur]^-{\id_B\otimes c_{A,C}}
  } 
\label{eq:hexagon1}
\end{equation}

\begin{equation}
  \xymatrix @R+20pt {
    & *{(U\otimes V)\otimes W}\ar@{=>}[rr]^-{c_{U\otimes V,W}}\ar[dl]^-{a_{U,V,W}} && 
      *{W\otimes (U\otimes V)} & \\
    *{U\otimes (V\otimes W)}\ar[dr]^-{\id_U\otimes c_{V,W}}
      &&&& *{(W\otimes U)\otimes V}\ar[ul]^-{a_{W,U,V}} \\
    & *{U\otimes (W\otimes V)}\ar[rr]^-{a_{U,W,V}^{-1}} && *{(U\otimes W)\otimes V}
      \ar[ur]^-{c_{U,W}\otimes\id_V}
  } 
\label{eq:hexagon2}
\end{equation}
\end{definition}
It is easy to check that
$\ribi^{\text{NS}}$ is a (non-strict) braided monoidal category
($\ribi^{\text{NS}}$ is only slightly more elaborate than $\ribi$).

\section{Balanced categories}
In this section we define categories with twisting (inspired by
ribbon graphs as in figure~(\ref{fig:twist})).  The definition is identical
in both the strict and non-strict cases.

\begin{definition}A (strict) \textbf{balanced category} is a
braided (strict) monoidal category equipped with a family
of natural isomorphisms (twists) for all objects:
\begin{equation}
  \lbrace \theta_{U}:U \rightarrow U \rbrace
\end{equation}
such that the following \textit{balancing diagram} commutes:
\begin{equation}
  \xymatrix @C+30pt @R+30pt {
    *{U\otimes V}\ar[r]^-{\theta_{U\otimes V}}\ar[d]_{\theta_U\otimes \theta_V} 
    & *{U\otimes V} \\
    *{U\otimes V}\ar[r]_{c_{U,V}}& *{V\otimes U}\ar[u]^{c_{V,U}}
  } 
\label{eq:twistbraid}
\end{equation}
This can be written as a formula for convenience:
\begin{equation}
 \theta_{U\otimes V}= c_{V\otimes U}\circ c_{U\otimes V} \circ (\theta_{U}\otimes\theta_V) 
\label{eq:balancing}
\end{equation}
\end{definition}

Since the inspiration for this construction comes from ribbon graphs it is
not surprising that $\ribi$ is a strict balanced category, and similarly 
$\ribi^{\text{NS}}$ is a (non-strict) balanced
category.  The balancing condition has a simple geometric interpretation
in $\ribi$ - it is highly recommended for the reader to draw this out independently.

\section{Right-Rigid monoidal categories}
It is possible to rewind the discussion back to monoidal categories and consider
a separate line of development (independent of braided monoidal and balanced categories).
In this section we define a notion of duality.  This is meant to mimic duality in
the category of vector spaces, however we note that there are many aspects of vector
spaces that do not necessarily have analogues in this more general theory 
(for example there is no 
canonical isomorphism $V\rightarrow V^{**}$).
\footnote{The connoiseur might be interested in following this branch further.  
Left duals can be defined similarly to right duals, and a right-left rigid monoidal 
category is simply called a \textit{rigid monoidal category}.  A \textit{tensor category} has
the simultaneous structure of a rigid monoidal category and an abelian category 
that has been enriched over
finite-dimensional vector spaces (i.e. the $\text{Hom}$ spaces are better than abelian groups -
they are finite-dimensional $\cplx$-vector spaces; any characteristic 0 field $k$ can be
substituted for $\cplx$).  
The abelian structure and the 
monoidal structure must be
compatible in the sense that $\otimes$ distributes over $\oplus$.  In addition we require
$\text{Hom}(\unitobj,\unitobj)\cong \cplx$.

A \textit{finite
tensor category} is a tensor category such that there are finitely-many \textit{simple objects} (see below),
each object can be decomposed as a finite-length list of simple objects, and
each simple object admits a projective cover.
If a finite tensor category is semisimple (stronger than the projective cover condition) then
the category is a \textit{fusion category}.
}
\subsection*{Right-rigid strict monoidal categories}
\begin{definition}
A \textbf{right-rigid strict monoidal category} $\Vcat$ is a strict monoidal category such
that for each object $V\in\text{Ob}(\Vcat)$ there is a distinguished \textbf{right dual} object
$V^*$ and morphisms (\textit{not necessarily isomorphisms})
\begin{align}
  b_V&:\unitobj\rightarrow V\otimes V^*\\  
  d_V&:V^*\otimes V\rightarrow\unitobj\notag
\end{align}
These are \textit{birth} and \textit{death} morphisms.  In addition we
require that the following maps must be equal to 
$\id_V$ and $\id_{V^*}$, respectively:
\begin{align}
  &V\xrightarrow{b_V\otimes\id_V} V\otimes V^*\otimes V\xrightarrow{\id_{V} 
    \otimes d_V} V \label{eq:rigidity} \\
  &V^*\xrightarrow{\id_{V^*}\otimes b_V} V^*\otimes V\otimes V^*
    \xrightarrow{d_V\otimes\id_{V^*}} V^*\notag
\end{align}
\end{definition}

\begin{figure}
  \centering
  \input{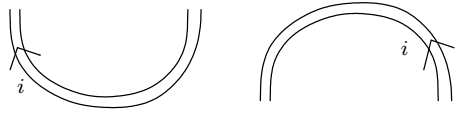}
  \caption{The birth $b_i:[]\rightarrow [[i,+1],[i,-1]]$ and death
    $d_i:[[i,-1],[i,+1]]\rightarrow []$ morphisms for the color $i$
    in the category $\ribi$.}
  \label{fig:birthdeath}
\end{figure}

$\ribi$ is a right-rigid strict monoidal category.  For a given object
\begin{equation}
 [[i_1,\pm 1],[i_2,\pm 1],\ldots,[i_k,\pm 1]] 
\end{equation}
the dual object is
\begin{equation}
 [[i_1,\mp 1],[i_2,\mp 1],\ldots,[i_k,\mp 1]] 
\end{equation}
(every $+1$ is changed to
a $-1$ and vica versa).  The birth and death morphisms are depicted in
figure~(\ref{fig:birthdeath}).  The conditions in equation~(\ref{eq:rigidity})
have simple geometric interpretations in $\ribi$ and again it is in the interest
of the reader to sketch these out.

\subsection*{(Non-strict) right-rigid monoidal categories}
\begin{definition}
A \textbf{right-rigid monoidal category} $\Vcat$ is a monoidal category such
that for each object $V\in\text{Ob}(\Vcat)$ there is a distinguished \textbf{right dual} object
$V^*$ and morphisms (\textit{not necessarily isomorphisms})
\begin{align}
  b_V&:\unitobj\rightarrow V\otimes V^*\\  
  d_V&:V^*\otimes V\rightarrow\unitobj\notag
\end{align}
These are \textit{birth} and \textit{death} morphisms.  Similar to the conditions above
we require that the following maps must be equal to 
$\id_V$ and $\id_{V^*}$, respectively:
\begin{align}
  &V\xrightarrow{l_V^{-1}}\unitobj\otimes V
    \xrightarrow{b_V\otimes\id_V} (V\otimes V^*)\otimes V 
    \xrightarrow{a_{V,V^*,V}} \notag\\ &\quad\quad\quad\quad\quad\quad V\otimes (V^*\otimes V)
    \xrightarrow{\id_{V} \otimes d_V} V\otimes \unitobj
    \xrightarrow{r_V} V \label{eq:nonstrictrigidity}\\
  &V^*\xrightarrow{r_{V^*}^{-1}}V^*\otimes\unitobj
    \xrightarrow{\id_{V^*}\otimes b_V} V^*\otimes (V\otimes V^*)
    \xrightarrow{a^{-1}_{V^*,V,V^*}} \notag\\ &\quad\quad\quad\quad\quad\quad (V^*\otimes V)\otimes V^*
    \xrightarrow{d_V\otimes\id_{V^*}} \unitobj \otimes V^*
    \xrightarrow{l_{V^*}} V^*\notag
\end{align}
The only difference is that the associativity maps appear.
\end{definition}

In a similar fashion to $\ribi$ it is easy to show that
$\ribi^{\text{NS}}$ is a (non-strict) right-rigid monoidal category. 

\section{Ribbon categories}
Ribbon categories were studied in \cite{shum}.
The definitions for strict and non-strict ribbon categories are nearly identical,
hence we define them simultaneously.

\begin{definition}
A (strict) ribbon category is a right-rigid (strict) monoidal category
that in addition is a (strict) balanced category. 

The balancing and rigidity must be compatible:
\begin{equation}
 (\theta_V\otimes \id_{V^*})\circ b_V=(\id_{V}\otimes\theta_{V^*})\circ b_V
\end{equation}
(again the geometric picture in $\ribi$ is illuminating). 
\end{definition}

We now describe some properties of ribbon categories.  First,
given an object $V$ in a ribbon category $\Vcat$ and
a morphism $f:V\rightarrow V$ we define
the \textbf{quantum trace} of $f$:
\begin{equation}
  \text{tr}_q(f:V\rightarrow V):=d_V\circ c_{V,V^*}\circ((\theta_V\circ f)\otimes\id_{V^*})\circ b_V
\label{eq:qtrace} 
\end{equation}
Furthermore the \textbf{quantum dimension} is defined by:
\begin{equation}
  \text{dim}_q(V):=\text{tr}_q(\id_V)=d_V\circ c_{V,V^*}\circ(\theta_V\otimes\id_{V^*})\circ b_V
\label{eq:qdim} 
\end{equation}
We note that if the objects in the underlying category have some underlying
\textit{intrinsic} notion of
trace and dimension (e.g. the objects are finite-dimensional vector spaces) 
then it is \textit{not true} that the
quantum trace and quantum dimension necessarily agree with the intrinsic notions.
For example the quantum dimension need \textit{not} even be an integer.

Every ribbon category is \textbf{pivotal}, that is for each object $V$ there is
a distinguished isomorphism $V\tilde{\rightarrow} V^{**}$ determined by the
composition:
\footnote{This composition makes sense for strict ribbon categories.  There is a similar composition
for non-strict ribbon categories.  We note that it is \textit{not} obvious that this
composition of morphisms is an \textit{iso}morphism.  This can be proven using the
functor $F$ introduced in the next section (see \cite{turaev} pg. 40).}
\begin{equation}
 \xymatrix{
  V \ar@{|->}[r]^-{id_V\otimes b_{V^*}} & 
    V\otimes V^*\otimes (V^*)^*  \ar@{|->}[rrr]^-{\theta_V\otimes\id_{V^*}\otimes\id_{V^{**}}} &&& \\
    V\otimes V^*\otimes (V^*)^*\ar@{|->}[rr]^-{c_{V,V^*}\otimes id_{V^{**}}} &&
    V^*\otimes V\otimes (V^*)^*\ar@{|->}[rr]^-{d_{V}\otimes id_{V^{**}}} && V^{**} 
 }
\end{equation}
Again (if the objects are finite-dimensional vector spaces) 
this isomorphism is typically \textit{not} the
same as the canonical vector space isomorphism $V\tilde{\rightarrow} V^{**}$.

It is also a fact that ribbon categories are \textbf{spherical}, that is
$\text{dim}_q(V)=\text{dim}_q(V^*)$ for every object.  The proof requires the
functor $F$ discussed in the next section.

\section{Invariants of colored $(k,l)$-ribbon graphs using ribbon categories}
\label{sec:ribboninvts}
In the last several sections we have been considering the category $\ribi$
where $I$ is some arbitrary labeling set.  Suppose that
we replace $I$ with a right-rigid strict monoidal category $\Vcat$ and consider $\ribv$,
i.e. we color the oriented ribbons (and annuli) with \textit{objects} in
$\Vcat$.  Because of the right-rigid strict monoidal structure we can go 
further and color the \textit{coupons} with \textit{morphisms} as well.
We discuss this now.

\begin{figure}
  \centering
  \input{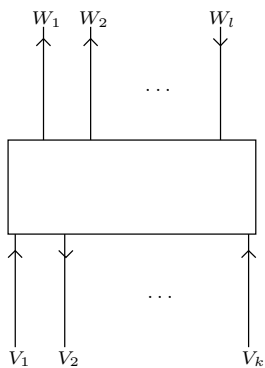}
  \caption{An elementary $(k,l)$-ribbon graph.}
  \label{fig:elementaryrg}
\end{figure}

First consider an \textit{elementary} $(k,l)$-ribbon graph in standard
drawing position as depicted in
figure~(\ref{fig:elementaryrg}).  The graph is
called ``elementary'' because there is neither braiding nor twisting
in any of the ribbons (neither birth nor death), there is a single coupon, and all of the
ribbons terminate on the coupon.

Denote $V^{+1}:=V$ and $V^{-1}:=V^*$.  Then it
makes sense to color the coupon in figure~(\ref{fig:elementaryrg}) 
with a morphism
\begin{equation}
  \phi\in\text{Hom}(V^{\pm 1}_1\otimes\ldots\otimes V^{\pm 1}_k,
    W^{\pm 1}_1\otimes\ldots\otimes W^{\pm 1}_l)
\end{equation}
where we use $+1$ for ribbons pointing ``up'' and $-1$ for ribbons pointing
``down''.  Note that both the monoidal and rigidity properties of $\Vcat$ have been used.
In this way we can color coupons in any arbitrary colored $(k,l)$-ribbon graph.

Let us introduce the terminology 
\textbf{fully colored $(k.l)$-ribbon graphs}
for colored ribbon graphs where in addition all of the coupons are colored with morphisms.
Using this enrich $\ribv$ by replacing the morphisms (colored $(k,l)$-ribbon graphs)
with \textit{fully colored} $(k,l)$-ribbon graphs.

Generalizing the above construction to the non-strict case $\ribv^{\text{NS}}$ is
straightforward and left to the reader.

\subsection*{The main functor $F$}
We can go further and consider $\ribv$ where $\Vcat$ is now a \textit{strict ribbon category}.
Then we have two strict ribbon categories to consider: $\ribv$ (which is a strict ribbon category
since any $\ribi$ is) and $\Vcat$.  The main theorem for
ribbon categories is the following (proven by Reshetikhin and Turaev in
the language of quantum groups):
\begin{figure}
  \centering
  \input{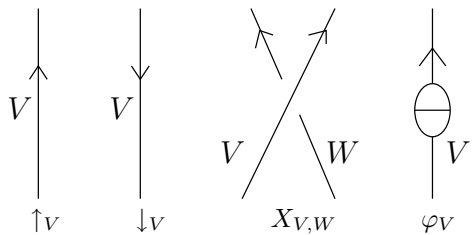}
  \caption{A list of primitive ribbons graphs.  The notation will be used in
    theorem~(\ref{thm:functorF})}
  \label{fig:allribbonmoves}
\end{figure} 

\begin{theorem}[Reshetikhin, Turaev]
\label{thm:functorF}
Let $\Vcat$ be a \textbf{strict} ribbon category.  Consider the enriched \textbf{strict} ribbon category
$\ribv$ (enriched means the morphisms are fully colored $(k,l)$-ribbon graphs).  Set notation
for primitive ribbon graphs as in figure~(\ref{fig:allribbonmoves}).  
Then there is a unique \textbf{strict} monoidal functor
\begin{equation}
  F:\ribv\rightarrow\Vcat
\end{equation}
such that
\begin{align}
  &F([[V,+1]])=V\\
  &F([[V,-1]])=V^*\notag\\
  &F(\uparrow_V)=\id_V\notag\\
  &F(\downarrow_V)=\id_{V^*}\notag\\
  &F(X_{V,W})=c_{V,W}\notag\\
  &F(\varphi_V)=\theta_V\notag
\end{align}
\end{theorem}

We have not seen a non-strict version of this theorem stated and proven in the literature.
We conjecture the following (and we implicitly use it
in the remainder of this paper):
\begin{figure}
  \centering
  \input{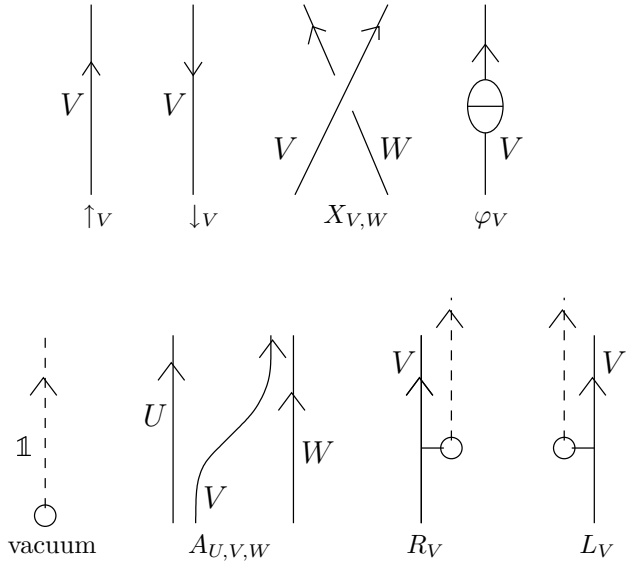}
  \caption{A list of primitive ``non-associative'' ribbons graphs.  The notation will be used in
    conjecture~(\ref{conj:functorFnonstrict})}
  \label{fig:allribnonstrict}
\end{figure} 

\begin{conjecture}
\label{conj:functorFnonstrict}
Let $\Vcat$ be a ribbon category.  Consider the enriched ribbon category
$\ribv^{\text{NS}}$ (enriched means the morphisms are fully colored non-associative
$(k,l)$-ribbon graphs).  
Set notation
for primitive ribbon graphs as in figure~(\ref{fig:allribnonstrict}).  
Then there is a unique monoidal functor
\begin{equation}
  F:\ribv^{\text{NS}}\rightarrow\Vcat
\end{equation}
such that
\begin{align}
  &F([[V,+1]])=V\\
  &F([[V,-1]])=V^*\notag\\
  &F(\uparrow_V)=\id_V\notag\\
  &F(\downarrow_V)=\id_{V^*}\notag\\
  &F(X_{V,W})=c_{V,W}\notag\\
  &F(\varphi_V)=\theta_V\notag\\
  &F(A_{U,V,W})=a_{U,V,W}\notag\\
  &F(R_V)=r_V\notag\\
  &F(L_V)=l_V\notag
\end{align}
\end{conjecture}

\section{Modular tensor categories}
In this section we define modular tensor categories.  We shall make no
reference to strict/non-strict categories, leaving it to the
reader to make the appropriate substitutions where necessary.

We begin with a preliminary definition:
\begin{definition}
Consider a preadditive category $\Vcat$ that also is enriched so that the
$\text{Hom}$ sets are $\cplx$-vector spaces (rather than just abelian groups).  
Then a \textbf{simple object} $V_x$ 
is an object such that
\begin{equation}
  \text{Hom}(V_x,V_x)\cong \cplx 
\end{equation}
\end{definition}

Suppose that $\Vcat$ is an enriched preadditive category and in addition
is a ribbon category.  We require that the preadditive structure
be compatible with the monoidal structure (i.e. $\otimes$ distributes over $+$
of morphisms).  Then it is straightforward to check that the dual $V^*_x$ of a simple
object is also simple.  
It is also straightforward to prove that
$\unitobj$ is a simple object.

The definition of a modular tensor category in \cite{turaev} is based on
preadditive ribbon categories and is slightly more general than what
is presented below.
We restrict attention to \textit{additive} ribbon categories:

\begin{definition}
A \textbf{modular tensor category} is a category with the following structure:
\begin{enumerate}
 \item Ribbon category
 \item Additive category enriched over $\cplx$-vector spaces
 \item Ribbon/additive compatibility ($\otimes$ distributes over $\oplus$)
 \item Semisimple with \textit{finitely}-many simple objects
 \item The \textbf{$S$-matrix} is invertible, where $S$ is defined
   by (using the ribbon structure on simple objects $V_x$ and $V_y$):
\begin{equation}
 S_{x,y}:=\text{tr}_q(c_{V_y,V^*_x} \circ c_{V^*_x,V_y})
\label{eq:smatrixdef}
\end{equation}
  \item A choice of square root
\begin{equation}
  \mathscr{D}:=\sqrt{\sum_{\text{simple objects}} (\text{dim}_q(V_x))^2} 
\end{equation}
\end{enumerate}
\end{definition}

Since if $V_x$ is a simple object then $\text{Hom}(V_x,V_x)\cong \cplx$ 
we see that the twist isomorphism $\theta_{V_x}:V_x\rightarrow V_x$ is given by
a complex number (denoted $\theta_x$).

The following expressions will be used often in the sequel:
\begin{align}
  p_+&:=\sum_{\text{simple objects}} (\text{dim}_q(V_x))^2 \theta_x\\
  p_-&:=\sum_{\text{simple objects}} (\text{dim}_q(V_x))^2 \theta^{-1}_x\notag
\end{align}
It is a fact (see \cite{bakalov_kirillov}) that
\begin{equation}
  \mathscr{D}^2=p_+p_-
\end{equation}

\section{Invariants of 3-manifolds, 2+1-dimensional TQFTs from MTCs}
We mentioned in section~(\ref{sec:ribboninvts}) that associated to any 
ribbon category $\Vcat$ is a monoidal functor
\begin{equation}
  F:\ribv\rightarrow \Vcat 
\end{equation}
Using this functor it is straightforward to assign to any 
fully-colored $(k,l)$-ribbon graph in $\mathbb{R}^3$ 
a morphism
$V^{\pm 1}_1\otimes\ldots\otimes V^{\pm 1}_k\rightarrow W^{\pm 1}_1\otimes\ldots\otimes W^{\pm 1}_l$ between the object coloring the bottom of the graph and the object
coloring the top.  It is proven in \cite{turaev} that the resulting morphism
is invariant under regular isotopy of the ribbon graph.

Now we turn our attention to modular tensor categories.  We shall see that the
stronger structure
allows us to define invariants of closed oriented 3-manifolds (and, eventually,
2+1 TQFTs).
Before we begin suppose first that we have a ribbon graph in $S^3$.
It is easy to isotope any ribbon graph in $S^3=\mathbb{R}^3\cup\{\infty\}$ appropriately
to ``miss'' the point $\{\infty\}$, hence we can consider the ribbon graph as embedded
in $\mathbb{R}^3$ (where we can apply the functor $F$).

Since we wish to study closed oriented 3-manifolds $X$ the following standard theorem is 
useful:
\footnote{Actually the original theorem requires \textit{rational} surgery, but there is
a well-known algorithm to reduce from rational surgery to integer surgery 
(see, e.g., \cite{prasolov_sossinsky}).  Since we will not
require rational surgery we do not bother here.}

\begin{theorem}[Dehn, Lickorish] Any orientable closed 3-manifold $X$ can be obtained
from $S^3$ by drilling out solid tori and gluing them back in along different diffeomorphisms
(up to isotopy)
of their boundaries.  Furthermore, each such surgery can be assumed to be an ``integer surgery''
(see below).
\end{theorem}

\subsection*{Surgery}
\begin{figure}
  \centering
  \input{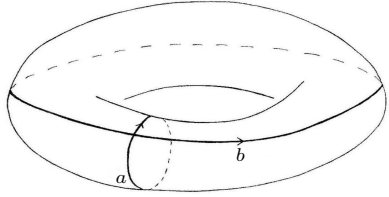}
  \caption{A torus with oriented meridian and longitude.}
  \label{fig:torus}
\end{figure}

The diffeomorphisms along which we reglue the solid tori can be neatly encoded
in terms of framed links in $S^3$.  This can be seen by considering each solid torus
individually.  Before drilling out the solid torus pick a reference longitude $b$ and
meridian $a$ on the boundary as in figure~(\ref{fig:torus}).

\begin{figure}
  \centering
  \input{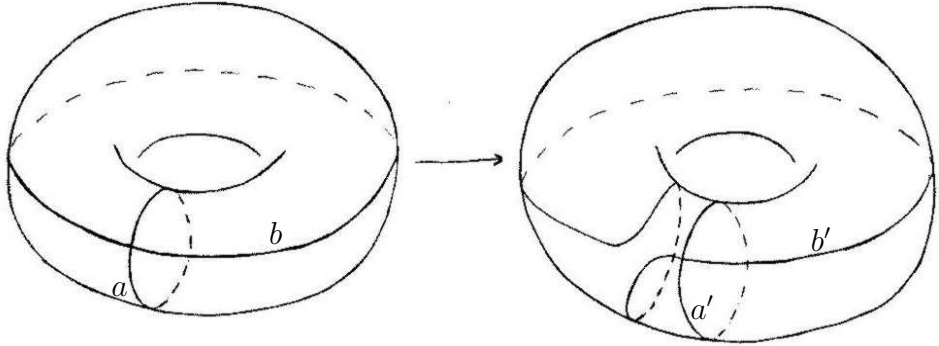}
  \caption{A Dehn twist on the curve $a$.}
  \label{fig:dehn_torus}
\end{figure}

From chapter~(\ref{ch:tqft}) we know that $\text{MCG}(T^2)\cong \text{PSL}(2,\mathbb{Z})$.
In particular
a diffeomorphism is determined by the action on homology generators
\footnote{This is not true in higher genus.}
\begin{equation}
  a=\begin{pmatrix}1\\ 0\end{pmatrix}\quad\quad\quad b=\begin{pmatrix}0\\ 1\end{pmatrix} 
\end{equation}
Consider the effect of drilling out a single torus and gluing it
back in along the diffeomorphism determined by the matrix
\begin{equation}
  T=\begin{pmatrix}1 & 1 \\ 0 & 1 \end{pmatrix} 
\end{equation}
This is depicted in figure~(\ref{fig:dehn_torus}).
It is not difficult to convince oneself that this surgery does \textit{not change}
the topology of the 3-manifold (removing a solid torus, cutting it, twisting it, gluing it together,
and replacing it in the hole is the same as simply filling in the hole).  More generally the
surgery determined by the boundary diffeomorphism
\begin{equation}
  T^m=\begin{pmatrix}1 & m \\ 0 & 1 \end{pmatrix} 
\end{equation}
also does not change the topology of the 3-manifold.

Because of this observation we have the following common fact (we could not find
the simple argument written down, hence we write it here for completeness):
\begin{fact}
A surgery on a single solid torus is determined by specifying two relatively-prime integers
$q$ and $p$.  We say that the ratio $\frac{q}{p}$ determines a \textbf{rational}
surgery.  In fact we only have to specify the image of $a$
\begin{equation}
  a\mapsto q\cdot a + p\cdot b
\end{equation}
\end{fact}
\begin{proof}
We construct a matrix
\begin{equation}
  \begin{pmatrix}q & -r \\ p & s \end{pmatrix}\in\text{SL}(2,\mathbb{Z})
\end{equation}
for some integers $r$ and $s$.
Since the determinant must be 1, we want to find integers $r$ and $s$ such
that
\begin{equation}
  qs+pr = 1
\end{equation}
However since $q$ and $p$ are relatively prime the Euclidean algorithm can
be used to find suitable integers $r$ and $s$ that satisfy the above equation.  
The choice is not unique since $r-kq$ and $s+kp$ also works for any integer $k$.

We need to know how the surgeries determined by the matrices
\begin{equation}
 \begin{pmatrix}q & -r \\ p & s \end{pmatrix}\quad\quad\quad\quad
 \begin{pmatrix}q & -(r-kq) \\ p & s+kp \end{pmatrix}
\end{equation}
differ.  It is easy to check that
\begin{equation}
 \begin{pmatrix}q & -(r-kq) \\ p & s+kp \end{pmatrix}=
 \begin{pmatrix}q & -r+kq \\ p & s+kp \end{pmatrix}=
 \begin{pmatrix}q & -r \\ p & s \end{pmatrix}\begin{pmatrix}1 & k \\ 0 & 1 \end{pmatrix}=
 \begin{pmatrix}q & -r \\ p & s \end{pmatrix}T^k
\end{equation}
Hence the surgeries differ by precomposing with a $T^k$ surgery (which we already
argued does not change the topology of the 3-manifold).

This proves that a surgery along a single solid torus is determined by two
relatively prime integers $q$ and $p$.
\end{proof}
When $p=1$ this is \textit{integer} surgery.  There is a standard
algorithm that reduces rational surgery to integer surgery (by continued fraction
expansion and drilling out more solid tori, see 
\cite{prasolov_sossinsky}) hence we set $p=1$ from now on.  Therefore
a surgery along a single solid torus is determined by a single integer $q$ 
and we have the following corollary:

\begin{corollary}
Any closed oriented 3-manifold $X$ can be presented as a surgery along \textbf{framed}
links in $S^3$.
\end{corollary}
\begin{proof}
Dehn-Lickorish implies that any closed oriented 3-manifold $X$ can be obtained
by drilling out/regluing solid tori in $S^3$.  
If we consider the cores of the tori this determines a
link in $S^3$ (from the link components we could recover the solid
tori by thickening).  The only issue is how to encode the regluing diffeomorphism. 
We have seen that any integer surgery (along a single solid torus)
is determined by a single integer
$q$, hence we can frame the corresponding link component with the appropriate framing number
$q$.  Repeating this for all of the solid tori produces a framed link in $S^3$ that
determines the surgery completely.
\end{proof}

\begin{example}
The most important example is the \textit{torus switch}, i.e. surgery along
a framed \textit{unknot} with framing number 0 (see the left side of 
figure~(\ref{fig:0and1framing})).
\begin{figure}
  \centering
  \input{0and1framing.eps_t}
  \caption{Framed unknots with $0$ and $1$ framing, respectively.  The lower
    diagrams are the skeletal schematic diagrams.}
  \label{fig:0and1framing}
\end{figure}

Since $q=0$ we have that the following matrix determines the surgery
\begin{equation}
 S = \begin{pmatrix} 0 & -1 \\ 1 & 0 \end{pmatrix}
\end{equation}
This sends
\begin{equation}
  a\mapsto b\quad\quad\quad b\mapsto -a
\end{equation}
i.e. the longitude and meridian swap roles (this is an orientation preserving map).

On the other hand consider the Heegaard decomposition of $S^3$ into two solid tori depicted
in figure~(\ref{fig:s3_torus}).
We have the identifications 
\begin{align}
  &a\leftrightarrow \tilde{b} \\
  &b\leftrightarrow \tilde{a}\notag 
\end{align}
\begin{figure}
  \centering
  \input{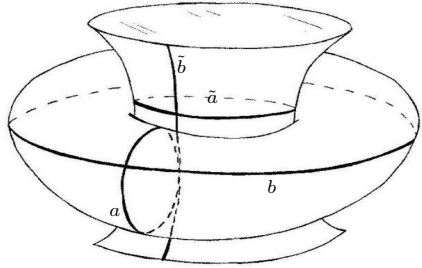}
  \caption{Heegaard decomposition of $S^3$ into two solid tori.  The ``plug'' is a
    solid torus that has been cut.  Imagine deforming the plug (as shown) and 
    enveloping completely the other
    solid torus to form a 3-ball with boundary $S^2$ (i.e. identify the longitude of the solid torus
    with the meridian of the plug, and the meridian of the solid torus with the
    longitude of the plug).  Since the plug is actually a cut solid torus we know that the
    top hemisphere of the boundary $S^2$ should be identified with the bottom
    hemisphere.  Topologically this is the same as crushing the entire
    $S^2$ to a point.  Hence we obtain $S^3$ (the 3-ball with boundary $S^2$ 
    crushed to a point).  
    We do not draw the orientations for $a$, $b$, $\tilde{a}$, and $\tilde{b}$,
    however a quick check verifies that $a\leftrightarrow \tilde{b}$
    and $b\leftrightarrow \tilde{a}$, which is an orientation \textit{reversing} gluing
    diffeomorphism as
    expected since we can only glue outgoing boundaries to incoming boundaries.}
  \label{fig:s3_torus}
\end{figure}
If we drill out one of the tori from figure~(\ref{fig:s3_torus}), 
apply the
self-diffeomorphism determined by the $S$ matrix given above,
and reglue then we have the identifications
\footnote{We need to be careful with orientations,
i.e. the $S$ matrix is an orientation preserving self-diffeomorphism, 
but the cutting and regluing are orientation reversing operations.} 
\begin{align}
  &a\mapsto b \Rightarrow b\leftrightarrow \tilde{b} \\
  &b\mapsto -a \Rightarrow -a\leftrightarrow \tilde{a}\notag 
\end{align}

In other words we have two solid tori that are glued together (longitude to longitude,
meridian to meridian).  Since a solid torus is just $D^2\times S^1$ where the
$S^1$ factor can be identified with $b$, and the boundary of the
disk $D^2$ can be identified with $a$, we see that gluing the two solid tori
together gives $S^2\times S^1$ (for a fixed point on the longitude $S^1$ both solid tori
look like $D^2\times\{\text{pt}\}$ - gluing two disks together along the boundary
gives us a 2-sphere $S^2\times\{ \text{pt}\}$).

Summarizing, a surgery along a $0$-framed unknot in $S^3$ gives the closed oriented
3-manifold $S^2\times S^1$.  Iterating the surgery again we recover $S^3$. 
\end{example}

\begin{example}
\label{ex:1framedsurgery}
It is shown in \cite{prasolov_sossinsky} that a surgery with framing number
$\pm 1$  (see the right side of figure~(\ref{fig:0and1framing})) 
along an isolated unknot is trivial, i.e. the 
3-manifold topology does not change.  
For example the diffeomorphism for the $+1$ framing is
\begin{equation}
  \begin{pmatrix} 1 & 0 \\ 1 & 1\end{pmatrix}
\end{equation}
and the proof that this does not alter the topology of the 3-manifold is
similar to the proof that the $T$ matrix diffeomorphism
\begin{equation}
  \begin{pmatrix} 1 & 1 \\ 0 & 1\end{pmatrix}
\end{equation}
does not alter the 3-manifold.
We note that this only applies to isolated unknots.  For contrast
$\pm 1$-framed surgery along a component that is \textit{linked} is
\textit{nontrivial}.
\end{example}

In general, given an oriented closed 3-manifold presented by some other means
(say, a Heegaard decomposition),
it may be difficult to provide a surgery presentation of framed links in $S^3$.
Furthermore, the surgery presentation is certainly not unique (considering the
example above, we could add as many $\pm 1$-framed isolated unknots to the diagram as
desired and not change the resulting 3-manifold).

However, any two surgery presentations of the same 3-manifold can be related
by the \textbf{Kirby moves} (see, e.g., \cite{prasolov_sossinsky}).
Since we do not require these moves explicitly (and since they are standard)
we omit their description.  However, we note that the proof that a modular
tensor category gives 3-manifold invariants essentially reduces to 
showing invariance under the Kirby moves.

\subsection*{Invariants of closed 3-manifolds from MTCs}
Once a surgery presentation is specified for $X$ the computation of the 3-manifold
invariant is straightforward.
The strategy is to \textit{average} over
all possible colorings of the framed link $L$ in $S^3$.
\footnote{Hence the necessity for finitely-many simple objects.}
We pick an orientation on each of
the components of $L=\{L_1,\ldots,L_m\}$.  The chosen orientation does not affect the
invariant because we are summing over \textit{all} colorings.
\footnote{Recall that we can switch orientation if we replace a coloring $V$ with
the dual $V^*$.}

Note that in general we may allow the 3-manifold $X$ to also 
contain some embedded oriented \textit{fixed colored} ribbon graph $\Omega$
in addition to the oriented framed link $L$.  It is understood that $\Omega$ does not
participate in the surgery.

If we pick a coloring for $L$ by \textbf{simple} objects $\{V_i\}_{i\in I}$ then we can compute the
ribbon graph invariant $F(L\cup\Omega)$.  Denote by $V_{\lambda_i}$ the coloring of the
link component $L_i$.  

We require a normalization convention.
Every \textit{oriented} framed link $L=\{L_1,\ldots,L_m\}$ has an $m\times m$ linking number matrix $B$
where an off-diagonal element is given by
\begin{equation}
  B_{ij}=\text{lk}(L_i,L_j)=\frac{\text{\# positive crossings}-\text{\# negative crossings}}{2}
\end{equation}
and a diagonal element is just
\begin{equation}
  B_{ii}=\text{framing number of }L_i
\end{equation}
Denote the signature of this matrix by $\sigma(L)$.  Then, given a surgery presentation for $X$
as a framed link $L$ in $S^3$ we compute the 3-manifold invariant
\begin{equation}
 \tau(X):=p_-^{\sigma(L)}\mathscr{D}^{-\sigma(L)-m-1}
  \sum_{\text{col of }L}\left(\prod_{i=1}^m \text{dim}_q(V_{\lambda_i})\right)F(L\cup\Omega)
\label{eq:3mnfldinvt}
\end{equation}
The components of $L$ can only be colored by
simple objects.

\subsection*{$(2+1)$-dimensional topological quantum field theory}
The 3-manifold invariant provided in the last subsection can be exploited
further to produce an extended $(2+1)$-dimensional TQFT in the sense of chapter~(\ref{ch:tqft}).
Consider an oriented 3-manifold $X$ with boundary $\partial X=-\Sigma_-\sqcup \Sigma_+$.
For simplicity we assume that $\Sigma_-$ and $\Sigma_+$ are connected \textit{closed}
2 surfaces
\footnote{In general colored ribbon graphs $\Omega$ can terminate on the boundary forming \textit{marked
arcs}.  As mentioned in chapter~(\ref{ch:tqft}) these can also be viewed as parameterized
boundary circles (from the perspective of conformal field theory).}

It was stated in chapter~(\ref{ch:tqft}) that an \textit{extended} structure is
required on 2-surfaces and 3-bordisms in order to define an anomaly-free $(2+1)$-dimensional
TQFT.  However for \textit{closed} 3-manifolds there is a canonical choice for this
extended structure (see \cite{atiyah}) and hence we did not need to mention it
in the previous subsection concerning 3-manifold invariants.

We now place a strong structure
on a boundary 2-surface $\Sigma$.
We say that $\Sigma$ is \textbf{parameterized} if it is equipped with a
fixed diffeomorphism
\begin{equation}
  \phi:\partial H_g\rightarrow \Sigma 
\end{equation}
where $H_g$ is a \textbf{standard} handlebody that we now specify.
\footnote{In the same spirit as the Segal modular functor
(where the complex structure turned out to be irrelevant when defining
a projective representation of $\text{MCG}(\Sigma)$)
the parameterization is irrelevant if we
are content with TQFTs with gluing anomaly, and the dependence
is weak for a full anomaly-free TQFT.}

\subsubsection*{Standard handlebodies}
\begin{figure}
  \centering
  \input{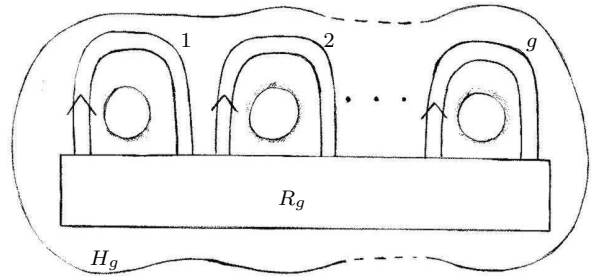}
  \caption{Standard handlebody $H_g$ with the standard embedded
    ribbon graph $R_g$.}
  \label{fig:H}
\end{figure}

We define the standard handlebody $H_g$ of genus $g$ as a thickening of
the standard uncolored ribbon graph $R_g$ (embedded in $\mathbb{R}^3$) depicted
in figure~(\ref{fig:H}).  The boundary
is a surface $\Sigma_g$ of genus $g$.  The handlebody $H_g$ inherits
an orientation from its embedding in $\mathbb{R}^3$.  We endow
$\Sigma_g$ with the orientation that \textit{agrees} with the boundary
orientation, i.e. $\Sigma_g=\partial H_g$.  In this sense
$\Sigma_g$ is \textit{outgoing}.

\begin{figure}
  \centering
  \input{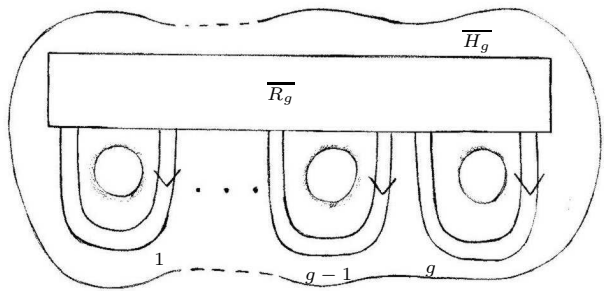}
  \caption{Standard handlebody $\overline{H_g}$ with the standard embedded
    ribbon graph $\overline{R_g}$.}
  \label{fig:Hbar}
\end{figure}

Likewise we define the standard handlebody $\overline{H_g}$ as a thickening
of the standard uncolored ribbon graph $\overline{R_g}$ depicted in 
figure~(\ref{fig:Hbar}).
\footnote{Note that $\overline{R_g}$ is not exactly a mirror image of $R_g$.}
Again the boundary is a surface
$\overline{\Sigma_g}$ of genus $g$.  $\overline{H_g}$ inherits an orientation
from its embedding in $\mathbb{R}^3$.  However here we supply $\overline{\Sigma_g}$ with
the opposite orientation from the boundary orientation, i.e.
$\overline{\Sigma_g}=-\partial \overline{H_g}$.  In this sense $\overline{\Sigma_g}$ is
\textit{incoming}.  There is a natural identification
\footnote{The construction is more complicated in the presence
of marked arcs.}
\begin{equation}
  \overline{\Sigma_g}=-\Sigma_g
\end{equation}

We can color the ribbons of $R_g$ (in order from left to right) with simple objects
$\{V_{\lambda_1},\ldots,V_{\lambda_g}\}$.  Denote the resulting ribbon graph
by
\begin{equation}
  R_g(V_{\lambda_1},\ldots,V_{\lambda_g})
\end{equation}
We can also then color the \textit{coupon} with
a morphism
\begin{equation}
  f:\unitobj\rightarrow V_{\lambda_1}\otimes V_{\lambda_1}^*\otimes\cdots\otimes 
    V_{\lambda_g}\otimes V_{\lambda_g}^*
\end{equation}
Denote the resulting fully-colored ribbon graph by
\begin{equation}
  R_g(V_{\lambda_1},\ldots,V_{\lambda_g}; f)
\end{equation}

Similarly we can color the ribbons of $\overline{R_g}$ with simple objects $\{V_{\zeta_1},\ldots,V_{\zeta_g}\}$ and denote the resulting ribbon graph
\begin{equation}
  \overline{R_g}(V_{\zeta_1},\ldots,V_{\zeta_g})
\end{equation}
Likewise can then color the \textit{coupon} with
a morphism
\begin{equation}
  h:V_{\zeta_1}\otimes V_{\zeta_1}^*\otimes\cdots\otimes 
    V_{\zeta_g}\otimes V_{\zeta_g}^*\rightarrow\unitobj
\end{equation}
Denote the fully colored ribbon graph by
\begin{equation}
  \overline{R_g}(V_{\zeta_1},\ldots,V_{\zeta_g}; h)
\end{equation}

If we color the ribbons of $R_g$ and $\overline{R_g}$ with the \textit{same}
ordered list of simple objects $\{V_{\lambda_1},\ldots,V_{\lambda_g}\}$ then we have
\begin{align}
  f&\in\text{Hom}(\unitobj,V_{\lambda_1}\otimes V_{\lambda_1}^*\otimes\cdots\otimes 
    V_{\lambda_g}\otimes V_{\lambda_g}^*) \\
  h&\in\text{Hom}(V_{\lambda_1}\otimes V_{\lambda_1}^*\otimes\cdots\otimes 
    V_{\lambda_g}\otimes V_{\lambda_g}^*,\unitobj) = \\
    &\quad =
    (\text{Hom}(\unitobj,V_{\lambda_1}\otimes V_{\lambda_1}^*\otimes\cdots\otimes 
      V_{\lambda_g}\otimes V_{\lambda_g}^*))^*
\end{align}
where the last line follows from the natural pairing
\begin{equation}
  h(f) := h\circ f\in\text{Hom}(\unitobj,\unitobj)\cong\cplx
\end{equation}
In this way we see that a colored coupon in $\overline{R_g}$ lives in the
dual space of a colored coupon in $R_g$.

\subsubsection*{Hilbert space of states} 
Now we describe how to associate a vector space (or Hilbert space if the theory 
is unitary - see \cite{turaev}) to an oriented closed surface $\Sigma$ of genus
$g$ equipped
with a parameterization $\phi:\partial H_g\rightarrow\Sigma$.  Again this construction
can be straightforwardly generalized to surfaces with marked arcs.

Since the surface is parameterized we identify it as the boundary of the
standard handlebody $H_g$.  The embedded ribbon $R_g$ is uncolored.  The idea
is to sum over all possible colorings of $R_g$.  Define
the associated vector space:
\begin{equation} 
 \modfunct(\Sigma):=\bigoplus_{\text{col }\{V_{\lambda_1},\ldots,V_{\lambda_g}\}}
    \text{Hom}(\unitobj,V_{\lambda_1}\otimes V_{\lambda_1}^*\otimes\cdots\otimes 
      V_{\lambda_g}\otimes V_{\lambda_g}^*)
\label{eq:modfunctvs}
\end{equation}
This defines part of the non-extended modular functor from chapter~(\ref{ch:tqft})).
It is still necessary to describe the action
of diffeomorphisms $\Sigma\rightarrow\Sigma$ on $\modfunct(\Sigma)$.

\subsubsection*{Operators associated to oriented 3-bordisms}
Recall that $X$ is an oriented 3-manifold 
with boundary $\partial X=\partial X_-\sqcup \partial X_+=-\Sigma_-\sqcup \Sigma_+$.
For simplicity we assume that $\Sigma_-$ and $\Sigma_+$ are connected \textit{closed}
2 surfaces of genus $g_-$ and $g_+$, respectively.  In addition 
assume that we have an extended structure on the boundaries, i.e. parameterizations
\begin{align}
  \phi_-&:\partial H_{g_-}\rightarrow\Sigma_- \\
  \phi_+&:\partial H_{g_+}\rightarrow\Sigma_+ \notag
\end{align}

From the axioms for a TQFT we expect to assign to the 3-bordism
$X$ an operator
\begin{equation}
 \tau(X):\modfunct(\Sigma_-)\rightarrow\modfunct(\Sigma_+)
\end{equation}

The matrix elements of the operator $\tau(X)$ are determined by the following
recipe: pick a basis for $\modfunct(\Sigma_-)$ (and for $\modfunct(\Sigma_+)$).
The vector space $\modfunct(\Sigma_-)$ is defined by equation~(\ref{eq:modfunctvs})
(and similarly for $\modfunct(\Sigma_+)$).
Using the parameterizations of $\Sigma_-$ and $\Sigma_+$ we ``cap off''
$X$ with the standard handlebodies $H_{g_-}$ and $\overline{H_{g_+}}$, respectively, 
to produce a \textit{closed}
3-manifold $\tilde{X}$ (with embedded \textit{uncolored} ribbon graphs $R_{g_-}$
$\overline{R_{g_+}}$ in the handlebodies).
 
Choosing a specific basis element out of $\modfunct(\Sigma_-)$
and a specific basis element out of $\modfunct(\Sigma_+)$ is the
same as specifying a
coloring for $R_{g_-}$ and
$R_{g_+}$ (which determines a dual coloring in the space $\overline{R_{g_+}}$).
Then calculating the 3-manifold invariant $\tau(\tilde{X})\in\cplx$ gives the 
corresponding matrix
element for the operator
\begin{equation}
 \tau(X):\modfunct(\Sigma_-)\rightarrow\modfunct(\Sigma_+)
\end{equation}

We only need to be careful about orientations to ensure that the
correct handlebodies are glued onto the correct boundary components.
Recall that in order to maintain an overall well-defined orientation
under gluing it is necessary to stipulate that incoming boundary components
can only be glued to outgoing boundary components (i.e. the gluing
diffeomorphisms must be orientation \textit{reversing}). 

Since the defined orientation of $\Sigma_-$ disagrees with its induced
orientation as part of the boundary $\partial X_-$ we can use the parameterization
$\phi_-:\partial H_{g_-}\rightarrow\Sigma_-=-\partial X_-$ to glue $H_{g_-}$ to
$X$.  This is an orientation-reversing diffeomorphism, and effectively caps off
$\Sigma_-$.

Now consider $\Sigma_+$.  Here the defined orientation agrees with the
boundary orientation $\partial X_+$, so we cannot glue using the parameterization
$\phi_+:\partial H_{g_+}\rightarrow\Sigma_+=\partial X_+$ since this
is orientation preserving.  However,
we can use the mirror standard handlebody $\overline{H_{g_+}}$ instead
since we have a natural identification $-\partial \overline{H_{g_+}}=\partial H_{g_+}$.
The same map $\phi_+$ is orientation \textit{reversing}
\begin{equation}
 \phi_+:\partial \overline{H_{g_+}}\rightarrow\Sigma_+=\partial X_+ 
\end{equation}
so we cap off $\Sigma_+$ with
the standard handlebody $\overline{H_{g_+}}$.

\subsubsection*{Mapping class group}
In particular the operator assignment $\tau(X)$ for a 3-bordism $X$ 
provides a (projective - see below) 
representation of the mapping class group
for any surface $\Sigma$ of genus $g$.

Consider the surface $\Sigma$ parameterized by
a fixed diffeomorphism $\phi:\partial H_g\rightarrow\Sigma$.  Form the
3-manifold $X=\Sigma\times I$ with boundary $\partial X=-\Sigma\sqcup \Sigma$
where both the incoming and outgoing boundary components have the \textit{same}
parameterization $\phi$.  The operator associated to $\Sigma\times I$ (using the
above procedure) is just
the identity
\begin{equation}
  \tau(X)=\id:\modfunct(\Sigma)\rightarrow\modfunct(\Sigma)
\end{equation}

Now assume the we have some isotopy class of diffeomorphisms $[f]\in\text{MCG}(\Sigma)$
that are \textit{not} isotopic to the identity.
Picking a representative diffeomorphism $f:\Sigma\rightarrow\Sigma$ we form a new 3-manifold
$X_f=\Sigma\times I$ where the outgoing boundary component $\Sigma$ is still parameterized
by $\phi$, however the incoming boundary component $\Sigma$ is parameterized instead by
the map $f\circ\phi$.  Clearly when we ``cap off'' with standard handlebodies the
resulting closed 3-manifold $\tilde{X_f}$ will be different, hence the operator
$\tau(X_f)$ will not be the identity, but instead a nontrivial operator
\begin{equation}
  \tau(X_f):\modfunct(\Sigma)\rightarrow\modfunct(\Sigma)
\end{equation}

In this way we can associate to any element $[f]$ of $\text{MCG}(\Sigma)$ a linear operator
$\tau(X_f):\modfunct(\Sigma)\rightarrow\modfunct(\Sigma)$.  The composition of
diffeomorphisms $g\circ f$ can be realized by gluing the outgoing boundary
component of $X_f$ to the incoming boundary component of $X_g$, so we
have
\begin{equation}
 \tau(X_{g\circ f})=\tau(X_g\cup_{\text{glued}} X_f)=k \tau(X_g)\circ\tau(X_f)
\end{equation}
using the gluing properties outlined in chapter~(\ref{ch:tqft}).  Notice the gluing
anomaly $k$, hence we have a \textit{projective} representation of $\text{MCG}(\Sigma)$.

\section{Trivial examples from $(\grp,q,c)$}
\label{sec:easymtcs}
In chapter~(\ref{ch:chernsimons}) it was shown 
that the quantum
data for (non-spin) toral Chern-Simons theories
is encoded in a finite abelian group $\grp$, a \textit{pure} quadratic
form $q:\grp\rightarrow\qmodz$, and $c$ (an integer mod 24) that encodes
a choice of cube root of the Gauss reciprocity formula.

An easy semisimple ribbon category that can be formed (mentioned in
the appendix of \cite{deloup1}) from $\trio$ is the group
algebra $\cplx[\grp]$ where we write $\grp$ multiplicatively:
\begin{enumerate}
 \item For each $x\in\grp$ we define a simple object $\cplx x$ (a 1-dimensional
    $\cplx$ vector space with basis element $x$).
    An arbitrary object
    is defined to be a formal $\oplus$ of simple objects.
 \item Define the tensor product using the group law,
    i.e. $\cplx x \otimes \cplx y = \cplx xy$ (extend to arbitrary
    objects using additivity).
 \item A morphism $k:\cplx x\rightarrow\cplx x$ from a simple object to itself is just
    multiplication by a complex number $k$.  The set $\text{Mor}(\cplx_x,\cplx_y)$ for
    $x\neq y$ contains only the zero morphism $0$.  Extend to arbitrary objects in
    the obvious way by additivity.
 \item It is easy to check that $\cplx[\grp]$ is a strict monoidal category.
\end{enumerate}
We define a ribbon category $\cplx[\grp]_\trio$ (recall $b:\grp\otimes\grp\rightarrow\qmodz$
is the bilinear form induced from $q$):
\begin{enumerate}
 \item The braiding $c_{x,y}:\cplx x\otimes\cplx y\rightarrow\cplx y\otimes\cplx x$ on
    two simple objects
    is defined as (since in this case $\cplx x\otimes\cplx y\cong \cplx y\otimes\cplx x\cong\cplx xy$)
\begin{equation}
  \cplx xy \rightarrow \cplx xy\quad\text{ multiplication by }\exp\left(2\pi i b(x,y)\right)
\end{equation}
 \item Enforcing the balancing condition (and using the fact that $b(x,y)=q(x+y)-q(x)-q(y)$)
    we see that the twist for a simple object is
\begin{equation}
  \cplx x\rightarrow\cplx x\quad\text{multiplication by }\exp\left(2\pi i 2q(x)\right)
\end{equation}
\end{enumerate}
 
It is easy to compute the $S$-matrix:
\begin{equation}
  S_{xy}=\exp\left(2\pi i 2b(-x,y)\right)=\exp\left(-2\pi i 2b(x,y)\right)
\end{equation}

However it is easy to see that for $\text{U}(1)$ at level $B$ where $B$ is
an even integer the above $S$-matrix is singular.  For example (see
chapter~(\ref{ch:chernsimons})), for $B=2$ the discriminant group is
$\grp\cong\mathbb{Z}_2=\{0,1/2\}$, and the only non-degenerate
bilinear form is determined by
\begin{equation}
  b(1/2,1/2)= 1/2\pmod 1
\end{equation}
Hence we see that $2b(1/2,1/2)=1=0\pmod 1$.  So the $S$ matrix is
\begin{equation}
 \begin{pmatrix}
   1 & 1 \\ 1 & 1 
 \end{pmatrix}
\end{equation}
which is clearly singular.  It is trivial to see that for any cyclic
group of even order there will always be two rows in the $S$-matrix with $1$'s in the
entries (the first row \textit{always} has $1$'s in the entries).  Hence the $S$-matrix
will be singular in these cases, i.e. $\cplx[\grp]_\trio$ is 
often not a modular tensor category.
These theories \textit{cannot} describe toral Chern-Simons.

%\section{Quantum groups as a guide: $SU(2)$ at level $k=1$}
%Rank-level duality

\chapter{Group Categories}
\label{ch:groupcategories}

\section{Introduction}
The goal of this chapter is to construct a family of modular tensor categories such
that the associated TQFTs are isomorphic to the TQFTs
arising from toral (non-spin) Chern-Simons theories.
\footnote{Belov and Moore produce only part of the data required for
an extended $(2+1)$-dim TQFT.  We prove an isomorphism of (non-extended) $2$-d modular
functors in this paper.}
We already saw an
easy family of examples in chapter~(\ref{ch:mtc})
but we argued that these categories do not correspond to toral Chern-Simons.

Here we formulate the underlying braided categories in terms of 
an explicit set of equations.  It turns out that these equations can be cast in the
language of abelian group cohomology formulated by Eilenberg and MacLane in
the 1940's, hence allowing the use of homology and homotopy theory techniques
\cite{eilenberg_maclane}.
This identification was studied (in various incarnations) by 
Fr\"olich and Kerler \cite{frolich_kerler},
Joyal and Street \cite{joyal_street}, and Quinn \cite{quinn}.
The resulting braided categories are \textit{group categories}.
Recently much more work
has been done concerning group categories
\footnote{We thank Victor Ostrik for useful comments that guided us toward these
examples.}  (see for example \cite{eno,dgno}).  The same braiding construction in slightly
altered language also appeared in appendix E of \cite{moore_seiberg} as
well as in \cite{polyak_reshetikhin}.

We point out that if the Belov-Moore construction had
provided an \textit{extended} $2$-d modular functor (see chapter~(\ref{ch:tqft})) then
we could reverse-engineer the corresponding modular tensor categories
As mentioned more completely
in chapter~(\ref{ch:tqft}) we have the following causal relationships:
\begin{equation}
  \xymatrix{
    *\txt{Modular\\Tensor\\Category}\ar@{=>}[r]\ar@{->}[dr] &  
      *\txt{Extended\\ $(2+1)$-dim TQFT} \ar@{=>}[d]\ar@{=>}[r] & 
      *\txt{$(2+1)$-dim TQFT} \ar@{=>}[d]\\
      & *\txt{Extended $2$-d Modular Functor} \ar@/^/@{.>}[ul]\ar@{=>}[ur]\ar@{=>}[r] &
      *\txt{$2$-d Modular Functor} 
    }
\end{equation}

In this limited sense
the modular tensor categories described here \textit{extend and complete} the 
partial theories introduced
in \cite{belov_moore} using a rather different approach.

From toral Chern-Simons
considerations in chapter~(\ref{ch:chernsimons}) it was shown that the quantum
data is encoded in a finite abelian group $\grp$, a \textit{pure} quadratic
form $q:\grp\rightarrow\qmodz$, and $c$ (an integer mod 24) that encodes
a choice of cube root of the Gauss reciprocity formula.  Hence we shall
use this data to construct a modular tensor category.  We remind the
reader that we are not considering the more general spin/odd theories
considered in \cite{belov_moore}, but rather we are restricted to the
even theories because modular tensor categories correspond to
\textit{ordinary} TQFTs.  We also mention that the third piece of data
$c$ will not be necessary.  However $c$ can play a
role depending on the type of extended structure placed on 3-bordisms
\cite{atiyah,walker,freed_gompf}.
\footnote{Two examples of an extended structure are a $2$-framing
and a $p_1$-structure.  The $2$-framing is related to the
$p_1$-structure by a factor of $1/3$, hence this explains why the treatment in
\cite{belov_moore} requires a cube root of the Gauss reciprocity
formula whereas we do not.  Compare equation (2.1) in \cite{atiyah} with
theorem (2.3) in \cite{freed_gompf}.}

\section{Category $\mathscr{C}_\grp$ of $\grp$-graded complex vector spaces}
Let $\grp$ be a finite 
\footnote{We limit ourselves to finite groups here, but this is not
necessary.}
group (not necessarily abelian, but abelian in our case).  Following
Fr\"olich and Kerler~\cite{frolich_kerler}, Quinn~\cite{quinn}, 
and Joyal and Street~\cite{joyal_street} we consider
the following category $\mathscr{C}_\grp$:
\begin{enumerate}
 \item $\text{Ob}(\mathscr{C}_\grp)$ consists of finite-dimensional $\grp$-graded complex vector
    spaces.  In other words each object $V\in\text{Ob}(\mathscr{C}_\grp)$ is
    a finite-dimensional complex vector space that can be decomposed into 
    homogeneously-graded summands $V=\oplus_{x\in\grp}V_x$.
 \item $\text{Mor}(\mathscr{C}_\grp)$ consists of $\mathbb{C}$-linear maps that respect
    the group grading (i.e. the only nonzero blocks in a linear map
    $L:(V=\oplus_{x\in\grp}V_x)\rightarrow (W=\oplus_{y\in\grp}W_y)$ are along
    the diagonal $x=y$). 
 \item $\mathscr{C}_\grp$ has a monoidal structure $\otimes$.  If $V_x$ and $W_y$ 
    are homogeneously-graded
    objects then the product is defined by: $V_x\otimes W_y\equiv(V\otimes W)_{xy}$
    (the tensor product on the RHS is the usual one for vector spaces, and the grading obeys
    the group law).  More generally, for non-homogeneously-graded objects 
    if we impose the condition that $\otimes$ distributes over
    $\oplus$ then the above multiplication formula becomes convolution:
\begin{equation}
  (V\otimes W)_z = \oplus_{x,y| xy=z}V_x\otimes V_y
\end{equation}
    The product of morphisms is defined similary.
\end{enumerate}
Now let us make explicit some of the properties of $\mathscr{C}_\grp$:
\begin{enumerate}
  \item Since the vector space tensor product is \textit{strictly} associative (see
    chapter~(\ref{ch:mtc})) and group multiplication is
    \textit{strictly} associative we have that $\mathscr{C}_\grp$ is strictly
    associative with the identity
\begin{equation}
  \oplus_{x,y,z|(xy)z=a}(V_x\otimes W_y)\otimes Z_z =
  \oplus_{x,y,z|x(yz)=a}V_x\otimes (W_y\otimes Z_z)
  \label{eq:associativityidentity}
\end{equation}
 \item The vector space tensor product always comes equipped with a canonical
  isomorphism $\text{Perm}_{V,W}:V\otimes W\tilde{\rightarrow}W\otimes V$ defined on
  vectors by $v\otimes w\mapsto w\otimes v$.  This product is \textit{symmetric},
  meaning that we have an involution $\text{Perm}_{V.W}\circ\text{Perm}_{V,W}=\id$.  More generally,
  the symmetric group $S_n$ acts on the tensor product of $n$ factors.  If we
  mod out by the action of $S_n$ then we obtain the symmetric tensor product.  In
  this sense the vector space tensor product is commutative.

  Now consider the graded picture.  
  For $x,y\in\grp$ it is not always true that $xy=yx$, hence $\text{Perm}_{V,W}$ does
  \textit{not} in general lift to a canonical isomorphism 
  $V_x\otimes W_y\nrightarrow W_y\otimes V_x$ (since morphisms by definition must
  preserve grading).  However, if $\grp$ is \textit{abelian} then $xy=yx$
  and we have an induced canonical isomorphism
\begin{equation}
  \overline{\text{Perm}_{V,W}}:V_x\otimes W_y\tilde{\rightarrow} W_y\otimes V_x
\end{equation}
  for any $x,y\in\grp$.
 \item Following the approach outlined in chapter~(\ref{ch:mtc}) 
   we will shortly abandon the above associativity and
   commutativity in favor of a nontrivial family of natural
   isomorphisms.
 \item $\mathscr{C}_\grp$ is an \textit{abelian} category enriched over
   $\cplx$-vector spaces.  This is easily verified
   as follows:  it is clearly preadditive (Ab-category)
   since the sets $\text{Mor}(\mathscr{C}_\grp)$ are abelian groups (even better they
   are $\cplx$-vector spaces, so we refer to the 
   morphism sets as \textit{Hom} sets from here on).

  The $\oplus$ operation makes $\mathscr{C}_\grp$ an additive category.  It is
  preabelian because any linear map in $\text{Hom}(V,W)$ has a kernel and
  a cokernel.  Finally, it is easy to verify that any injective map
  $L:V\rightarrow W$ is the kernel of some map (namely the projection
  $W\rightarrow W/L(V)$);
  also any surjective map $L:V\rightarrow W$ is the cokernel of the projection
  map $V\oplus W\rightarrow V$.  So $\mathscr{C}_\grp$ is an abelian category enriched
  over $\cplx$-vector spaces.
 \item The monoidal structure on $\mathscr{C}_\grp$ is compatible with the
    abelian category structure  (i.e. $\otimes$ distributes over $\oplus$).
 \item $\mathscr{C}_\grp$ is clearly semisimple (every short exact sequence splits).
  More plainly any object can be decomposed as the direct sum of \textit{simple objects}. 
  The simple objects are 1-dimensional homogeneously-graded vector spaces; we denote them
\begin{equation}
  \{\mathbb{C}_x\}_{x\in\grp}\quad\quad\text{simple objects}.
\end{equation}
 \item There are only finitely-many simple objects since $\grp$ is a finite group.  In fact
   it
   is easy to define left and right duals and interpret $\mathscr{C}_\grp$ as a fusion category, but
   we refrain from doing so (we shall only define a right dual below).
 \item $\mathscr{C}_\grp$ can be viewed 
  as the group ring $\text{Vect}_{\mathbb{C}}[\grp]$
  where the coefficients are finite dimensional complex vector spaces.
 \item Alternatively, $\mathscr{C}_\grp$ can be profitably
  interpreted as the category of finite dimensional complex vector bundles over $\grp$.
  The multiplication of two complex vector bundles is defined to be the pushforward
  along multiplication on the base space $\grp$ (i.e. convolution).
\end{enumerate}

The category $\mathscr{C}_\grp$ is the canonical example of a \textit{group category}:

\begin{definition}
A \textbf{group category}
\footnote{We follow Quinn's definition \cite{quinn} which has an additive structure
that does not appear in the ``categorical groups'' discussed in Joyal and Street 
\cite{joyal_street} (the \textit{only} objects in
\cite{joyal_street} are simple).  However by adding a formal $\oplus$ it is trivial
to recover Quinn's definition.}
is a category with the following additional structure:
\begin{enumerate}
 \item Additive $\oplus$
 \item Monoidal $\otimes$
 \item $\otimes$ distributes over $\oplus$.
 \item Each $\text{Hom}$ space is an complex vector space.
\footnote{Quinn points out that it is often necessary to work with
$R$-modules where $R$ is a commutative ground ring.  We do not need
that greater generality here.}
 \item An object $V$ is called \textit{simple} if $\text{Hom}(V,V)\cong \mathbb{C}$.  Group
   categories are required to be semisimple (any object can be decomposed as a \textit{finite} sum of
   simple objects - however there need \textit{not} be finitely-many simple objects).
 \item For each simple object $V$ we require a \textit{right dual} object $V^*$ and a
   distinguished \textit{iso}morphism $d_V:V^*\otimes V\rightarrow\unitobj$ where $\unitobj$
   is the unit object for the monoidal structure.
\footnote{This is the ``$d_V$'' map that is part of the definition of duality.  However here
it is an isomorphism rather than just a morphism.  We did not mention this for the
example $\mathscr{C}_\grp$, but we shall mention it below.} 
 \item If $V$ and $W$ are distinct simple objects then we require $\text{Hom}(V,W)\cong 0$.
\end{enumerate}
\end{definition}

We note that the existence of a distinguished \textit{iso}morphism 
$d_V$ for each simple object is a strong condition.
We say that the simple objects are \textit{invertible}.  It is straightforward to check that the
definition implies that if $V$ and $W$ are simple then $V\otimes W$ is simple.  In
other words the simple objects form a group - the \textit{underlying} group of the
group category.

From here on we limit ourselves to the situation where $\grp$ is a finite abelian
group.

\section{Twisted version $\grpcat$: nontrivial associativity and braiding}
In the last section we introduced the category $\mathscr{C}_\grp$.  We
mentioned that if the underlying group $\grp$ is abelian then $\mathscr{C}_\grp$
is commutative in the sense that the tensor product of $n$ objects admits
an action of the symmetric group $S_n$.  Furthermore the monoidal structure
is strict.  \textit{Since we are dealing with finite abelian groups from now on
we switch from multiplicative $xy$ to additive $x+y$ notation.}

In light of chapter~(\ref{ch:mtc}) we aim to twist the structure
described in the last section to produce a non-strict modular tensor category.
Since the quantum data for toral Chern-Simons is encoded in the trio
$\trio$ we expect to use this data to twist the structure
appropriately (however we shall not require $c$ in this chapter).  
In light of this we denote the resulting
twisted category $\grpcat$.  Interestingly, a fixed set of data $\triotrunc$ actually produces
a \textit{family} of modular tensor categories.  We shall discuss how MTCs in
a given family are related to each other.

Since $\grpcat$ is an additive category it suffices to confine our study to the simple
objects
\begin{equation}
  \{\mathbb{C}_x\}_{x\in\grp}
\end{equation}
(we can extend to arbitary objects by additivity).
The fusion rules are trivial because of the strong structure imposed by a group category:
\begin{equation}
   \mathbb{C}_x \otimes \mathbb{C}_y\cong \mathbb{C}_{x+y}
\end{equation}
Let us first consider relaxing the associativity \textit{identity} in 
equation~(\ref{eq:associativityidentity}) and allow instead a family of
natural isomorphisms
\begin{equation}
  \{a_{x,y,z}:(\cplx_x\otimes\cplx_y)\otimes\cplx_z\tilde{\rightarrow}
    \cplx_x\otimes(\cplx_y\otimes\cplx_z)\}_{x,y,z\in\grp}
\end{equation}
Since the tensor product of simple objects is simple, for fixed $x,y,z\in\grp$ this is 
just an endomorphism
\begin{equation}
  a_{x,y,z}:\cplx_{x+y+z}\tilde{\rightarrow}\cplx_{x+y+z}
\end{equation}
In other words for each $x,y,z\in\grp$ it suffices to specify a complex number
$a_{x,y,z}$ (we have reused notation) such for $v\in\cplx_{x+y+z}$ we have $v\mapsto a_{x,y,z}v$.

It is clear that the unit object is just $\unitobj\equiv\cplx_0$.  In order to
find the coefficients $a_{x,y,z}$ we impose the pentagon identity 
(equation~(\ref{eq:pentagon})) and the triangle identity (equation~(\ref{eq:triangle})).
Since all isomorphisms involved are merely multiplication by complex numbers we
need not be concerned with ordering.  Explicity, for $v_x\in\cplx_x$, $v_y\in\cplx_y$,
$v_z\in\cplx_z$, and $v_w\in\cplx_w$ we follow the upper part of the pentagon diagram:
\begin{multline}
  ((v_x\otimes v_y)\otimes v_z)\otimes v_w\mapsto
  a_{x+y,z,w}(v_x\otimes v_y)\otimes (v_z\otimes v_w)\mapsto \\
  a_{x,y,z+w}a_{x+y,z,w}v_x\otimes (v_y\otimes (v_z\otimes v_w))
\end{multline}
Following the lower part of the pentagon diagram gives us
\begin{multline}
  ((v_x\otimes v_y)\otimes v_z)\otimes v_w\mapsto
  a_{x,y,z}(v_x\otimes (v_y\otimes v_z))\otimes v_w\mapsto \\
  a_{x,y+z,w}a_{x,y,z}v_x\otimes ((v_y\otimes v_z)\otimes v_w)\mapsto
  a_{y,z,w}a_{x,y+z,w}a_{x,y,z}v_x\otimes (v_y\otimes (v_z\otimes v_w))
\end{multline}
Comparing these we see that
\begin{equation}
  a_{x,y,z+w}a_{x+y,z,w}=a_{y,z,w}a_{x,y+z,w}a_{x,y,z}
  \label{eq:associativitycoefficients}
\end{equation}

If we restrict ourselves to solutions living in the unit circle then
we can write
\begin{equation}
  a_{x,y,z}:=\text{exp}(2\pi i h(x,y,z))
\end{equation}
for a phase function $h:\grp^3\rightarrow \qmodz$.  Equation~(\ref{eq:associativitycoefficients}) 
becomes
\begin{equation}
  h(x,y,z+w)+h(x+y,z,w)\equiv h(y,z,w)+h(x,y+z,w)+h(x,y,z)\pmod 1
  \label{eq:pentagonqmodz}
\end{equation}

Now let us consider the
triangle diagram in equation~(\ref{eq:triangle}).  If we set the right and
left identity maps in equations~(\ref{eq:rightidentity}) and (\ref{eq:leftidentity})
to be just multiplication by $1$, then the triangle diagram implies
\begin{equation}
  a_{x,0,y} = 1 
\end{equation}
In terms of $h$ this is just (mod 1)
\begin{equation}
  h(x,0,y)=0
\end{equation}
It is easy to exploit equation~(\ref{eq:pentagonqmodz}) to then prove that (mod 1)
\begin{equation}
  h(x,0,y)=h(0,x,y)=h(x,y,0)=0
  \label{eq:triangleqmodz}
\end{equation}

Now we wish to consider the hexagon relations depicted in equations~(\ref{eq:hexagon1})
and (\ref{eq:hexagon2}).  For simple objects $\cplx_x$ and $\cplx_y$ we postulate
a braiding isomorphism meant to replace the involution $\text{Perm}$:
\footnote{this can be extended to arbitrary objects by linearity}
\begin{equation}
  c_{x,y}:\cplx_x\otimes \cplx_y\tilde{\rightarrow}\cplx_y\otimes\cplx_x
\end{equation}
Again, because of the trivial fusion rules $\cplx_x\otimes \cplx_y\cong\cplx_{x+y}$
this is effectively an isomorphism
\begin{equation}
  c_{x,y}:\cplx_{x+y}\tilde{\rightarrow}\cplx_{x+y}
\end{equation}
and hence is determined by a $1\times 1$ complex matrix $[c_{x,y}]$.  Continuing
with our previous restriction to coefficients living in the unit circle
\begin{equation}
  c_{x,y}:=\text{exp}(2\pi i s(x,y))\quad\quad s:\grp^2\rightarrow\qmodz
\end{equation}
we see that the hexagon relations imply (mod 1)
\begin{align}
  s(x,y+z)&=-h(x,y,z)+s(x,y)+h(y,x,z)+s(x,z)-h(y,z,x)\label{eq:hexagonqmodz} \\
  s(x+y,z)&=h(x,y,z)+s(y,z)-h(x,z,y)+s(x,z)+h(z,x,y)\notag
\end{align}
As was the case for the function $h$, it is easy to calculate using these 
identities that
\begin{equation}
 s(0,y)=s(x,0)=0
\end{equation}

Summarizing, we can twist the category $\mathscr{C}_\grp$ into
a braided group category $\grpcat$ by
relaxing the associativity and commutativity identities.  We still expect
that any reasonable theory should obey the pentagon, triangle, and hexagon
relations as described in chapter~(\ref{ch:mtc}).  Since the
fusion rules are rather simple these relations can be cast
into the form of equations~(\ref{eq:pentagonqmodz}), (\ref{eq:triangleqmodz}),
and (\ref{eq:hexagonqmodz}) which are valued in $\qmodz$.

An interesting observation is that typically there
are multiple solutions to these equations (that turn out to be braided monoidal equivalent).
Since there are multiple solutions we denote the group category associated to
a solution $(h,s)$ by the notation
\begin{equation}
 \grpcat(h,s)
\end{equation}

This provides a richer structure
than one might naively expect.  In the next section following
Fr\"olich and Kerler \cite{frolich_kerler}, Quinn \cite{quinn}, and
Joyal and Street \cite{joyal_street} we identify these equations as
cocycles in group cohomology of abelian groups and provide explicit solutions.

\section{Connection with group cohomology}
In this section we provide a brief outline of \textit{abelian} group cohomology
as introduced by Eilenberg and MacLane (see 
\cite{eilenberg_maclaneintro} for a brief introduction and 
\cite{eilenberg_maclane} for a more detailed account).
\footnote{Warning: the conventions used by Quinn \cite{quinn} do not follow
those of Eilenberg and MacLane.  In particular the dimensions of the cells in the
relevant complex are defined to be 1 dimension higher in Quinn's paper.  Hence there
he studies $H^4$ whereas the \textit{same} cohomology classes are in $H^3$ in the
other references.}

Before we begin fix an underlying group $\Pi$ (in our case we will be considering the
finite abelian group $\grp$).  Fix an integer $m$ and an abelian coefficient
group $H$ (in our case $H=\qmodz$).

Consider a path-connected topological space $X$ such that $\pi_m(X)\cong \Pi$ and
all other homotopy groups are trivial (clearly if $m>1$ then
$\Pi$ must be abelian).  We wish to study the homology and cohomology
groups of this space.  One of the fundamental results of Eilenberg
and Maclane is that if $Y$ is a different topological space with the
same homotopy groups then the homology (cohomology) groups are also the
same:
\begin{equation}
  H(X;H)\cong H(Y;H)
\end{equation} 
This implies that it suffices to study the homology and cohomology groups of
the \textit{standard} Eilenberg-MacLane space $K(\Pi,m)$ (a cell complex
explicitly constructed below 
such that $\pi_m(X)\cong \Pi$ and all other homotopy groups are trivial).

On the other hand the main point of \cite{eilenberg_maclaneintro} and
\cite{eilenberg_maclane} is that if $\Pi$ is abelian (it is in our case)
then the cell complex $K(\Pi,m)$ can be replaced by a cell complex $A(\Pi)$ such
that the cohomology groups $H^k(A(\Pi);H)$ are much simpler to compute.  By
``replace'' we mean that the following isomorphism holds
(\cite{eilenberg_maclaneintro} article II,
Theorem 6):
\begin{equation}
  H^{m-1+k}(K(\Pi,m);H)\cong H^k(A(\Pi);H)\quad\quad k=1,\ldots,m
\end{equation}
We note that $m$ does not appear on the RHS (and $A(\Pi)$ is independent of $m$).
However, the isomorphism only holds for $k\leq m$.

We will eventually be interested in the case when $m=2$ and $k=3$, which clearly does not
satisfy the requirement $k\leq m$.  However, a more general statement can be made as follows.  The
space $A(\Pi)$ is constructed iteratively using the \textit{bar}
construction $B$.  That is we have a sequence of embedded spaces
\begin{equation}
  A^0(\Pi) \subset A^1(\Pi) \subset A^2(\Pi) \subset \ldots \subset A^\infty(\Pi)
\end{equation}
where we start with $A^0(\Pi)=K(\Pi,1)$ and apply the iterated bar construction
(see below)
$A^1(\Pi)=B(A^0(\Pi))$, $A^2(\Pi)=B(A^1(\Pi))=B(B(A^0(\Pi)))$, etc.  We define
$A(\Pi)=A^\infty(\Pi)$.

Now for \textit{arbitrary} $k$ the following isomorphism holds:
\begin{equation}
  H^{m-1+k}(K(\Pi,m);H)\cong H^k(A^{m-1}(\Pi);H)
\end{equation}
which is compatible with the previous isomorphism in the sense that
\begin{equation}
  H^k(A^{m-1}(\Pi);H)\cong H^k(A(\Pi);H)\quad\quad k=1,\ldots,m
\end{equation}

For $m=2$ and $k=3$ (our case of interest) this is just
\begin{equation}
  H^{4}(K(\Pi,2);H)\cong H^3(A^{1}(\Pi);H)
\end{equation}
Now let us discuss the iterated bar construction which will demonstrate why we
are interested in $H^3(A^{1}(\grp);\qmodz)$.

\subsection*{Iterated bar construction}
Since the iterated bar construction bootstraps using $K(\Pi,1)$ we construct this
cell complex first.  
Provide a $q$-dimensional cell labelled $[x_1,\ldots,x_q]$ for each $q$-tuple of
elements $x_1,\ldots,x_q\in\Pi$.  This cell attaches to the $(q-1)$-skeleton using
the boundary operator
\begin{multline}
 \partial [x_1,\ldots,x_q]= [x_2,\ldots,x_q] + 
   \sum_{i=1}^{q-1}(-1)^i [x_1,\ldots,x_i x_{i+1},\ldots,x_q]\\
   +(-1)^q[x_1,\ldots,x_{q-1}]
\end{multline}
(for 1-cells the boundary formula is defined as $\partial [x]=0$ since
each endpoint will attach to the unique 0-cell $[\,]$).

From now on we refer to $K(\Pi,1)$ as $A^0(\Pi)$.  We define a product
$*_0$ on the cells of $A^0(\Pi)$ via \textit{shuffling} (extend this to
chains by bilinearity):
\begin{equation}
  [x_1,\ldots,x_q]*_0 [y_1,\ldots,y_r]=\sum (-1)^\epsilon [z_1,\ldots,z_{q+r}]
\end{equation}
Here we are summing over all of the shuffles of the list $\{x_1,\ldots,x_q,y_1,\ldots,y_r\}$
where the $x_i$'s must stay in order relative to each other, 
and likewise for the $y_i$'s (i.e. $x$'s can only
swap with $y$'s).  The sign $(-1)^\epsilon$ is $1$ if the total number of transpositions
is even, and $-1$ if the total number of transpositions is odd.

For abelian $\Pi$ the operation $*_0$ defines a \textit{product of excess 0}.  
In general a \textit{product
of excess} $k$ on a chain complex is a bilinear function $*_k$ on chains $a$ and $b$ such that
if $d(a)$ denotes the \textit{cell} dimension of $a$ then
\begin{equation}
  d(a*_k b)=d(a)+d(b)+k
\end{equation}
If we define $d_k(a)=d(a)+k$ then this can be written more suggestively as
\begin{equation}
   d_k(a*_k b)=d_k(a)+d_k(b)
\end{equation}
In addition we require a product of excess $k$ to be associative, graded commutative, and
behave as usual with respect to the boundary operator:
\begin{align}
 & a*_k(b*_k c)=(a*_k b)*_k c\\
 & b*_k a = (-1)^\epsilon a*_k b\quad\quad\epsilon=d_k(a)d_k(b)\\
 & \partial(a*_k b)=(\partial a)*_k b + (-1)^{d_k(a)}a*_k(\partial b)
\end{align}

We iteratively define the complexes $A^*(\Pi)$ as follows:
from the complex $A^{k-1}(\Pi)$ with the product $*_{k-1}$ of excess ${k-1}$ we can produce
a complex $A^k(\Pi)$ which contains $A^{k-1}(\Pi)$ and in addition contains new
cells written
\begin{equation}
  [a_1|_k\ldots|_ka_p]\quad\quad a_i\text{ are cells of }A^k(\Pi) 
\end{equation}
These cells are defined to have cell dimension
\begin{equation}
 d([a_1|_k\ldots|_ka_p])=d(a_1)+\ldots+d(a_p) + (p-1)k
\end{equation}
In practice we write $|_1=|$, $|_2=||$, etc.

The boundary operator is defined as
\begin{multline}
 \partial[a_1|_k\ldots|_ka_p]=\sum_{i=1}^p (-1)^{\epsilon_{i-1}}
  [a_1|_k\ldots|_k a_{i-1} |_k \partial a_i |_k a_{i+1}|_k\ldots  a_p] +\\
 \sum_{i=1}^{p-1} (-1)^{\epsilon_{i}}
  [a_1|_k\ldots|_k a_{i-1} |_k a_i*_{k-1}a_{i+1} |_k a_{i+2}|_k\ldots  a_p]
\end{multline}
where $\epsilon_i=d_k(a_1)+\ldots+d_k(a_i)$.

We can also define a product of excess $k$ on $A^k(\Pi)$ using a similar shuffle
construction
\begin{equation}
  [a_1|_k\ldots|_ka_p]*_k[b_1|_k\ldots|_k b_r]=\sum (-1)^\epsilon[z_1|_k\ldots|_k z_{p+r}]
\end{equation}
where $\epsilon$ can be determined via the rule: a transposition of $a$ and $b$
multiplies by a factor $(-1)^{d_k(a)d_k(b)}$.
\footnote{again the $a$'s must stay in order relative to each other, and likewise
for the $b$'s.}

\subsection*{$H^3(A^1(\Pi);H)$}
Using these constructions it is simple to write down the cells in
$A^1(\Pi)$ (we will only bother up through cell dimension 4):

\begin{itemize}
 \item dimension 0: $[\,]$
 \item dimension 1: $[x]$ where $x\in\Pi$
 \item dimension 2: $[x,y]$ where $x,y\in\Pi$
 \item dimension 3: $[x,y,z]$ and $[x|y]$ where $x,y,z\in\Pi$
 \item dimension 4: $[x,y,z,w]$, $[x,y|z]$, and $[x|y,z]$ where $x,y,z,w\in\Pi$
\end{itemize}
The boundaries are easily computed:
\begin{itemize}
 \item dimension 0: $\partial[\,]=0$
 \item dimension 1: $\partial[x]=0$
 \item dimension 2: $\partial[x,y]=[y]-[x+y]+[x]$
 \item dimension 3: 
\begin{align}
  &\partial[x,y,z]=[y,z]-[x+y,z]+[x,y+z]-[x,y]\label{eq:3boundary}\\
  &\partial[x|y]=[x,y]-[y,x]\notag
\end{align}
 \item dimension 4:
\begin{align}
  \partial[x,y,z,w]&=[y,z,w]-[x+y,z,w]\label{eq:4boundary}\\
    &+[x,y+z,w]-[x,y,z+w]+[x,y,z]\notag\\
  \partial[x,y|z]&=[\partial[x,y]|z]-[[x,y]*_0 z]\notag\\
    &=[y|z]-[x+y|z]+[x|z]-[x,y,z]+[x,z,y]-[z,x,y]\notag\\
  \partial[x|y,z]&=[x|\partial[y,z]]-[x*_0 [y,z]]\notag\\
    &=[x|z]-[x|y+z]+[x|y]-[x,y,z]+[y,x,z]-[y,z,x]\notag
\end{align}
\end{itemize}

This provides a characterization of homology.  Now let
us compute cohomology.  We are only interested in $H^3(A^1(\Pi);H)$.
Consider a 3-cochain (a homomorphism)
\begin{equation}
  f:\text{3-chains}\rightarrow H
\end{equation}
When restricted to 3-cells of the form $[x,y,z]$ we use the notation
\begin{equation}
  h(x,y,z) := f([x,y,z])
\end{equation}
When restricted to 3-cells of the form $[x|y]$ we use the notation
\begin{equation}
  s(x,y) := f([x|y])
\end{equation}
To compute the cocycle condition $\delta f = 0$ it is easy to
write out the condition
\begin{equation}
  (\delta f)([\text{4-chain}]):= f(\partial[\text{4-chain}])=0
\end{equation}
and then use the boundary formulas in equation~(\ref{eq:4boundary}).  If
we consider the case where $\Pi=\grp$ and the coefficient group $H=\qmodz$
then this obviously reproduces equations~(\ref{eq:pentagonqmodz}) and (\ref{eq:hexagonqmodz}).

The only condition left to encode is the triangle identity (and its consequences)
in equation~(\ref{eq:triangleqmodz}).
For convenience we copy the conditions again:
\begin{equation}
  h(x,0,z)=h(x,y,0)=h(0,y,z)=s(x,0)=s(0,y)=0
\end{equation}
This is straightforward to achieve with cohomology of \textit{normalized chains}.  Let
$A_N^1(\Pi)$ be the subcomplex of $A^1(\Pi)$ consisting of cells
$[x_1,\ldots,x_q]$ with at least one $x_i=0$ (any of the commas
may be replaced with bars $|$ as well).  Then all of the identities are
satisfied by the cohomology of \textit{normalized} 3-cochains
\footnote{A similar subcomplex $K_N(\Pi,m)$ of $K(\Pi,m)$ can be defined and
cohomology can be studied there as well.}
\begin{equation}
 H^3(A^1(\grp)/A_N^1(\grp);\qmodz)
\end{equation}
We will refrain from over-decorating the notation with
cohomology of normalized chains since it does not affect the outcome.

\subsection*{Explicit cocycles}
The groups $H^3(A^{1}(\Pi);H)$ 
were computed in the original Eilenerg-MacLane articles
(see \cite{eilenberg_maclane} article II pg 92 and pg 130).  For $\Pi$ cyclic
an explicit computation is performed in \cite{joyal_street} and the full
computation for general finite abelian groups can be found in \cite{quinn}.
For the reader who wishes to compare the different references 
we emphasize again the following isomorphism:
\begin{equation}
  H^{4}(K(\Pi,2);H)\cong H^3(A^{1}(\Pi);H)
\end{equation}

Let $q_1:\grp\rightarrow\qmodz$ and $q_2:\grp\rightarrow\qmodz$ be two
pure quadratic forms.  Then it is easy to verify that $q_1+q_2$ is also
a pure quadratic form.  It is also trivial to verify that for a pure quadratic
form $q$ its inverse $-q$ is also a pure quadratic form.  Finally the constant
function $q=0$ is also a pure quadratic form.  Hence the set of pure quadratic
forms
\begin{equation}
  \{q:\grp\rightarrow\qmodz \}
\end{equation}
forms a group which we denote by $\text{Quad}(\grp,\qmodz)$.  It is shown
in \cite{eilenberg_maclane} pg 130 that there is a canonical isomorphism
\begin{equation}
 H^3(A^{1}(\grp);\qmodz)\tilde{\rightarrow} \text{Quad}(\grp,\qmodz)
\end{equation}
determined by defining $q(x):=s(x,x)$.

What we are missing is a recipe that produces an
explicit representative cocycle $(h,s)$ from a finite abelian
group $\grp$ equipped with a quadratic form $q:\grp\rightarrow\qmodz$.
Following Quinn \cite{quinn} we have the following (family of) explicit solutions:

\begin{enumerate}
 \item Pick a set of generators $1_i$ for $\grp$ ($\grp$ is a finite abelian group,
   hence can be decomposed into cyclic factors of order $n_i$)
 \item Pick an \textit{ordering} of the generators $1_1<1_2<\ldots$
 \item Write any arbitrary element $x\in\grp$ as $x=a_1 1_1+a_2 1_2+\ldots$
   \textit{such that $0\leq a_i<n_i$ for every $i$}
\end{enumerate}
We emphasize that this construction is \textit{not} well defined on the group $\grp$, but
is well defined on the group $\grp$ \textit{equipped with ordered generators}.  For
further emphasis we repeat that the coefficients $a_i$ \textbf{must always be written as integers
$0\leq a_i <n_i$} (i.e. we do \textit{not} write $-x=-a_1 1_1-a_2 1_2-\ldots$, but rather
$-x=(n_1-a_1) 1_1+(n_2-a_2) 1_2+\ldots$).

Since $\grp$ is equipped with a quadratic form $q$ we denote $q_i:=q(1_i)\in\qmodz$.
Also a pure quadratic form $q:\grp\rightarrow\qmodz$ determines a bilinear form
$b:\grp\otimes\grp\rightarrow\qmodz$ defined by $b(x,y):=q(x+y)-q(x)-q(y)$.
We denote $b_{ij}:=b(1_i,1_j)$.

Then if $x=\sum_i a_i 1_i$, $y=\sum_i b_i 1_i$, and $z=\sum_i c_i 1_i$ then the
associativity is defined by
\begin{equation}
  h(x,y,z)=\sum_i
  \begin{cases}
     0 & \text{if }b_i+c_i<n_i\\
     n_i a_i q_i & \text{if }b_i+c_i\geq n_i
  \end{cases}
\label{eq:explicitassociativity}
\end{equation}
and the braiding is given by
\begin{equation}
  s(x,y)=\sum_{i<j} a_i b_j b_{ij} +\sum_{i} a_i b_i q_i
\label{eq:explicitbraiding}
\end{equation}
Some quick calculations confirm that this solution satisfies equations~(\ref{eq:pentagonqmodz}),
(\ref{eq:triangleqmodz}), and (\ref{eq:hexagonqmodz}).

\subsection*{Coboundaries and braided monoidal equivalence}
We mentioned in the last section that a cohomology class $[h,s]$ is determined by
a quadratic form $q:\grp\rightarrow\qmodz$, and to find an explicit representative
$(h,s)$ we are forced to pick an ordered set of generators.

The isomorphism (proved in \cite{eilenberg_maclane})
\begin{equation}
 H^3(A^{1}(\grp);\qmodz)\tilde{\rightarrow} \text{Quad}(\grp,\qmodz)
\end{equation}
means that if we have two representatives $(h,s)$ and
$(h^\prime,s^\prime)$ that are determined by
different choices of ordered generators then their difference $(h,s)-(h^\prime,s^\prime)$
must be a coboundary.  This is easy to show directly: if we consider the
homology boundary maps in equation~(\ref{eq:3boundary}) then passing to cohomology
the expression $(h,s)-(h^\prime,s^\prime)$ 
should be the coboundary of \textit{some} function $k:\grp^2\rightarrow\qmodz$, i.e.
\begin{align}
  (h-h^\prime)(x,y,z)&=k(y,z)-k(x+y,z)+k(x,y+z)-k(x,y)\label{eq:cohomologousk}\\
  (s-s^\prime)(x,y)&=k(x,y)-k(y,x)\notag
\end{align}
A tedious calculation shows that for $(h,s)$, $(h^\prime,s^\prime)$ determined by
different choices of ordered generators there is such a function $k$.

Now we must answer how
two group categories $\grpcat(h,s)$ and $\grpcat(h^\prime,s^\prime)$
constructed from cohomologous $(h,s)$ and $(h^\prime,s^\prime)$
are related.  It turns out that the resulting group categories are
\textit{braided monoidal equivalent}.  This was proven by Joyal and Street
\cite{joyal_street} (the proof is written in slightly greater detail below).

In order to define a braided monoidal equivalence we start with some preliminaries.
\begin{definition}
Let $\Vcat$, $\Vcat^\prime$ be two monoidal categories.  A
\textbf{monoidal functor} is a triple $(F,\phi_2,\phi_0)$ given by \cite{joyal_street}
\begin{enumerate}
 \item A functor $F:\Vcat\rightarrow \Vcat^\prime$.
 \item A family of natural isomorphisms (one for each pair of objects $A,B\in\Vcat$):
\begin{equation}
 \phi_{2,A,B}:FA\otimes FB\tilde\rightarrow F(A\otimes B)
\end{equation}
 \item An isomorphism
\begin{equation}
 \phi_0:\unitobj^\prime \tilde{\rightarrow} F\unitobj
\end{equation}
\end{enumerate}
In addition we require that the following diagrams commute:
\begin{equation}
\xymatrix {
      & FA\otimes (FB\otimes FC)\ar[dr]^-{\id_A\otimes \phi_{2,B,C}}& \\
  (FA\otimes FB)\otimes FC \ar[ur]^-{a_{A,B,C}}\ar[d]^-{\phi_{2,A,B}\otimes\id_C}& &
   FA\otimes F(B\otimes C)\ar[d]^-{\phi_{2,A,B\otimes C}} \\
  F(A\otimes B)\otimes FC\ar[dr]^-{\phi_{2,A\otimes B,C}} & & F(A\otimes(B\otimes C)) \\
  & F((A\otimes B)\otimes C)\ar[ur]^-{F(a_{A,B,C})}&
}
\label{eq:monoidalfunctorhexagon}
\end{equation}
\begin{equation}
\xymatrix {
 FA\otimes\unitobj^\prime\ar[r]^-{r_{FA}}\ar[d]^-{\id_{FA}\otimes \phi_0} & FA
   & & \unitobj^\prime\otimes FA\ar[r]^-{l_{FA}}\ar[d]^-{\phi_0\otimes\id_{FA}} & FA \\
 FA\otimes F\unitobj\ar[r]^-{\phi_{2,A,\unitobj}} & F(A\otimes\unitobj)\ar[u]^-{F(r_A)} 
 & & F\unitobj\otimes FA\ar[r]^-{\phi_{2,\unitobj,A}} & F(\unitobj\otimes A)\ar[u]^-{F(l_A)}
}
\end{equation}
\end{definition}
We can have natural transformations (and natural isomorphisms) between 
ordinary functors; we want to extend to 
a notion of \textit{monoidal natural transformation} between two \textit{monoidal} 
functors.
\begin{definition} Let $F:\Vcat\rightarrow\Vcat^\prime$
and $G:\Vcat\rightarrow\Vcat^\prime$ be monoidal functors.  
A \textbf{monoidal natural transformation} is an ordinary
natural transformation $\theta:F\rightarrow G$ that in addition is required
to satisfy the following commutative diagrams:
\begin{equation}
\xymatrix {
  FA\otimes FB\ar[r]^-{\phi^F_{2,A,B}}\ar[dd]^-{\theta_A\otimes \theta_B} & 
    F(A\otimes B)\ar[dd]^-{\theta_{A\otimes B}} & & & F\unitobj\ar[dd]^-{\theta_\unitobj} \\
  & & & \unitobj^\prime \ar[ur]^-{\phi^F_0} \ar[dr]^-{\phi^G_0} & \\
  GA\otimes GB\ar[r]^-{\phi^G_{2,A,B}} & G(A\otimes B) & & & G\unitobj
}
\end{equation}
This defines a \textbf{monoidal natural isomorphism} if all of the
arrows $\theta_A$ are isomorphisms.  We denote a monoidal natural isomorphism
by the symbol $\cong$.
\end{definition}

Now define a notion of equivalence between two monoidal categories:
\begin{definition}
Let 
$(F,\phi^F_2,\phi^F_0):\Vcat\rightarrow \Vcat^\prime$
and $(F^\prime,\phi^{F^\prime}_2,\phi^{F^\prime}_0):\Vcat^\prime\rightarrow \Vcat$
be monoidal functors.  Then these are said to be a \textbf{monoidal equivalence}
if
\begin{equation}
  F^\prime\circ F\cong I_\Vcat\quad\quad\quad F\circ F^\prime\cong I_{\Vcat^\prime}
\end{equation}
where $I_{\Vcat}, I_{\Vcat^\prime}$ are the identity monoidal functors.
\end{definition}

Now we are ready to consider \textit{braided} monoidal
categories.
\begin{definition}
Let $\Vcat$ and $\Vcat^\prime$ be braided monoidal categories with braidings
$c$ and $c^\prime$, respectively (in the sense
of chapter~(\ref{ch:mtc})).  A \textbf{braided monoidal functor}
$F:\Vcat\rightarrow \Vcat^\prime$ is a monoidal functor that in addition must
make the following compatibility diagram commute:
\begin{equation}
\xymatrix{
  FA\otimes FB\ar[r]^-{\phi_{2,A,B}} \ar[d]^-{c^\prime_{FA,FB}} & 
    F(A\otimes B) \ar[d]^-{F(c_{A,B})} \\
  FB\otimes FA \ar[r]^-{\phi_{2,B,A}} & F(B\otimes A)
}
\label{eq:braidedmonoidalfunctorsquare}
\end{equation}
\end{definition}

\begin{definition}
A \textbf{braided natural transformation} between two
braided monoidal functors $F:\Vcat\rightarrow\Vcat^\prime$ and 
$G:\Vcat\rightarrow\Vcat^\prime$ is a monoidal natural transformation
$\theta:F\rightarrow G$ that satisfies the following compatibility
commutative diagram:
\begin{equation}
\xymatrix{
  FA\otimes FB \ar[r]^-{c^\prime_{FA,FB}} \ar[d]^-{\theta_A\otimes\theta_B}
    & FB \otimes FA \ar[d]^-{\theta_B\otimes\theta_A} \\
  GA\otimes GB \ar[r]^-{c^\prime_{GA,GB}} & GB\otimes GA 
} 
\end{equation}
Obviously this defines a \textbf{braided monoidal natural isomorphism} if
all of the arrows $\theta_A$ are isomorphisms.  We reuse notation and
denote this $\cong$.
\end{definition}

\begin{definition}
Let 
$(F,\phi^F_2,\phi^F_0):\Vcat\rightarrow \Vcat^\prime$
and $(F^\prime,\phi^{F^\prime}_2,\phi^{F^\prime}_0):\Vcat^\prime\rightarrow \Vcat$
be braided monoidal functors.  Then these are said to be a \textbf{braided monoidal equivalence}
if
\begin{equation}
  F^\prime\circ F\cong I_\Vcat\quad\quad\quad F\circ F^\prime\cong I_{\Vcat^\prime}
\end{equation}
where $I_{\Vcat}, I_{\Vcat^\prime}$ are the identity braided monoidal functors.
\end{definition}

Two braided monoidal categories that are braided monoidal equivalent are (in the above sense) 
the same.  This is the appropriate way to interpret the following theorem which answers
how to relate group categories constructed by choosing different ordered lists of generators.

\begin{theorem}[Joyal and Street] The group categories $\grpcat(h,s)$ and
$\grpcat(h^\prime,s^\prime)$ are braided monoidal equivalent iff $(h,s)$
and $(h^\prime,s^\prime)$ are cohomologous 3-cocycles in $H^3(A^1(\grp);\qmodz)$.
\end{theorem}
\begin{proof}
$(\Leftarrow)$
Suppose that $(h,s)$ and $(h^\prime,s^\prime)$ are cohomologous, i.e. let
$k:\grp^2\rightarrow \qmodz$ be as in equation~(\ref{eq:cohomologousk}).
Since both categories share the same underlying ordinary category
we consider the identity functor
\begin{equation}
  I:\grpcat(h,s)\rightarrow\grpcat(h^\prime,s^\prime)
\end{equation}
This functor is not yet a monoidal functor because the associativity structures
$h$ and $h^\prime$ are different.  We need to construct $\phi_2$ and $\phi_0$.

It is enough to consider the simple objects
and extend by linearity.  Let $x,y\in\grp$.  Then the map 
\footnote{the source and
target and the same since we are using the identity functor}
\begin{equation}
 \phi_{2,x,y}:\cplx_x\otimes\cplx_y\tilde{\rightarrow}\cplx_x\otimes\cplx_y
 \quad\quad\text{multiplication by }\text{exp}(2\pi i k(x,y))
\end{equation}
and the map
\begin{equation}
 \phi_0:\unitobj\tilde{\rightarrow}\unitobj
 \quad\quad\text{multiplication by }1
\end{equation}
define a monoidal functor $(I,\phi_2,\phi_0)$ since it is straightforward to
verify that the diagram in equation~(\ref{eq:monoidalfunctorhexagon}) is encoded
in the first line of equation~(\ref{eq:cohomologousk}) (and the other diagrams
are trivial).

In fact $(I,\phi_2,\phi_0)$ also defines a braided monoidal functor because the
diagram in equation~(\ref{eq:braidedmonoidalfunctorsquare}) is seen to be
encoded in the second line of equation~(\ref{eq:cohomologousk}).

Finally,
$(I,\phi_2,\phi_0)$ and its obvious inverse $(I,\phi_2^{-1},\phi_0^{-1})$ are
verified (trivially) to form a braided monoidal equivalence.

$(\Rightarrow)$
Straightforward using essentially the reverse argument to produce $k$ (left to the reader
since we shall not use this result).
\end{proof}

\section{Modular tensor category}
The categories $\grpcat(h,s)$ are braided (non-strict) monoidal categories.
In addition we have seen that they are finitely-semisimple abelian categories.
\footnote{They are also enriched over $\cplx$-vector spaces, 
i.e. the $\text{Hom}$ sets are $\cplx$-vector spaces.  Furthermore the
monoidal structure and the abelian structure are compatible in the sense
that $\otimes$ distributes over $\oplus$.}
In this section we slightly extend the categories $\grpcat(h,s)$ to produce
modular tensor categories (we use the same notation since no additional data
is required).  We do not know if this appears explicitly elsewhere in the literature.

\subsection*{Ribbon structure}
First, it is necessary to form a ribbon structure on $\grpcat(h,s)$.  We
start with the twist.  

\subsubsection*{Twisting}
Note that the quadratic form satisfies $q(x):= s(x,x)$.
For a simple object $\cplx_x$ we define the twist to be
\begin{align}
 &\theta_x:\cplx_x\rightarrow\cplx_x\\
 &v \mapsto \text{exp}(2\pi i q(x)) v\notag
\end{align}
This can be extended to arbitrary objects by linearity.
We need to check the balancing identity in equation~(\ref{eq:balancing}).

\begin{proposition}
The braided monoidal category $\grpcat(h,s)$ with twisting defined on the
simple objects $\cplx_x$ by
\begin{align}
 &\theta_x:\cplx_x\rightarrow\cplx_x\\
 &v \mapsto \text{exp}(2\pi i q(x)) v\notag
\end{align}
is balanced.
\end{proposition}
\begin{proof}
We check this only on the simple objects.  Let $\cplx_x$ and $\cplx_y$
be two simple objects.  Since $\cplx_x\otimes\cplx_y\cong\cplx_{x+y}$
what we are trying to verify is the equation
\begin{equation}
  \theta_{x+y}\theta^{-1}_{x}\theta^{-1}_{y}=c_{y,x}\circ c_{x,y}
\end{equation}
The LHS is easy to write out as
\begin{equation}
  \exp[2\pi i(q(x+y)-q(x)-q(y))]
\end{equation}
However, because $q$ is a quadratic form we have $q(x+y)-q(x)-q(y)=b(x,y)$ where
$b:\grp\otimes\grp\rightarrow\qmodz$ is the induced bilinear form.

In terms of the generators $1_i$ for $\grp$ we can write
\begin{align}
  x&=\sum_i a_i 1_i\\
  y&=\sum_j b_j 1_j\notag
\end{align}
in which case $b(x,y)$ becomes
\begin{equation}
  \sum_{i,j} a_i b_j b(1_i,1_j) 
\end{equation}
which in the notation preceding equation~(\ref{eq:explicitbraiding}) is
\begin{equation}
  \sum_{i,j} a_i b_j b_{ij}
\end{equation}
which is
\begin{equation}
    =2\sum_{i<j} a_i b_j b_{ij} + \sum_i a_i b_i b_{ii}
\end{equation}
(we have used the symmetry of $b(\cdot,\cdot)$).  However the general relation
$q(x+y)-q(x)-q(y)=b(x,y)$ specializes when $x=y$ to $q(2x)-2q(x)=b(x,x)$, and
since $q$ is a pure quadratic form we see that this is just $4q(x)-2q(x)=2q(x)=b(x,x)$.
In particular $b_{ii}=2q_i$.  In light of this the expression above becomes
\begin{equation}
  2\sum_{i<j} a_i b_j b_{ij} + 2\sum_i a_i b_i q_i
\end{equation}
which is clearly equal (after taking the exponent) 
to the RHS $c_{y,x}\circ c_{x,y}$ using the braiding in equation~(\ref{eq:explicitbraiding}).
\end{proof}

\subsubsection*{Rigidity}
Now let us address rigidity.  Again, by linearity it suffices to restrict our
attention to the simple objects $\cplx_x$.  Given a simple object $\cplx_x$ the
right dual is
\begin{equation}
  (\cplx_x)^*:=\cplx_{-x}
\end{equation}

Pick a basis $v_x$ for
each $\{\cplx_x\}_{x\in\grp}$ (the construction does not depend this choice).  
Define the birth morphism via the formula
\begin{align}
  b_x:&\unitobj\rightarrow \cplx_x\otimes\cplx_{-x}\\
    &v_0\mapsto v_x\otimes v_{-x}\notag
\end{align}
We do \textit{not} define the death morphism via the obvious formula
\begin{align}
  d_x:&\cplx_{-x}\otimes\cplx_{x}\rightarrow\unitobj\\
    &v_{-x}\otimes v_{x}\nrightarrow v_0\notag
\end{align}
Instead we are obligated to enforce the rigidity conditions in equation~(\ref{eq:nonstrictrigidity}).
Consider the first sequence of maps in equation~(\ref{eq:nonstrictrigidity}) (the second
sequence is similar and provides identical information).  For a simple object $\cplx_x$
the sequence (which must equal $\id_x$) is:
\begin{multline}
  v_x\overset{l_x^{-1}}{\longmapsto} v_0\otimes v_x
    \overset{b_x\otimes\id_x}{\longmapsto} (v_x\otimes v_{-x})\otimes v_x
    \overset{a_{x,-x,x}}{\longmapsto} \\ [a_{x,-x,x}]\cdot v_x\otimes (v_{-x}\otimes v_x)
    \overset{\id_{x}\otimes d_x}{\longmapsto} \\ [a_{x,-x,x}\cdot d_x]\cdot v_x\otimes v_0
    \overset{r_V}{\longmapsto} [a_{x,-x,x}\cdot d_x]\cdot v_x   
\end{multline}
this implies that
\begin{equation}
  a_{x,-x,x}\cdot d_x = 1
\end{equation}
i.e.
\begin{equation}
  \text{exp}(2\pi i h(x,-x,x))\cdot d_x = 1
\end{equation}
Hence we define the death morphism by the formula 
\begin{align}
  d_x&:\cplx_{-x}\otimes\cplx{x}\rightarrow\unitobj \\
    & v_{-x}\otimes v_x\mapsto \text{exp}(-2\pi i h(x,-x,x))v_0
\end{align}

Collecting these facts, we have proven:
\begin{proposition}
\label{prop:ribbongroupcat}
The group category $\grpcat(h,s)$ extended by the above twisting and rigidity
is a finitely-semisimple ribbon category.
\end{proposition}

\subsubsection*{Quantum dimension}
The quantum dimension is defined by equation~(\ref{eq:qdim}).  We reuse
the following lemma several times in the sequel:

\begin{lemma}
\label{lem:twistbraiddeathtrivial}
Let $\cplx_x$ be a simple object in $\grpcat(h,s)$.  Then the
map
\begin{equation}
  d_x\circ c_{x,-x} \circ (\theta_x\otimes\id_{-x}):\cplx_x\otimes\cplx_{-x}\cong\unitobj
    \rightarrow\unitobj
\end{equation}
is just multiplication by $1$.
\end{lemma}
\begin{proof}
This is a calculation (with a fairly tricky point that has confused the
author more than once).  $\theta_x$ is multiplication by the
coefficient $\exp(2\pi i q(x))$.  The braiding $c_{x,-x}$ is multiplication
by $\exp(2\pi i s(x,-x))$, and the death operator $d_x$ is multiplication
by $\exp(-2\pi i h(x,-x,x))$.  Hence the total coefficient is just
$\exp[2\pi i(q(x)+s(x,-x)-h(x,-x,x))]$.  In terms of ordered generators
for $\grp$ we write $x=\sum_i a_i 1_i$.  The tricky point is that
it is \textit{not} true that $-x=\sum_i (-a_i)1_i$.  In view of the
commentary above equations~(\ref{eq:explicitassociativity}) and 
(\ref{eq:explicitbraiding}) we see instead that $-x=\sum_i (n_i-a_i)1_i$, so
\begin{multline}
  q(x)+s(x,-x)= \sum_{i<j} a_i a_j b_{ij} +\sum_{i} a_i a_i q_i \\
    + \sum_{i<j} a_i (n_j-a_j) b_{ij} +\sum_{i} a_i (n_i-a_i) q_i =\\
    \sum_{i<j} a_i n_j b_{ij} +\sum_{i} a_i n_i q_i
\end{multline}
However $\sum_{i<j} a_i n_j b_{ij}=\sum_{i<j} b(a_i 1_i,n_j 1_j)=
\sum_{i<j} b(a_i 1_i,0)=0$.  Hence we are left with
\begin{equation}
  q(x)+s(x,-x)= \sum_{i} a_i n_i q_i
\end{equation}
The death operator gives
\begin{equation}
  h(x,-x,x) = \sum_i n_i a_i q_i
\end{equation}
so
\begin{equation}
  q(x)+s(x,-x)-h(x,-x,x)= \sum_{i} a_i n_i q_i - \sum_{i} a_i n_i q_i = 0
\end{equation}
Taking the exponent we get that the map is just multiplication by $1$.
\end{proof}

This easily implies the following (note: this result has nothing to do with the
fact that the simple objects are $1$-dimensional vector spaces $\cplx_x$;
the quantum dimension is not related):

\begin{corollary}
\label{cor:simpleobjectsdim1}
The simple objects $\cplx_x$ in $\grpcat(h,s)$ all have quantum dimension
$\text{dim}_q(\cplx_x)=1$.  
\end{corollary}

\subsection*{Modular tensor category}
In light of proposition~(\ref{prop:ribbongroupcat}) we only need to mention
the rank $\mathscr{D}$ and verify that the $S$ matrix is invertible.  Then
we will have a modular tensor category.  The rank is
\begin{equation}
 \mathscr{D}=\sqrt{\sum_{x\in\grp} (\text{dim}_q(\cplx_x))^2}=\sqrt{|\grp|}
\end{equation}

The coefficients of the $S$ matrix are determined by equation~(\ref{eq:smatrixdef}).
Recall that the quadratic form $q:\grp\rightarrow\qmodz$ induces a bilinear 
form $b:\grp\otimes\grp\rightarrow\qmodz$.  
A quick calculation using equation~(\ref{eq:explicitbraiding})
shows that
\begin{equation}
  S_{x,y}=\exp\left(2\pi ib(-x,y)\right)=\exp\left(-2\pi ib(x,y)\right)
\end{equation}
This proves:

\begin{theorem}
The group category $\grpcat(h,s)$ extended with the twist and rigidity structure
defined above is a modular tensor category iff the quadratic form $q:\grp\rightarrow\qmodz$
is a refinement of a bilinear form $b:\grp\otimes\grp\rightarrow\qmodz$ such
that the matrix $S_{x,y}=\exp\left(-2\pi ib(x,y)\right)$ is invertible.
\end{theorem}

We believe that the following proposition is true for all finite abelian groups,
but we have only been able to prove it for cyclic groups:

\begin{proposition}
Let $\grp$ be a cyclic group.  Then the matrix
\begin{equation}
S_{x,y}=\exp\left(-2\pi ib(x,y)\right) 
\end{equation}
is invertible iff 
$b:\grp\otimes\grp\rightarrow\qmodz$ is non-degenerate.
\end{proposition}
\begin{proof}
If $b$ is degenerate then the matrix $b(x,y)$ has two rows consisting
of zeros: the top row (since $b(0,y)=0$) and another row
$b(x,y)=0$ for some $x\neq 0$.  Hence the matrix
\begin{equation}
 S_{x,y}=\exp\left(-2\pi ib(x,y)\right) 
\end{equation}
has two rows filled with $1$'s, hence $S_{x,y}$ is not invertible.

Conversely, suppose that $b$ is non-degenerate.  Let $1$ be a generator
for the cyclic group $\grp$ of order $n$, 
and define $X:=\exp\left(-2\pi ib(1,1)\right)$.
Then for integers $k,l=0,1,2,\ldots,n-1$ we have the $S$-matrix
\begin{equation}
  S_{k,l}:=X^{kl}
\end{equation}

A \textit{Vandermonde determinant} is a determinant of a matrix of
the form
\begin{equation}
  \begin{pmatrix}
    1 & x_1 & x^2_1 & x^3_1 & \ldots \\
    1 & x_2 & x^2_2 & x^3_2 & \ldots \\
    1 & x_3 & x^2_3 & x^3_3 & \ldots \\
    \vdots 
  \end{pmatrix}
\end{equation}
It is well-known that the determinant of this matrix is just
\begin{equation}
  \prod_{0\leq k<l\leq n-1}(x_l-x_k)
\end{equation}
The $S$-matrix is of the Vandermonde form
\begin{equation}
  \begin{pmatrix}
    1 & 1 & 1 & 1 & \ldots \\
    1 & X & X^2 & X^3 & \ldots \\
    1 & X^2 & X^4 & X^6 & \ldots \\
    1 & X^3 & X^6 & X^9 & \ldots \\
    \vdots 
  \end{pmatrix}
\end{equation}
Since for non-degenerate $b$ we have that $X^k\neq X^l$ when $k\neq l$ we
see that the determinant of $S$ is non-zero.
\end{proof}

%\subsection*{Mirror category}
%Every modular tensor category $\grpcat(h,s)$ induces a mirror (opposite
%chirality) modular tensor category $\overline{\grpcat(h,s)}$ \cite{turaev}.
%It is in fact the \textit{mirror} category and associated TQFT that corresponds
%to the toral Chern-Simons TQFT (this is proven in the next chapter).
%In our case the mirror category is defined by
%\begin{equation}
% \overline{\grpcat(h,s)}:=\mathscr{C}_{(\grp,-q,-c)}(-h,-s)
%\end{equation}

%We mention in particular (restricting attention to simple objects) for 
%$\overline{\grpcat(h,s)}$ the twist is given by
%\begin{align}
% \theta_x&:\cplx_x\rightarrow\cplx_x\\
%  &v\mapsto \text{exp}(-2\pi i q(x)) \notag
%\end{align}
%and the braiding is given by
%\begin{align}
% c_{x,y}&:\cplx_x\otimes\cplx_y\rightarrow\cplx_y\otimes\cplx_x \\
% &v_x\otimes v_y\mapsto \text{exp}(-2\pi i s(x,y)) v_y\otimes v_x \notag
%\end{align}
%The $S$-matrix is
%\begin{equation}
%  S_{x,y}=-b(-x,y)=b(x,y)
%\end{equation}

\chapter{Main Theorem}
\label{ch:correspondence}

\section{Introduction}
 The goal of this chapter is to provide a correspondence between the toral 
(non-spin) Chern-Simons
theories classified by Belov and Moore (see chapter~(\ref{ch:chernsimons})) and
the group categories
described in chapters~(\ref{ch:mtc}) and (\ref{ch:groupcategories}).  We achieve
this by showing that the respective projective representations of the mapping class group
\footnote{we restrict ourselves to closed surfaces}
are isomorphic.

Let $\Sigma$ be a \textit{closed} surface.
The toral Chern-Simons projective representation of $\text{MCG}(\Sigma)$
factors through the symplectic group $\text{Sp}(2g,\mathbb{Z})$.
This is explicitly given in equations~(\ref{eq:Atransformeven}),
(\ref{eq:Btransformeven}), and (\ref{eq:Stransformeven}).  The bulk of the work in this chapter
concerns deriving the projective representation of $\text{MCG}(\Sigma)$ induced
from $\grpcat(h,s)$ using surgery.
The main work involves converting a Heegaard decomposition into a surgery
presentation.

\section{Projective representation of $\text{MCG}(\Sigma)$ from $\grpcat(h,s)$}
As a first step we outline briefly some standard constructions from low-dimensional topology
(see, for example, \cite{prasolov_sossinsky}).

\subsection*{Presentation of the mapping class group via Dehn-Lickorish twists}
Since we wish to consider the group $\text{MCG}(\Sigma)$ we require an efficient
presentation for it.  It is well known that $\text{MCG}(\Sigma)$ is
generated by compositions of Dehn twists around simple closed curves
(see, for example, \cite{farb_margalit}).
We use the standard ``turn left'' Dehn twist convention as depicted in 
figure~(\ref{fig:dehn_torus}).  
We note that ``turn left'' makes sense
independent of any choice of orientation of the curves.

\begin{figure}
  \centering
  \input{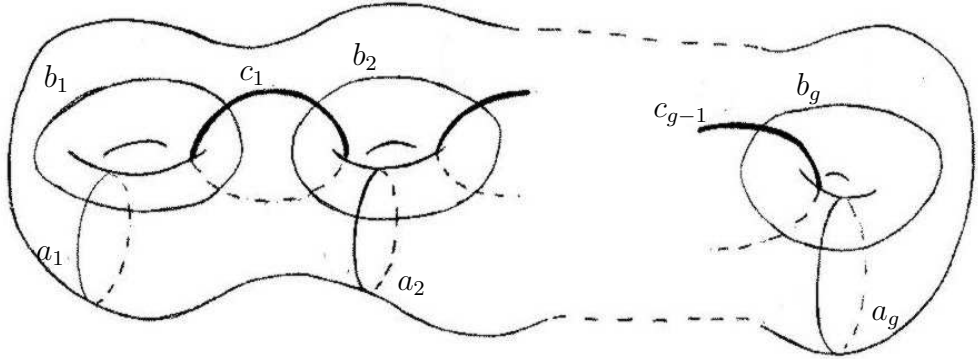}
  \caption{The $3g-1$ Lickorish generators.}
  \label{fig:lickorish}
\end{figure}

It is equally well known that for a closed surface $\Sigma$ of genus $g$ 
it suffices to consider only Dehn twists along the $3g-1$ \textit{Lickorish}
generators depicted in figure~(\ref{fig:lickorish}).
In what follows we will limit our study to the Lickorish generators.

\subsection*{Motivation}
It was mentioned in chapter~(\ref{ch:mtc}) (and studied thoroughly in \cite{turaev})
that a modular tensor category associates to any
oriented 3-manifold $X$ with boundary $\partial X=-\Sigma_-\sqcup \Sigma_+$
an operator
\begin{equation}
 \tau(X):\modfunct(\Sigma_-)\rightarrow\modfunct(\Sigma_+)
\end{equation}
In general $X$ needs to be endowed with some \textit{extended structure} in order
to construct a theory free from gluing anomalies.  The boundary surfaces
 $\Sigma_-$ and $\Sigma_+$ must be endowed with some
extra structure (a parameterization here) to \textit{construct} a theory at all.  However,
if we are satisfied with a TQFT \textit{with anomaly} then the parameterization is irrelevant, and
for an \textit{anomaly-free} TQFT the dependence on parameterization is
very weak (see chapter~(\ref{ch:tqft})).
The matrix elements of
$\tau(X)$ are defined by first ``capping off'' $\Sigma_-$ and $\Sigma_+$ with
the standard handlebodies $H_{g_-}$ and $\overline{H_{g_+}}$, respectively.
\footnote{using the parameterizations}
We then choose a coloring for the embedded ribbon graphs $R_{g_-}$ and
$\overline{R_{g_+}}$.  This gives a closed 3-manifold $\tilde{X}$ with
colored embedded ribbons.  The matrix element (corresponding to the
chosen coloring) is defined to be the 3-manifold
invariant $\tau(\tilde{X})\in\cplx$.  Varying over all choices of coloring
gives all of the matrix elements of the operator
\begin{equation}
 \tau(X):\modfunct(\Sigma_-)\rightarrow\modfunct(\Sigma_+)
\end{equation}

In particular recall that this procedure provides a (projective) 
representation of the mapping class group
for any surface $\Sigma$ of genus $g$ equipped with a parameterization
$\phi:\partial H_g\rightarrow\Sigma$.  We start by considering
the cylinder $\Sigma\times I$ where both boundary components
$\Sigma\times\{0\}$ and $\Sigma\times\{1\}$ have the same
parameterization $\phi$.

Then given an isotopy class of diffeomorphisms $[f]\in\text{MCG}(\Sigma)$
pick a representative diffeomorphism $f:\Sigma\rightarrow\Sigma$.
Then \textit{alter} the parameterization of the boundary component
$\Sigma\times\{0\}$ to be
\begin{equation}
  f\circ \phi
\end{equation}
Denote $\Sigma\times I$ (with the altered parameterization
of $\Sigma\times\{0\}$) by $X_f$.  Then the operator
\begin{equation}
  \tau(X_f):\modfunct(\Sigma)\rightarrow\modfunct(\Sigma)
\end{equation}
defines a projective representation of $\text{MCG}(\Sigma)$.

\subsection*{Converting Heegaard decomposition to integer surgery presentation}
We just saw that in order to study the projective action of the
$\text{MCG}(\Sigma)$ we cap off the 3-manifold $X_f$
with standard handlebodies to form $\tilde{X_f}$.  
However, since $\Sigma\times I$ deformation retracts onto
$\Sigma$ by collapsing the interval $I$, we can view the closed manifold
$\tilde{X_f}$ as two solid handlebodies
glued along $f$.  This provides a Heegaard decomposition
for $\tilde{X_f}$ (however the standard handlebodies contain the embedded ribbon graphs $R_g$ and
$\overline{R_g}$, respectively).

To find the matrix elements we are required to calculate the 3-manifold
invariant $\tau(\tilde{X_f})$.  However, the machinery described
in chapter~(\ref{ch:mtc}) relies on an integer surgery presentation
instead.  Hence we are left with the task of converting a Heegaard decomposition
into a surgery presentation.
Our task is greatly simplified since $\text{MCG}(\Sigma)$ is generated by
the Lickorish generators.

First suppose $f=\id$ (so we have two genus $g$ standard handlebodies glued 
together along the identity boundary diffeomorphism).  
We want to obtain this manifold from integer surgery
along links in $S^3$.  In genus $1$ this is straightforward and already described
in chapter~(\ref{ch:mtc}).  Two solid tori glued together along the identity boundary
diffeomorphism is just $S^2\times S^1$.  This
can be obtained from $S^3$ (see figure~(\ref{fig:s3_torus})) by a single torus switch,
i.e. a $0$-framed surgery (see figure~(\ref{fig:0and1framing})).

\begin{figure}
  \centering
  \input{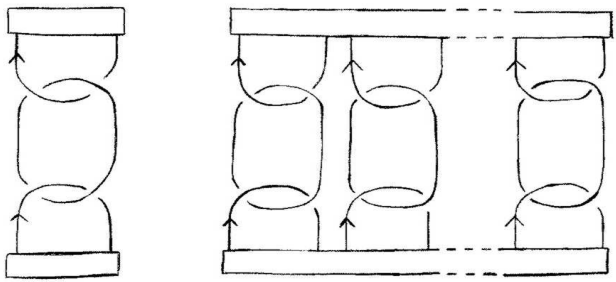}
  \caption{A link diagram in $S^3$ that reproduces the identity map that glues two genus $1$
    standard handlebodies together (left) or more generally two genus $g$ standard
    handlebodies (right).  The bottom component is $R_g$ and the top component is
    $\overline{R_g}$.  The middle link(s) encode the surgery.}
  \label{fig:id_surgery}
\end{figure}

If we remember to place the ribbon graphs $R_g$ and $\overline{R_g}$ into the
handlebodies then we obtain a surgery presentation in $S^3$ as in 
figure~(\ref{fig:id_surgery}) (left side).  Note that the ribbon graphs
$R_g$ and $\overline{R_g}$ do \textit{not} participate in the surgery.
\begin{figure}
  \centering
  \input{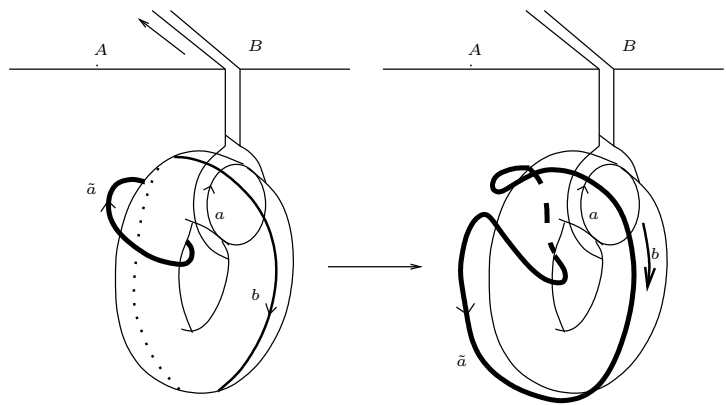}
  \caption{A cross-section of the handlebody.  The Dehn twist takes place on a 
    simple closed curve (not shown) separating $A$ and $B$.  
    The curve is pushed into the handlebody,
    slicing it.  It is then thickened up to a solid torus, and then is drilled out.  this
    leaves a torus-shaped hole (not drawn) in the handlebody.  The
    region labelled $A$ is rotated past $B$ one full turn (making sure that any necessary deformation
    is restricted to the torus-shaped hole).  The solid torus is then glued back in.  The curve
    $\tilde{b}$ is not shown.}
  \label{fig:heegaardtosurgery}
\end{figure}

\begin{figure}
  \centering
  \input{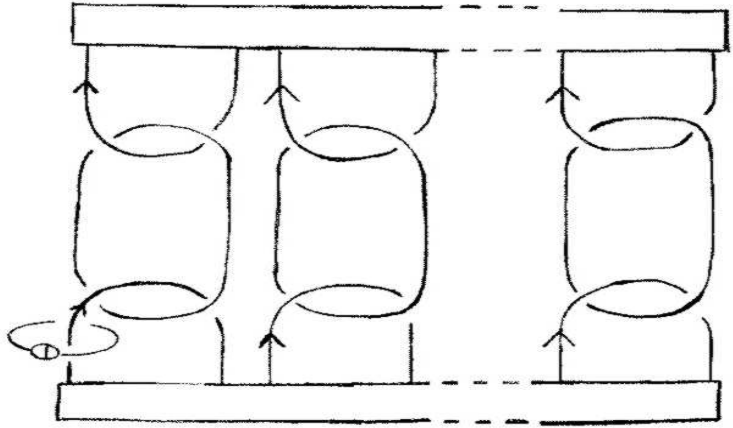}
  \caption{A link diagram in $S^3$ that reproduces the Dehn twist along $a_1$ as
    in figure~(\ref{fig:lickorish}).  The bottom component is $R_g$ and the top component is
    $\overline{R_g}$.  The unoriented links encode the surgery.}
  \label{fig:dehna}
\end{figure}

Now consider a Dehn twist along one of the Lickorish simple closed curves
in figure~(\ref{fig:lickorish}).  There is a surgery that
is equivalent to performing this Dehn twist.  The trick is sketched 
in \cite{prasolov_sossinsky}
on pg. 85.  The appropriate surgery entails the following steps.  First
push the curve slightly into the handlebody $H_{g}$.  As the
curve is pushed let it slice the handlebody (see figure~(\ref{fig:heegaardtosurgery})).  
Now thicken up the curve to
a solid torus and drill it out (this leaves a torus-shaped ``hole'').  Next draw some
markings $a$ and $b$ on the solid torus and
matching markings $\tilde{a}$ and $\tilde{b}$ on
the complementary hole (choose orientations as in 
figure~(\ref{fig:torus})).  Perform the Dehn twist by
sliding $A$ past $B$ one complete revolution and then regluing
(we can confine any necessary
stretching to the torus-shaped hole).
Now glue the solid torus back in.
This produces the following identifications:
\begin{align}
  \tilde{a}= a-b\\
  \tilde{b}=b\notag
\end{align}
This procedure can be viewed equivalently as \textit{not} stretching in the
hole, but rather stretching the solid torus in the opposite direction and gluing it back in.  
In other words we can equivalently solve for $a$ and
$b$ in terms of $\tilde{a}$ and $\tilde{b}$ to obtain
\begin{align}
  a=\tilde{a}+\tilde{b}\\
  b=\tilde{b}\notag
\end{align}
which is just the surgery matrix
\begin{equation}
  \begin{pmatrix} 1 & 0 \\ 1 & 1 \end{pmatrix}
\end{equation}
i.e. a 1-framed surgery (as in example~(\ref{ex:1framedsurgery})).  This shows
that we can perform a Dehn twist along a simple closed curve as in
figure~(\ref{fig:lickorish}) by replacing it with a 1-framed surgery along the
same simple closed curve.  Let us exploit this by providing surgery
presentations for the Lickorish generators 
$\{a_1,\ldots,a_g,b_1,\ldots,b_g,c_1,\ldots,c_{g-1}\}$ as in figures~(\ref{fig:dehna}),
(\ref{fig:dehnb}), and (\ref{fig:dehnc}).
\begin{figure}
  \centering
  \input{dehnb.eps_t}
  \caption{A link diagram in $S^3$ that reproduces the Dehn twist along $b_1$ as
    in figure~(\ref{fig:lickorish}).  The bottom component is $R_g$ and the top component is
    $\overline{R_g}$.  The unoriented links encode the surgery.}
  \label{fig:dehnb}
\end{figure}
\begin{figure}
  \centering
  \input{dehnc.eps_t}
  \caption{A link diagram in $S^3$ that reproduces the Dehn twist along $c_1$ as
    in figure~(\ref{fig:lickorish}).  The bottom component is $R_g$ and the top component is
    $\overline{R_g}$.  The unoriented links encode the surgery.}
  \label{fig:dehnc}
\end{figure}

\subsection*{Representation of Lickorish generators from $\grpcat(h,s)$}
Given the simplistic fusion rules for the simple objects of $\grpcat(h,s)$
(that were specified in chapter~(\ref{ch:groupcategories})) it is easy to see
that for $x_1,\ldots,x_g\in\grp$ the following tensor product is $1$-dimensional:
\begin{equation}
  \cplx_{x_1}\otimes\cplx_{-x_1}\otimes\cdots\otimes\cplx_{x_g}\otimes\cplx_{-x_g}\cong\unitobj
\end{equation}
Since $\text{Hom}(\unitobj,\unitobj)\cong\cplx$ we see that
\begin{equation}
 \text{Hom}(\unitobj,
    \cplx_{x_1}\otimes\cplx_{-x_1}\otimes\cdots\otimes\cplx_{x_g}\otimes\cplx_{-x_g})\cong\cplx
\end{equation}
is $1$-dimensional.

Therefore given a coloring $x_1,\ldots,x_g\in\grp$ for the ribbons in figure~(\ref{fig:H})
(embedded in $H_g$)
the coloring of the coupon  $\in\text{Hom}(\unitobj,
    \cplx_{x_1}\otimes\cplx_{-x_1}\otimes\cdots\otimes\cplx_{x_g}\otimes\cplx_{-x_g})$
is essentially unique (up to a complex constant).
In terms of a basis $v_x\in\cplx_x$ for the simple objects
let us (for example) color the coupon with the linear
map
\begin{equation}
  v_0\mapsto v_{x_1}\otimes v_{-x_1}\otimes\cdots\otimes v_{x_g}\otimes v_{-x_g}
\end{equation}
However, since the associativity isomorphisms are non-trivial we
should be careful with parenthesis (we choose the convention to group from the left):
\begin{equation}
  v_0\mapsto (\cdots\left(\left([v_{x_1}\otimes v_{-x_1}]\otimes[v_{x_2}\otimes v_{-x_2}]\right)
    \otimes[v_{x_3}\otimes v_{-x_3}]\right)
    \otimes\cdots\otimes [v_{x_g}\otimes v_{-x_g}]
\end{equation}

Similarly, for the handlebody $\overline{H_g}$ the space 
\begin{equation}
 \text{Hom}(\cplx_{x_1}\otimes\cplx_{-x_1}\otimes\cdots\otimes\cplx_{x_g}\otimes\cplx_{-x_g},
    \unitobj)\cong\cplx
\end{equation}
is $1$-dimensional (the associativity parenthesis have been omitted to avoid confusion).
Given a coloring $x_1,\ldots,x_g\in\grp$ for the ribbons in figure~(\ref{fig:Hbar})
we color the coupon with the linear morphism (for example)
\begin{equation}
  (\cdots\left(\left([v_{x_1}\otimes v_{-x_1}]\otimes[v_{x_2}\otimes v_{-x_2}]\right)
    \otimes[v_{x_3}\otimes v_{-x_3}]\right)
    \otimes\cdots\otimes [v_{x_g}\otimes v_{-x_g}]\mapsto v_0
  \label{eq:easydeath}
\end{equation}

The computed matrix elements depend on the choices made above, however
it is easy to see (see equation~(\ref{eq:modfunctvs})) 
that all choices made above are equivalent to choosing a basis
for the Hilbert space $\modfunct(\Sigma)$.  Hence the operator is
well-defined independent of these choices.

\subsubsection*{The identity diffeomorphism $\id:\Sigma\rightarrow\Sigma$ (sanity check)}
\begin{figure}
  \centering
  \input{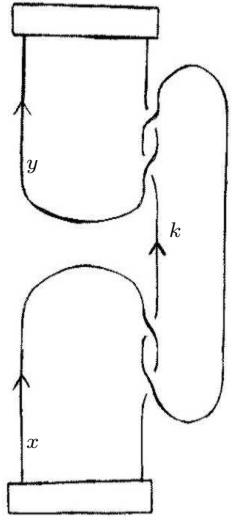}
  \caption{Surgery presentation of identity diffeomorphism $\id:\Sigma\rightarrow\Sigma$
    in genus $1$.  The surgery is performed on the link component colored by $\cplx_k$.}
  \label{fig:id_surgerygenus1}
\end{figure}

Let us proceed to calculate the matrix corresponding to the identity 
diffeomorphism
\begin{equation}
  \id:\Sigma\rightarrow\Sigma
\end{equation}
The surgery presentation for this is given in
figure~(\ref{fig:id_surgery}).  In genus $g$ the different vertical braid sections
do not interact (on the right side of figure~(\ref{fig:id_surgery})), hence
we restrict ourselves to genus $1$ and the genus $g$ calculation will
be $g$ copies of the genus $1$ calculation tensored
together.  Consult figure~(\ref{fig:id_surgerygenus1}).  
It is understood that $x\in\grp$ and $y\in\grp$ are
fixed, and $k\in\grp$ is summed over since that component performs the
surgery.

We note that we are required to explicitly write the associativity
maps since they are nontrivial (see chapter~(\ref{ch:groupcategories})).
However, we shall see shortly that they cancel each other (this is
only true because the category is abelian), hence we will drop the
explicit associativity maps quickly.

Also we recall lemma~(\ref{lem:twistbraiddeathtrivial}).  When we
annihilate a simple object $\cplx_x$ and its dual $\cplx_{-x}$ we
do not bother to write the map $d_x\circ c_{x,-x}\circ(\theta_{x}\otimes \id_{-x})$
since it is trivial.

Following the diagram from the bottom to the top we compute
\begin{align}
  v_0&\mapsto v_x\otimes v_{-x}\\
    &\mapsto (v_x\otimes v_{-x})\otimes(v_k\otimes v_{-k})\notag\\
    &\mapsto [a_{x,-x,k-k}]v_x\otimes (v_{-x}\otimes(v_k\otimes v_{-k}))\notag\\
    &\mapsto [a_{x,-x,k-k}][a^{-1}_{-x,k,-k}]v_x\otimes ((v_{-x}\otimes v_k)\otimes v_{-k})\notag\\
    &\mapsto [a_{x,-x,k-k}][a^{-1}_{-x,k,-k}][c^{-1}_{k,-x}]
      v_x\otimes ((v_{k}\otimes v_{-x})\otimes v_{-k})\notag\\
    &\mapsto [a_{x,-x,k-k}][a^{-1}_{-x,k,-k}][c^{-1}_{k,-x}][c^{-1}_{-x,k}]
      v_x\otimes ((v_{-x}\otimes v_k)\otimes v_{-k})\notag\\
    &\mapsto [a_{x,-x,k-k}][a^{-1}_{-x,k,-k}][c^{-1}_{k,-x}][c^{-1}_{-x,k}][a_{-x,k,-k}]
      v_x\otimes (v_{-x}\otimes (v_k\otimes v_{-k}))\notag\\
    &\mapsto [a_{x,-x,k-k}][a^{-1}_{-x,k,-k}][c^{-1}_{k,-x}][c^{-1}_{-x,k}][a_{-x,k,-k}]
      [a^{-1}_{x,-x,k-k}] (v_x\otimes v_{-x})\otimes (v_k\otimes v_{-k})\notag
\end{align}
Clearly the associativity coefficients cancel each other.  Annihilating
$\cplx_x\otimes\cplx_{-x}$ we obtain
\begin{equation}
    \mapsto [c^{-1}_{k,-x}][c^{-1}_{-x,k}] v_k\otimes v_{-k}
\end{equation}
It is easy to convince ourselves that the associativity maps are always going
to appear in cancelling pairs, hence we omit them from here on to simplify
notation.  Note that, in principle, the associativity maps must be included.
Continuing up the diagram, there is a birth of $\cplx_y\otimes\cplx_{-y}$:
\begin{align}
    &\mapsto [c^{-1}_{k,-x}][c^{-1}_{-x,k}] (v_y\otimes v_{-y})\otimes(v_k\otimes v_{-k})\\
    &\mapsto [c^{-1}_{k,-x}][c^{-1}_{-x,k}][c_{-y,k}] v_y\otimes v_{k}\otimes v_{-y}\otimes v_{-k}\notag\\
    &\mapsto [c^{-1}_{k,-x}][c^{-1}_{-x,k}][c_{-y,k}][c_{k,-y}] (v_y\otimes v_{-y})\otimes 
      (v_{k}\otimes v_{-k})\notag
\end{align}
Annihilate $\cplx_k\otimes\cplx_{-k}$, then apply the map in equation~(\ref{eq:easydeath})
to annihilate $\cplx_y\otimes\cplx_{-y}$:
\begin{equation}
  \mapsto [c^{-1}_{k,-x}][c^{-1}_{-x,k}][c_{-y,k}][c_{k,-y}]v_0
\end{equation}

Hence the ribbon invariant $F(L\cup\Omega)$ is just
\begin{equation}
[c^{-1}_{k,-x}][c^{-1}_{-x,k}][c_{-y,k}][c_{k,-y}]
\end{equation}
To calculate the 3-manifold invariant we use equation~(\ref{eq:3mnfldinvt}).  We
note that the quantum dimension $\text{dim}_q(\cplx_x)=1$ for all simple
objects, hence we omit the dimension factor.  The $L$ surgery link is
the one colored by $\cplx_k$, and the fixed ribbon $\Omega$ is the
two component ribbon graph colored by $\cplx_x$ and $\cplx_y$.

Summing over colorings is the same as summing over $k\in\grp$.  So we have
\begin{equation}
 \tau(X_{\id})=(p_-)^{\sigma(L)}\mathscr{D}^{-\sigma(L)-m-1}\sum_{k\in\grp} F(L\cup\Omega)
\end{equation}
We calculate using equation~(\ref{eq:explicitbraiding})
\begin{align}
 &[c^{-1}_{k,-x}][c^{-1}_{-x,k}] = \text{exp}(2\pi i b(x,k))\\
 &[c_{-y,k}][c_{k,-y}] = \text{exp}(-2\pi i b(y,k))\notag
\end{align}
hence
\begin{equation}
 \tau(X_{\id})=(p_-)^{\sigma(L)}\mathscr{D}^{-\sigma(L)-m-1}
  \sum_{k\in\grp}\text{exp}(2\pi i b(x,k))\text{exp}(-2\pi i b(y,k))
\end{equation}
Using the bilinearity and symmetry of $b$ this becomes
\begin{equation}
 \tau(X_{\id})=(p_-)^{\sigma(L)}\mathscr{D}^{-\sigma(L)-m-1}
   \sum_{k\in\grp}\text{exp}(2\pi i b(x-y,k))
\end{equation}

Now we appeal to lemma~(\ref{lem:sumdeltafunction}) (see below).  The 3-manifold invariant becomes
\begin{equation}
 \tau(X_{\id})=(p_-)^{\sigma(L)}\mathscr{D}^{-\sigma(L)-m-1}
  \mathscr{D}^2\delta_{x,y}
\label{eq:id_surgerygenus1}
\end{equation}
The signature of the linking matrix for $L$ is just $\sigma(L)=0$, and in genus
$g=1$ we have $m=1$ component of $L$.  So the 3-manifold invariant is
\begin{equation}
 \tau(X_{\id})=\mathscr{D}^{-2}
  \mathscr{D}^2\delta_{x,y} = \delta_{x,y} 
\end{equation}
as we expect for the identity diffeomorphism $\id:\Sigma\rightarrow\Sigma$.

In genus $g$ (see right side of figure~(\ref{fig:id_surgery})) we have $m=g$ components
of $L$ (it is still true that $\sigma(L)=0$) and the 3-manifold invariant becomes
$g$ copies of $\mathscr{D}^2\delta_{x,y}$ tensored together (the
normalization must be considered separately):
\begin{align}
 \tau(X_{\id})&=\mathscr{D}^{-g-1}
  \mathscr{D}^{2g}\delta_{x_1,y_1}\cdots\delta_{x_g,y_g}\\ 
  &=\mathscr{D}^{g-1}\delta_{x_1,y_1}\ldots\delta_{x_g,y_g}\notag
\end{align}
The projective factor in front is a symptom that we only have a projective
representation of $\text{MCG}(\Sigma)$.

\begin{lemma}
\label{lem:sumdeltafunction}
\begin{equation}
  \sum_{k\in\grp}\text{exp}(2\pi i b(g,k))=\mathscr{D}^2 \delta_{g,0}
\end{equation}
\end{lemma}
\begin{proof}
Clearly if $g=0$ then the LHS will just be $|\grp|$, i.e. $\mathscr{D}^2$
for the special case $\grpcat(h,s)$ since $\text{dim}_q \cplx_x=1$ for all
simple objects.

Suppose $g\neq 0$.  In terms of generators $1_1,\ldots,1_p$ for the group
$\grp$ write $g=\sum_i g_i 1_i$ and
write any arbitrary element $k=\sum_i k_i 1_i$.  The sum becomes
\begin{multline}
  \sum_{k\in\grp}\text{exp}(2\pi i \sum_{i,j} g_i k_j b_{ij})= \\
  \sum_{k_1=0}^{k_1=n_1-1}\cdots\sum_{k_p=0}^{k_p=n_p-1}
  \prod_i\text{exp}(2\pi i g_i k_1 b_{i1})\ldots\text{exp}(2\pi i g_i k_p b_{ip})
\end{multline}
Consider the last sum by itself.  We intend to show that this vanishes.
\begin{equation}
 \sum_{k_p=0}^{k_p=n_p-1}
  \prod_i\text{exp}(2\pi i g_i k_1 b_{i1})\ldots\text{exp}(2\pi i g_i k_p b_{ip})
\end{equation}
This can be written
\begin{multline}
 \sum_{k_p=0}^{k_p=n_p-1}
  \prod_i\text{exp}(2\pi i g_i k_1 b_{i1})\ldots\text{exp}(2\pi i g_i k_{p-1} b_{i,(p-1)})
  \prod_i\text{exp}(2\pi i g_i k_{p} b_{ip}) = \\
  \prod_i\text{exp}(2\pi i g_i k_1 b_{i1})\ldots\text{exp}(2\pi i g_i k_{p-1} b_{i,(p-1)})
   \sum_{k_p=0}^{k_p=n_p-1}\prod_i\text{exp}(2\pi i g_i k_{p} b_{ip})
\end{multline}
Again restricting attention to the last sum this is
\begin{equation}
  \sum_{k_p=0}^{k_p=n_p-1}[\prod_i\text{exp}(2\pi i g_i b_{ip})]^{k_{p}}
\end{equation}
However $n_p b_{ip}=b(1_i,n_p 1_p)=b(1_i,0)=0$ so we see that
$[\prod_i\text{exp}(2\pi i g_i b_{ip})]$ is an $n_p$-th root of unity.
Hence the terms $[\prod_i\text{exp}(2\pi i g_i b_{ip})]^0$, 
$[\prod_i\text{exp}(2\pi i g_i b_{ip})]^1$, $\ldots$, $[\prod_i\text{exp}(2\pi i g_i b_{ip})]^{n_p-1}$
will be symmetrically distributed around the unit circle, so the sum will be $0$.
\end{proof}

\subsubsection*{Dehn twist along $a_1$}
Let us proceed to calculate the matrix corresponding to the Dehn twist
along the curve $a_1$ depicted in figure~(\ref{fig:lickorish}).  Dehn
twists along the other $a_i$ curves are similar.  Denote the Dehn
twist diffeomorphism as 
\begin{equation}
  T_{a_i}:\Sigma\rightarrow\Sigma
\end{equation}
The surgery presentation for this is given in
figure~(\ref{fig:dehna}).  In genus $g$ the different vertical braid sections
again do
not interact (see figure~(\ref{fig:dehna})), hence
we restrict ourselves to genus $1$ since the genus $g$ calculation can
be recovered by tensoring the genus $1$ calculation here with
$g-1$ copies of the genus $1$ $\id$ calculation as in equation~(\ref{eq:id_surgerygenus1}) (the
normalization must be considered separately).
Consult figure~(\ref{fig:a_surgerygenus1}).
It is understood that $x\in\grp$ and $y\in\grp$ are
fixed, and $k,l\in\grp$ are summed over since those components perform the
surgery.
\begin{figure}
  \centering
  \input{a_surgerygenus1.eps_t}
  \caption{Surgery presentation of Dehn twist along $a$
    in genus $1$.  The surgery is performed on the link components colored by $\cplx_k$
    and $\cplx_l$.}
  \label{fig:a_surgerygenus1}
\end{figure}

In genus $1$ we write
\begin{equation}
  T_a=\begin{pmatrix}1 & 1 \\ 0 & 1\end{pmatrix}\in \text{MCG}(\text{torus})
\end{equation}

Again, as in the last calculation we drop the explicit associativity
maps since they cancel each other.  Technically they should be 
written, however.

Also again recall lemma~(\ref{lem:twistbraiddeathtrivial}).  When we
annihilate a simple object $\cplx_x$ and its dual $\cplx_{-x}$ we
do not bother to write the map $d_x\circ c_{x,-x}\circ(\theta_{x}\otimes \id_{-x})$
since it is trivial.

Following the diagram from the bottom to the top we compute
\begin{align}
  v_0&\mapsto v_x\otimes v_{-x} \\
   &\mapsto (v_l\otimes v_{-l})\otimes(v_x\otimes v_{-x})\notag\\
   &\mapsto [\theta_l]v_l\otimes v_{-l}\otimes v_x\otimes v_{-x}\notag\\
   &\mapsto [\theta_l][c_{-l,x}]v_l\otimes v_{x}\otimes v_{-l}\otimes v_{-x}\notag\\
   &\mapsto [\theta_l][c_{-l,x}][c_{x,-l}](v_l\otimes v_{-l})\otimes(v_x\otimes v_{-x})\notag\\
\end{align}
Now annihilating $\cplx_l\otimes\cplx_{-l}$ gives
\begin{equation}
   \mapsto [\theta_l][c_{-l,x}][c_{x,-l}]v_x\otimes v_{-x}\notag\\
\end{equation}
The remainder of the calculation proceeds exactly as for the genus $1$ $\id$ braid used
to calculate equation~(\ref{eq:id_surgerygenus1}).  This implies that the
ribbon invariant is
\begin{equation}
 F(L\cup\Omega)=[c^{-1}_{k,-x}][c^{-1}_{-x,k}][c_{-y,k}][c_{k,-y}]
   [\theta_l][c_{-l,x}][c_{x,-l}]
\end{equation}
Using equation~(\ref{eq:explicitbraiding}) we compute
\begin{align}
 &[c^{-1}_{k,-x}][c^{-1}_{-x,k}] = \text{exp}(2\pi i b(x,k))\\
 &[c_{-y,k}][c_{k,-y}] = \text{exp}(-2\pi i b(y,k))\notag\\
 &[c_{-l,x}][c_{x,-l}]= \text{exp}(-2\pi i b(l,x))\notag\\
 &[\theta_l] = \text{exp}(2\pi i q(l))\notag
\end{align}
which implies that the 3-manifold invariant $\tau(X_{T_a})$ given by
equation~(\ref{eq:3mnfldinvt}) is
\begin{multline}
 \tau(X_{T_a})=(p_-)^{\sigma(L)}\mathscr{D}^{-\sigma(L)-m-1} \\
  \sum_{k,l\in\grp}\text{exp}(2\pi i b(x,k))\text{exp}(-2\pi i b(y,k))
  \text{exp}(-2\pi i b(l,x))\text{exp}(2\pi i q(l))
\end{multline}
Breaking the sum up
\begin{multline}
  \sum_l\text{exp}(-2\pi i b(l,x))\text{exp}(2\pi i q(l))
   \sum_k\text{exp}(2\pi i b(x,k))\text{exp}(-2\pi i b(y,k))
\end{multline}
But by lemma~(\ref{lem:sumdeltafunction}) the sum over $k$ becomes
$\mathscr{D}^2\delta_{xy}$.  Hence the 3-manifold invariant is
\begin{equation}
 \tau(X_{T_a})=(p_-)^{\sigma(L)}\mathscr{D}^{-\sigma(L)-m-1}
    \mathscr{D}^2\delta_{xy}
    \sum_l\text{exp}(-2\pi i b(l,x))\text{exp}(2\pi i q(l))
\end{equation}
Now we use the properties of the bilinear form
$-b(l,x)=b(l,-x)=q(l-x)-q(-x)-q(l)$ and substitute to obtain
\begin{equation}
 \tau(X_{T_a})=(p_-)^{\sigma(L)}\mathscr{D}^{-\sigma(L)-m-1}
    \mathscr{D}^2\delta_{xy}
    \text{exp}(-2\pi i q(x))
    \sum_l\text{exp}(2\pi i q(l-x))
\end{equation}
We have used the fact that $q(-x)=q(x)$ for a pure quadratic form.
The last sum is just $p_+$ from chapter~(\ref{ch:mtc}), so the
3-manifold invariant is
\begin{equation}
 \tau(X_{T_a})=p_+ (p_-)^{\sigma(L)}\mathscr{D}^{-\sigma(L)-m-1}\cdot
    \mathscr{D}^2\text{exp}(-2\pi i q(x))\delta_{xy}
\end{equation}
In genus $1$ we see that the signature of the linking matrix for $L$
is just $\sigma(L)=1$ (the component colored by $\cplx_l$ has a 1-framing,
the component colored by $\cplx_k$ has a zero framing, and the components are not
linked with each other).  The number of components of $L$ is $m=2$. 
Hence the 3-manifold invariant is
\begin{equation}
 \tau(X_{T_a})=\text{exp}(-2\pi i q(x))\delta_{xy}
\end{equation}
where we have used the fact that $p_+p_-=\mathscr{D}^2$.

In genus $g$ (as in figure~(\ref{fig:dehna})) this computation is tensored
with $g-1$ genus $1$ $\id$ calculations.  We recall that from equation~(\ref{eq:id_surgerygenus1})
each genus $1$ $\id$ computation (without normalization) gives a factor of $\mathscr{D}^2\delta_{x_i y_i}$.
There are $m=g+1$ link components, and the signature is still $\sigma(L)=1$.
Thus the 3-manifold invariant for the Dehn twist $T_{a_i}$ is
\begin{align}
 \tau(X_{T_{a_i}})&=p_+ p_-\mathscr{D}^{-1-(g+1)-1}\mathscr{D}^2\mathscr{D}^{2(g-1)}
    \text{exp}(-2\pi i q(x_i))\delta_{x_1 y_1}\ldots\delta_{x_g y_g}\label{eq:dehna_operator}\\
    &=\mathscr{D}^{g-1}\text{exp}(-2\pi i q(x_i))\delta_{x_1 y_1}\ldots\delta_{x_g y_g}\notag
\end{align}

\subsubsection*{Dehn twist along $b_1$}
The computation for a Dehn twist along $b_1$ is nearly identical.  Again
we can restrict to genus $1$ as in figure~(\ref{fig:b_surgerygenus1}). 
\begin{figure}
  \centering
  \input{b_surgerygenus1.eps_t}
  \caption{Surgery presentation of Dehn twist along $b$
    in genus $1$.  The surgery is performed on the link components colored by $\cplx_k$
    and $\cplx_l$.}
  \label{fig:b_surgerygenus1}
\end{figure}

In genus $1$ we write
\begin{equation}
  T_b=\begin{pmatrix}1 & 0 \\ -1 & 1\end{pmatrix}\in \text{MCG}(\text{torus})
\end{equation}
Rather than follow a similar tedious computation
we skip to the result
\begin{equation}
 F(L\cup\Omega)=[c^{-1}_{k,-x}][c^{-1}_{-x,k}][c_{-y,k}][c_{k,-y}]
   [\theta_l][c_{-l,k}][c_{k,-l}]
\end{equation}
Using equation~(\ref{eq:explicitbraiding}) compute
\begin{align}
 &[c^{-1}_{k,-x}][c^{-1}_{-x,k}] = \text{exp}(2\pi i b(x,k))\\
 &[c_{-y,k}][c_{k,-y}] = \text{exp}(-2\pi i b(y,k))\notag\\
 &[c_{-l,k}][c_{k,-l}]= \text{exp}(-2\pi i b(l,k))\notag\\
 &[\theta_l] = \text{exp}(2\pi i q(l))\notag
\end{align}
which implies that the 3-manifold invariant $\tau(X_{T_b})$ given by
equation~(\ref{eq:3mnfldinvt}) is
\begin{multline}
 \tau(X_{T_b})=(p_-)^{\sigma(L)}\mathscr{D}^{-\sigma(L)-m-1} \\
  \sum_{k,l\in\grp}\text{exp}(2\pi i b(x,k))\text{exp}(-2\pi i b(y,k))
  \text{exp}(-2\pi i b(l,k))\text{exp}(2\pi i q(l))
\end{multline}
Summing over $k$ and using lemma~(\ref{lem:sumdeltafunction}) this becomes
\begin{equation}
 \tau(X_{T_b})=(p_-)^{\sigma(L)}\mathscr{D}^{-\sigma(L)-m-1}
  \sum_{l\in\grp}\mathscr{D}^2\text{exp}(2\pi i q(l))\delta_{x-y-l,0}
\end{equation}
which is just
\begin{equation}
 \tau(X_{T_b})=(p_-)^{\sigma(L)}\mathscr{D}^{-\sigma(L)-m-1}
  \mathscr{D}^2\text{exp}(2\pi i q(x-y))
\end{equation}
In genus $1$ there are $m=2$ components of $L$.  The $\cplx_l$-colored
component has framing $1$.  The $\cplx_k$-colored component has framing $0$.
These two components have linking number $-1$ with respect to each other.  
Hence the linking matrix is
\begin{equation}
  B=\begin{pmatrix}1 & -1 \\ -1 & 0\end{pmatrix}
\end{equation}
We see that $\text{det}(B)=-1$, hence there is $1$ positive and $1$ negative
eigenvalue.  So the signature is $\sigma(L)=0$.

Thus, in genus $1$ we see that
\begin{equation}
 \tau(X_{T_b})=\mathscr{D}^{-2-1}
  \mathscr{D}^2\text{exp}(2\pi i q(x-y))=\frac{1}{\mathscr{D}}\text{exp}(2\pi i q(x-y))
\end{equation}

In genus $g$ if we perform a Dehn twist along $b_i$ and tensor with $g-1$
copies of the genus $1$ $\id$ computation
then we have $m=g+1$ surgery link components as in figure~(\ref{fig:dehnb}).  It
is easy to verify that the signature remains $\sigma(L)=0$.  Thus the
3-manifold invariant is just
\begin{multline}
 \tau(X_{T_{b_i}})=\mathscr{D}^{-(g+1)-1}
  \mathscr{D}^2\text{exp}(2\pi i q(x_i-y_i)) \\ \mathscr{D}^{2(g-1)}
  \delta_{x_1 y_1}\ldots\delta_{x_{i-1}y_{i-1}}\delta_{x_{i+1}y_{i+1}}\ldots\delta_{x_g y_g}
\end{multline}
which is
\begin{equation}
  \tau(X_{T_{b_i}})=\mathscr{D}^{g-2}\text{exp}(2\pi i q(x_i-y_i))
    \delta_{x_1 y_1}\ldots\delta_{x_{i-1}y_{i-1}}\delta_{x_{i+1}y_{i+1}}\ldots\delta_{x_g y_g}
  \label{eq:dehnb_operator} 
\end{equation}

\subsubsection*{Dehn twist along $c_1$}
\begin{figure}
  \centering
  \input{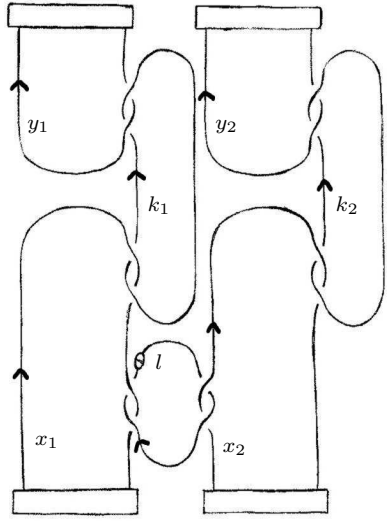}
  \caption{Surgery presentation of Dehn twist along $c$
    in genus $2$.  The surgery is performed on the link components colored by $\cplx_{k_1}$,
    $\cplx_{k_2}$, and $\cplx_{l}$.}
  \label{fig:c_surgerygenus1}
\end{figure}

The computation for a Dehn twist along $c_1$ is only slightly more involved.
In this example there is no genus $1$
case because two vertical braid sections interact (see figure~(\ref{fig:dehnc})).
Consider the genus
$2$ case as in figure~(\ref{fig:c_surgerygenus1}).  
As usual we can consider the genus $g$ case by tensoring with
$g-2$ copies of the genus $1$ $\id$ computation as in equation~(\ref{eq:id_surgerygenus1})
(the normalization must be considered separately).
The case of a Dehn twist
\begin{equation}
  T_{c_i}:\Sigma\rightarrow\Sigma 
\end{equation}
along an arbitrary $c_i$ is similar.  We drop the explicit associativity maps.

Following the diagram up we compute:
\begin{align}
  v_0&\mapsto (v_{x_1}\otimes v_{-x_1})\otimes(v_{x_2}\otimes v_{-x_2})\\
    &\mapsto (v_{x_1}\otimes v_{-x_1})\otimes(v_{l}\otimes v_{-l})
      \otimes(v_{x_2}\otimes v_{-x_2})\notag\\
    &\mapsto [c^{-1}_{l,-x_1}][c_{-l,x_2}]v_{x_1}\otimes v_{l}\otimes v_{-x_1}\otimes v_{x_2}
      \otimes v_{-l}\otimes v_{-x_2}\notag\\
    &\mapsto [c^{-1}_{l,-x_1}][c^{-1}_{-x_1,l}][c_{-l,x_2}][c_{x_2,-l}]
      v_{x_1}\otimes v_{-x_1}\otimes v_{l}\otimes v_{-l}
      \otimes v_{-x_2}\otimes v_{-x_2}\notag\\
    &\mapsto [c^{-1}_{l,-x_1}][c^{-1}_{-x_1,l}][c_{-l,x_2}][c_{x_2,-l}][\theta_l]
      v_{x_1}\otimes v_{-x_1}\otimes v_{l}\otimes v_{-l}
      \otimes v_{-x_2}\otimes v_{-x_2}\notag\\
    &\mapsto [c^{-1}_{l,-x_1}][c^{-1}_{-x_1,l}][c_{-l,x_2}][c_{x_2,-l}][\theta_l]
      v_{x_1}\otimes v_{-x_1}\otimes v_{-x_2}\otimes v_{-x_2}\notag\\
\end{align}
where in the last line the pair $\cplx_l\otimes\cplx_{-l}$ has been annihilated.

From here the diagram proceeds as two copies of the genus $1$ $\id$ computation.
Hence (copying the results before equation~(\ref{eq:id_surgerygenus1})) we obtain that
the ribbon graph invariant $F(L\cup\Omega)$ is
\begin{multline}
 [c^{-1}_{l,-x_1}][c^{-1}_{-x_1,l}][c_{-l,x_2}][c_{x_2,-l}][\theta_l]\times\\
 [c^{-1}_{k_1,-x_1}][c^{-1}_{-x_1,k_1}][c_{-y_1,k_1}][c_{k_1,-y_1}]\times\\
 [c^{-1}_{k_2,-x_2}][c^{-1}_{-x_2,k_2}][c_{-y_2,k_2}][c_{k_2,-y_2}]
\end{multline}
Writing this out using equation~(\ref{eq:explicitbraiding}) this becomes
\begin{multline}
 \text{exp}(2\pi i b(l,x_1))\text{exp}(2\pi i b(-l,x_2))\text{exp}(2\pi i q(l))\times\\
 \text{exp}(2\pi i b(k_1,x_1))\text{exp}(-2\pi i b(y_1,k_1))\times\\ 
 \text{exp}(2\pi i b(k_2,x_2))\text{exp}(-2\pi i b(y_2,k_2))
\end{multline}

The 3-manifold invariant is calculated using equation~(\ref{eq:3mnfldinvt}):
\begin{multline}
 \tau(X_{T_{c}})=(p_-)^{\sigma(L)}\mathscr{D}^{-\sigma(L)-m-1}
 \sum_{l,k_1,k_2}\\
 \text{exp}(2\pi i b(l,x_1))\text{exp}(2\pi i b(-l,x_2))\text{exp}(2\pi i q(l))\times\\
 \text{exp}(2\pi i b(k_1,x_1))\text{exp}(-2\pi i b(y_1,k_1))\times\\ 
 \text{exp}(2\pi i b(k_2,x_2))\text{exp}(-2\pi i b(y_2,k_2)) 
\end{multline}
Performing the sum over $k_1$ and $k_2$ the expression picks up a factor
of $\mathscr{D}^2\delta_{x_1,y_1}$ and $\mathscr{D}^2\delta_{x_2,y_2}$
according to lemma~(\ref{lem:sumdeltafunction}).  Hence this simplifies:
\begin{multline}
 \tau(X_{T_{c}})=(p_-)^{\sigma(L)}\mathscr{D}^{-\sigma(L)-m-1}
 \sum_{l}\\
 \text{exp}(2\pi i b(l,x_1))\text{exp}(2\pi i b(-l,x_2))\text{exp}(2\pi i q(l))\times\\
 \mathscr{D}^2\delta_{x_1,y_1}\mathscr{D}^2\delta_{x_2,y_2}
\end{multline}
Combine the two factors containing $b$ by bilinearity and symmetry: 
\begin{multline}
 \tau(X_{T_{c}})=(p_-)^{\sigma(L)}\mathscr{D}^{-\sigma(L)-m-1}
 \sum_{l}\\
 \text{exp}(2\pi i b(l,x_1-x_2))\text{exp}(2\pi i q(l))\times
  \mathscr{D}^4\delta_{x_1,y_1}\delta_{x_2,y_2}
\end{multline}
Now rewrite $b(l,x_1-x_2)=q(l+x_1-x_2)-q(l)-q(x_1-x_2)$ to obtain
\begin{multline}
 \tau(X_{T_{c}})=(p_-)^{\sigma(L)}\mathscr{D}^{-\sigma(L)-m-1}
 \sum_{l}\\
 \text{exp}(2\pi i q(l+x_1-x_2))\text{exp}(-2\pi i q(x_1-x_2))\times
  \mathscr{D}^4\delta_{x_1,y_1}\delta_{x_2,y_2}
\end{multline}
However we have that $\sum_{l}\text{exp}(2\pi i q(l+x_1-x_2))=p_+$
hence we finally obtain (in genus $2$)
\begin{equation}
 \tau(X_{T_{c}})=p_+(p_-)^{\sigma(L)}\mathscr{D}^{-\sigma(L)-m-1}
 \text{exp}(-2\pi i q(x_1-x_2))\times
 \mathscr{D}^4\delta_{x_1,y_1}\delta_{x_2,y_2}
\end{equation}
The link $L$ has $m=3$ components and the signature of the linking
matrix is $\sigma(L)=1$.  Properly normalized the genus $2$
invariant is:
\begin{equation}
 \tau(X_{T_{c}})=\mathscr{D}\text{exp}(-2\pi i q(x_1-x_2))\delta_{x_1,y_1}\delta_{x_2,y_2}
\end{equation}
We have used the fact that $p_+p_=\mathscr{D}^2$.

In genus $g\geq 2$ it is necessary to tensor with $g-2$ copies of the
genus $1$ $\id$ computation as in equation~(\ref{eq:id_surgerygenus1}) (however the
normalization is not included).
This gives a surgery link $L$ with signature
$\sigma(L)=1$ and $m=g+1$ components.  The 3-manifold invariant corresponding
to a Dehn twist along $c_i$ ($1\leq i\leq g-1$) is then:
\begin{equation}
 \tau(X_{T_{c_i}})=p_+p_-\mathscr{D}^{-1-(g+1)-1}
 \text{exp}(-2\pi i q(x_i-x_{i+1}))\times
 \mathscr{D}^{2g}\delta_{x_1,y_1}\ldots\delta_{x_g,y_g}
\end{equation}
which simplies to
\begin{equation}
 \tau(X_{T_{c_i}})=\mathscr{D}^{g-1}\text{exp}(-2\pi i q(x_i-x_{i+1}))\times
 \delta_{x_1,y_1}\ldots\delta_{x_g,y_g}
  \label{eq:dehnc_operator}
\end{equation}

\subsection*{Lickorish generators and $\text{Sp}(2g,\mathbb{Z})$}
In the last subsection the Lickorish generators $\{a_1,\ldots,a_g,b_1,\ldots,b_g,c_1,\ldots,c_{g-1}\}$
were studied and their associated projective representations on the Hilbert space
$\modfunct(\Sigma)$
\footnote{see equation~(\ref{eq:modfunctvs})}
were produced (the matrix elements 
$\tau(X_{T_{a_i}})$, $\tau(X_{T_{b_i}})$, and $\tau(X_{T_{c_i}})$ were computed explicitly).
So we have constructed a map
\begin{equation}
  \text{MCG}(\Sigma)\rightarrow \mathbb{P}\text{GL}(\modfunct(\Sigma))
\end{equation}

If $\Sigma$ is a closed genus $g$ surface then there is a map
\begin{equation}
  \text{Sp}:\text{MCG}(\Sigma)\rightarrow \text{Sp}(2g,\mathbb{Z})
\end{equation}
determined by recording the action of $\text{MCG}(\Sigma)$ 
only on homology $H_1(\Sigma,\mathbb{Z})$.
The kernel of this map is the \textit{Torelli} group, i.e. there is a
short exact sequence
\begin{equation}
  1\rightarrow \text{Torelli}(\Sigma)\rightarrow\text{MCG}(\Sigma)\rightarrow\text{Sp}(2g,\mathbb{Z})
\end{equation}
The map $\text{MCG}(\Sigma)\rightarrow \mathbb{P}\text{GL}(\modfunct(\Sigma))$ \textit{factors
through} $\text{Sp}(2g,\mathbb{Z})$ if there is a map (broken line) that
makes the following diagram commute:
\begin{equation}
  \xymatrix {
    \text{MCG}(\Sigma)\ar[rr]\ar[dr]^{\text{Sp}} & & \mathbb{P}\text{GL}(\modfunct(\Sigma))\\
    & \text{Sp}(2g,\mathbb{Z})\ar@{.>}[ur] &
  }
\end{equation}

In genus $1$ the Torelli group is trivial.  The mapping class group is generated
by the $s$ and $t$ matrices:
\begin{equation}
 s=\begin{pmatrix} 0 & -1 \\ 1 & 0 \end{pmatrix} \quad\quad
 t=\begin{pmatrix} 1 & 1 \\ 0 & 1 \end{pmatrix} 
\end{equation}
which satisfy the relations $(st)^3=s^2$ and $s^4=1$.
The Lickorish generators $\{a,b\}$ provide another basis
\begin{equation}
 \text{Sp}(T_a)=\begin{pmatrix} 1 & 1 \\ 0 & 1 \end{pmatrix}=t \quad\quad
 \text{Sp}(T_b)=\begin{pmatrix} 1 & 0 \\ -1 & 1 \end{pmatrix}=s^3ts 
\end{equation}

In genus $g$ the Torelli group is not usually trivial.  However we can
still analyze the image of the Lickorish generators in $\text{Sp}(2g,\mathbb{Z})$.
The symplectic matrices are as follows:
\begin{equation}
  \text{Sp}(T_{a_i})=\begin{pmatrix} \unitobj_g & \Delta_i \\ 0 & \unitobj_g \end{pmatrix}
  \quad\text{ where }(\Delta_i)_{\alpha\beta}=\begin{cases}
                               1 & \text{ if }\alpha=\beta=i \\
                               0 & \text{ otherwise} 
                             \end{cases}
  \label{eq:sp_a_lickorish}
\end{equation}
\begin{equation}
  \text{Sp}(T_{b_i})=\begin{pmatrix} \unitobj_g & 0 \\ -\Delta_i & \unitobj_g \end{pmatrix}
  \label{eq:sp_b_lickorish}
\end{equation}
\begin{equation}
  \text{Sp}(T_{c_i})=\begin{pmatrix} \unitobj_g & \Gamma_i \\ 0 & \unitobj_g \end{pmatrix}
  \quad\text{ where }(\Gamma_i)_{\alpha\beta}=\begin{cases}
                               1 & \text{ if }\alpha=\beta=i \\
                               1 & \text{ if }\alpha=\beta=i+1 \\
                               -1 & \text{ if }\alpha=i,\beta=i+1 \\
                               & \text{ or }\alpha=i+1,\beta=i \\
                               0 & \text{ otherwise} 
                             \end{cases}
  \label{eq:sp_c_lickorish}
\end{equation}
These matrices can be written in terms of the symplectic basis given in 
equation~(\ref{eq:symplecticgenerators}).  It is clear that $\text{Sp}(T_{a_i})$ and
$\text{Sp}(T_{c_i})$ are already in the symplectic basis by identifying
$B=\Delta_i$ and $B=\Gamma_i$, respectively.  Denoting the symplectic basis
element
\begin{equation}
 s_g:=  \begin{pmatrix}
     0 & -\idmat_g \\
     \idmat_g & 0
  \end{pmatrix}
\end{equation}
it is easy to check that $\text{Sp}(T_{b_i})=s^3_g\text{Sp}(T_{a_i})s_g$.

\section{Main theorem}
\begin{theorem}\textbf{(Main Theorem)}
\label{th:maintheorem}
The group category $\grpcat(h,s)$ constructed from the data $\triotrunc$
induces a projective representation of the mapping class group $\text{MCG}(\Sigma)$
that is isomorphic to the projective representation of $\text{MCG}(\Sigma)$ constructed
from toral Chern-Simons theory 
\end{theorem}
\begin{proof}
This is essentially a matter of writing the Lickorish generators 
(actually their images in the symplectic group, i.e.
equations~(\ref{eq:sp_a_lickorish}), (\ref{eq:sp_b_lickorish}), (\ref{eq:sp_c_lickorish}))
in terms of the symplectic generators in equation~(\ref{eq:symplecticgenerators}).
We can then use this basis change to compute explicitly what the
projective representation (from toral Chern-Simons) 
found in equations~(\ref{eq:Atransformeven}),
(\ref{eq:Btransformeven}), and (\ref{eq:Stransformeven}) are in
terms of the Lickorish generators.

Once we have this we can compare directly with equations~(\ref{eq:dehna_operator}),
(\ref{eq:dehnb_operator}), and (\ref{eq:dehnc_operator}) that were derived
from $\grpcat(h,s)$.

Manifestly equations~(\ref{eq:sp_a_lickorish}) and (\ref{eq:sp_c_lickorish})
are already in the form of equation~(\ref{eq:Btransformeven}), and it is
straightforward to check that
\begin{equation}
 \text{Sp}(T_{b_i})=s^3_g\text{Sp}(T_{a_i})s_g
\end{equation}

Now, using the toral Chern-Simons projective representation in
equation (\ref{eq:Btransformeven}) we see
that
\begin{equation}
  \widehat{\text{Sp}(T_{a_i})}=
     \{\Psi_{\gamma}(\omega)\rightarrow
     e^{2\pi i \phi(B)c/24}e^{-2\pi i \Sigma_j B^{jj}q_W(\gamma_j)}
    e^{-2\pi i\Sigma_{j<k}B^{jk}b(\gamma_j,\gamma_k)}
    \Psi_{\gamma}(\omega)\}
\end{equation}
where the hat denotes the operator corresponding to $\text{Sp}(T_{a_i})$.  
Using $B=\Delta_i$ from above
we calculate
\begin{equation}
  \widehat{\text{Sp}(T_{a_i})}=
     \{\Psi_{\gamma}(\omega)\rightarrow
     e^{2\pi i \phi(B)c/24}e^{-2\pi i q_W(\gamma_i)}\Psi_{\gamma}(\omega)\} 
\end{equation}
which agrees (up to a projective scalar) with the $\grpcat(h,s)$
projective representation in equation~(\ref{eq:dehna_operator}) (notice that
the delta functions in equation~(\ref{eq:dehna_operator}) agree with
$\gamma\mapsto\gamma$ here).

The toral Chern-Simons matrix in
equation~(\ref{eq:Btransformeven}) also implies
\begin{equation}
  \widehat{\text{Sp}(T_{c_i})}=
     \{\Psi_{\gamma}(\omega)\rightarrow
     e^{2\pi i \phi(B)c/24}e^{-2\pi i \Sigma_j B^{jj}q_W(\gamma_j)}
    e^{-2\pi i\Sigma_{j<k}B^{jk}b(\gamma_j,\gamma_k)}
    \Psi_{\gamma}(\omega)\}
\end{equation}
where we use $B=\Gamma_i$ from above.  This becomes
\begin{equation}
  \widehat{\text{Sp}(T_{c_i})}=
     \{\Psi_{\gamma}(\omega)\rightarrow
     e^{2\pi i \phi(B)c/24}e^{-2\pi i [q_W(\gamma_i)+q_W(\gamma_{i+1})]}
    e^{-2\pi i(-1)b(\gamma_{i},\gamma_{i+1})}
    \Psi_{\gamma}(\omega)\}
\end{equation}
Using the bilinearity of $b$ this is
\begin{equation}
  \widehat{\text{Sp}(T_{c_i})}=
     \{\Psi_{\gamma}(\omega)\rightarrow
     e^{2\pi i \phi(B)c/24}e^{-2\pi i [q_W(\gamma_i)+q_W(\gamma_{i+1})]}
    e^{-2\pi ib(\gamma_{i},-\gamma_{i+1})}
    \Psi_{\gamma}(\omega)\}
\end{equation}
Using $b(\gamma_{i},-\gamma_{i+1})=q_W(\gamma_{i}-\gamma_{i+1})-q_W(\gamma_{i})-q_W(-\gamma_{i+1})$
and the fact that for a pure quadratic form $q_W(-\gamma_{i+1})=q_W(\gamma_{i+1})$ we have
\begin{equation}
  \widehat{\text{Sp}(T_{c_i})}=
     \{\Psi_{\gamma}(\omega)\rightarrow
     e^{2\pi i \phi(B)c/24}e^{-2\pi i q_W(\gamma_{i}-\gamma_{i+1})}
    \Psi_{\gamma}(\omega)\}
\end{equation}
which agrees (up to a projective scalar) with the $\grpcat(h,s)$
projective representation in equation~(\ref{eq:dehnc_operator}).

It remains to compute the toral Chern-Simons matrix
$\widehat{\text{Sp}(T_{b_i})}$ using the fact that $\text{Sp}(T_{b_i})=s^3_g\text{Sp}(T_{a_i})s_g$
and equations~(\ref{eq:Btransformeven}) and (\ref{eq:Stransformeven}).
We have
\begin{multline}
  (\widehat{\text{Sp}(T_{b_i})})_\gamma^{\overline{\gamma}}=
    (|\grp|^{-g/2})^4
    \sum_{\gamma^{\prime},\gamma^{\prime\prime},\gamma^{\prime\prime\prime}\in\grp^g}
    e^{2\pi i b(\gamma_j,\gamma_j^{\prime})} 
    e^{2\pi i b(\gamma^{\prime}_j,\gamma_j^{\prime\prime})}
    e^{2\pi i b(\gamma^{\prime\prime}_j,\gamma_j^{\prime\prime\prime})} \times\\
    e^{2\pi i \phi(B)c/24}e^{-2\pi i q_W(\gamma^{\prime\prime\prime}_i)}\times
    e^{2\pi i b(\gamma^{\prime\prime\prime}_j,\overline{\gamma_j})}
\end{multline}
This is a map \textit{from} a basis of wavefunctions indexed by
$\overline{\gamma}$ \textit{to} a basis indexed by $\gamma$.
The index $j=1,\ldots,g$ counts the factors of $\grp^g$ (i.e. 
$\sum_{\gamma^{\prime}\in\grp^g}=\sum_{\gamma_1^{\prime}\in\grp}\ldots
\sum_{\gamma_g^{\prime}\in\grp}=\prod^{g}_{j=1}\sum_{\gamma_j^{\prime}\in\grp}$).
Note that $(|\grp|^{-g/2})^4=\mathscr{D}^{-4g}$.  Using lemma~(\ref{lem:sumdeltafunction})
we can sum over $\gamma^\prime$, and then sum over $\gamma^{\prime\prime}$ to obtain
\begin{multline}
  (\widehat{\text{Sp}(T_{b_i})})_\gamma^{\overline{\gamma}}=
    \mathscr{D}^{-4g}\mathscr{D}^{2g} e^{2\pi i \phi(B)c/24}
    \sum_{\gamma^{\prime\prime\prime}\in\grp^g}
    e^{2\pi i b(-\gamma_j,\gamma_j^{\prime\prime\prime})} \times\\
    e^{-2\pi i q_W(\gamma^{\prime\prime\prime}_i)}\times
    e^{2\pi i b(\gamma^{\prime\prime\prime}_j,\overline{\gamma_j})}
\end{multline}
The factor of $\mathscr{D}^{2g}=(\mathscr{D}^2)^g$ appears because the sum over
$\gamma^\prime$ is shorthand for $g$ separate sums $j=1,\ldots,g$.  We have
\begin{equation}
  (\widehat{\text{Sp}(T_{b_i})})_\gamma^{\overline{\gamma}}=
    \mathscr{D}^{-2g} e^{2\pi i \phi(B)c/24}
    \sum_{\gamma^{\prime\prime\prime}\in\grp^g}
    e^{2\pi i b(\overline{\gamma_j}-\gamma_j,\gamma_j^{\prime\prime\prime})} \times
    e^{-2\pi i q_W(\gamma^{\prime\prime\prime}_i)}
\end{equation}
Likewise the sum over $\gamma^{\prime\prime\prime}$ breaks up as
separate sums $j=1,\ldots,g$. We have
\begin{multline}
  (\widehat{\text{Sp}(T_{b_i})})_\gamma^{\overline{\gamma}}=
    \mathscr{D}^{-2g} e^{2\pi i \phi(B)c/24}
    \left(\sum_{\gamma^{\prime\prime\prime}_1\in\grp}
    e^{2\pi i b(\overline{\gamma_1}-\gamma_1,\gamma_1^{\prime\prime\prime})}\right)\times\\
    \left(\sum_{\gamma^{\prime\prime\prime}_2\in\grp}
    e^{2\pi i b(\overline{\gamma_2}-\gamma_2,\gamma_2^{\prime\prime\prime})}\right)\times\ldots\times
    \left(\sum_{\gamma^{\prime\prime\prime}_{i-1}\in\grp}
    e^{2\pi i b(\overline{\gamma_{i-1}}-\gamma_{i-1},\gamma_{i-1}^{\prime\prime\prime})}\right)\times\\
    \left(\sum_{\gamma^{\prime\prime\prime}_{i}\in\grp}
    e^{2\pi i b(\overline{\gamma_{i}}-\gamma_{i},\gamma_{i}^{\prime\prime\prime})}\times
    e^{-2\pi i q_W(\gamma^{\prime\prime\prime}_i)}\right)
    \left(\sum_{\gamma^{\prime\prime\prime}_{i+1}\in\grp}
    e^{2\pi i b(\overline{\gamma_{i+1}}-\gamma_{i+1},\gamma_{i+1}^{\prime\prime\prime})}\right)\times\ldots\\
    \left(\sum_{\gamma^{\prime\prime\prime}_{g}\in\grp}
    e^{2\pi i b(\overline{\gamma_{g}}-\gamma_{g},\gamma_{g}^{\prime\prime\prime})}\right)
\end{multline}
By lemma~(\ref{lem:sumdeltafunction}) each factor $j\neq i$ is just 
$\mathscr{D}^2\delta_{\overline{\gamma_j},\gamma_j}$.  Hence we obtain
\begin{multline}
  (\widehat{\text{Sp}(T_{b_i})})_\gamma^{\overline{\gamma}}=
    \mathscr{D}^{-2g} e^{2\pi i \phi(B)c/24}
    \sum_{\gamma_i^{\prime\prime\prime}\in\grp}
    e^{2\pi i b(\overline{\gamma_i}-\gamma_i,\gamma_i^{\prime\prime\prime})}\times
    e^{-2\pi i q_W(\gamma^{\prime\prime\prime}_i)}\\
    \mathscr{D}^{2(g-1)}\delta_{\overline{\gamma_1},\gamma_1}
    \delta_{\overline{\gamma_2},\gamma_2}\ldots\delta_{\overline{\gamma_{i-1}},\gamma_{i-1}}
    \delta_{\overline{\gamma_{i+1}},\gamma_{i+1}}\ldots\delta_{\overline{\gamma_g},\gamma_g}
\end{multline}
However $-b(\gamma_i-\overline{\gamma_i},\gamma^{\prime\prime\prime}_i)=
-q_W(\gamma_i-\overline{\gamma_i}+\gamma^{\prime\prime\prime}_i)+q_W(\gamma_i^{\prime\prime\prime})
+q_W(\gamma_i-\overline{\gamma_i})$ hence substituting we obtain
\begin{multline}
  (\widehat{\text{Sp}(T_{b_i})})_\gamma^{\overline{\gamma}}=
    \mathscr{D}^{-2} e^{2\pi i \phi(B)c/24}
    e^{2\pi i q_W(\gamma_i-\overline{\gamma_i})}
    \sum_{\gamma_i^{\prime\prime\prime}\in\grp}
    e^{-2\pi i q_W(\gamma_i-\overline{\gamma_i}+\gamma^{\prime\prime\prime}_i)}\\
    \delta_{\overline{\gamma_1},\gamma_1}
    \delta_{\overline{\gamma_2},\gamma_2}\ldots\delta_{\overline{\gamma_{i-1}},\gamma_{i-1}}
    \delta_{\overline{\gamma_{i+1}},\gamma_{i+1}}\ldots\delta_{\overline{\gamma_g},\gamma_g}
\end{multline}
The last sum is $p_-$ so we obtain
\begin{multline}
  (\widehat{\text{Sp}(T_{b_i})})_\gamma^{\overline{\gamma}}=
    p_-\mathscr{D}^{-2} e^{2\pi i \phi(B)c/24}
    e^{2\pi i q_W(\gamma_i-\overline{\gamma_i})}\\
    \delta_{\overline{\gamma_1},\gamma_1}
    \delta_{\overline{\gamma_2},\gamma_2}\ldots\delta_{\overline{\gamma_{i-1}},\gamma_{i-1}}
    \delta_{\overline{\gamma_{i+1}},\gamma_{i+1}}\ldots\delta_{\overline{\gamma_g},\gamma_g}
\end{multline}
Recall that $q_W$ is pure so $q_W(\gamma_i-\overline{\gamma_i})=q_W(\overline{\gamma_i}-\gamma_i)$.
Hence we obtain
\begin{multline}
  (\widehat{\text{Sp}(T_{b_i})})_\gamma^{\overline{\gamma}}=
    p_-\mathscr{D}^{-2} e^{2\pi i \phi(B)c/24}
    e^{2\pi i q_W(\overline{\gamma_i}-\gamma)}\\
    \delta_{\overline{\gamma_1},\gamma_1}
    \delta_{\overline{\gamma_2},\gamma_2}\ldots\delta_{\overline{\gamma_{i-1}},\gamma_{i-1}}
    \delta_{\overline{\gamma_{i+1}},\gamma_{i+1}}\ldots\delta_{\overline{\gamma_g},\gamma_g}
\end{multline}
which agrees (up to a projective factor) with equation~(\ref{eq:dehnb_operator}).
\end{proof}

\begin{corollary}
The projective
representation of $\text{MCG}(\Sigma)$ induced by $\grpcat(h,s)$ 
factors through the symplectic group, i.e.
there is a map $\text{CS}$ that makes the following diagram
commute:
\begin{equation}
  \xymatrix {
    \text{MCG}(\Sigma)\ar[rr]^{\grpcat(h,s)}\ar[dr]^{\text{Sp}} & & \mathbb{P}\text{GL}(\modfunct(\Sigma))\\
    & \text{Sp}(2g,\mathbb{Z})\ar@{.>}^{\text{CS}}[ur] &
  }
\end{equation}
Alternatively, the Torelli groups acts trivially,
\end{corollary}
\begin{proof}
This is a part of the proof of theorem~(\ref{th:maintheorem}) since the
toral Chern-Simons projective representation of $\text{Sp}(2g,\mathbb{Z})$
provides such a map $\text{CS}$.
\end{proof}

%\section{The role of $c$}
%NEED TO FIGURE THIS OUT

%\include{conclusion}

\appendix

\chapter{Remark on Nikulin's Lifting Theorem}
\label{appendix1}
The aim here is to slightly revise the main theorem in \cite{belov_moore}
to correct a small error in the statement.  The theorem should read
\begin{theorem}[Belov and Moore, 2005]
\label{thm:belovmoore}
Classification of quantum toral Chern-Simons:
\begin{enumerate}
  \item The set of ordinary quantum toral Chern-Simons theories
        is in one-to-one correspondence with trios of data
        $(\grp,q,c)$ where $\grp$ is a finite abelian group,
        $q$ is a \textbf{pure} quadratic form, and $c$ is
        a cube root of the Gauss reciprocity formula. 
  \item The set of spin quantum toral Chern-Simons theories
        is in one-to-one correspondence with trios of data
        $(\grp,q,c)$ where $\grp$ is a finite abelian group,
        $q$ is a \textbf{generalized} quadratic form, and $c$ is
        a cube root of the Gauss reciprocity formula. 
\end{enumerate}
\end{theorem}
We have replaced ``a quadratic form such that $q(0)=0$'' with ``a
pure quadratic form''.  Let us show that this cannot be relaxed.

It is obviously true that if a quadratic form $q$ is pure then
$q(0)=0$.  Hence one may wonder if the ``pure'' condition in
theorem~(\ref{thm:belovmoore}) (see corollary~(\ref{cor:nikulinpure}) for context) 
can be weakened to ``generalized'' along with the
extra condition that $q(0)=0$.  This is not true.
We achieve this
by proving a proposition that shows that the conditions in the
relevant theorem of Nikulin \cite{nikulin} are sharp.

Before we begin we require the following result
(see the appendix in \cite{milnor_husemoller}):

\begin{theorem}[Milgram]
\label{thm:gausssum}
Let $\Lambda$ be an \textbf{even} lattice, i.e. a lattice equipped with an
even symmetric nondegenerate bilinear form $B:\Lambda\otimes\Lambda\rightarrow\mathbb{Z}$.
Embed $\Lambda$ in the vector space $V=\Lambda\otimes\textbb{Q}$.  Then by bilinearity
$B$ extends to a symmetric nondegenerate bilinear form $B:V\otimes V\rightarrow\mathbb{Q}$. 
Let $Q:V\rightarrow\mathbb{Q}$ be the induced
quadratic refinement defined by $Q(v):=\frac{1}{2}B(v,v)$ for $v\in V$.  
Let $\text{sign}(B) $ be the signature of $(\Lambda,B)$.  Consider
the discriminant group $\grp:=\Lambda^*/\Lambda$.  $B$ descends to a bilinear
form $b:\grp\otimes\grp\rightarrow\qmodz$ and $Q$ descends to a \textbf{pure} 
quadratic form $q:\grp\rightarrow\qmodz$.  It is a fact that the following
Gauss formula is satisfied:
\begin{equation}
  \frac{1}{\sqrt{\lvert\grp\rvert}}\sum_{x\in\grp}{\exp{(2\pi i
    q(x))}}=\text{exp}(2\pi i\cdot \text{sign}(B)/8)
\end{equation}
\end{theorem}

Now for the main result of this appendix:

\begin{proposition}\ \par
\label{prop:counterexample}
\begin{enumerate}
  \item There exists a finite abelian group 
    $\grp$ equipped with a generalized quadratic form such that
    $q(0)=0$ but $q$ is not pure. 
  \item There exists a finite abelian group
    $\grp$ equipped with a generalized quadratic form such that
    $q(0)=0$ but the data $(\grp, q, C)$ does not lift to any
    \textbf{even} lattice (where $C$ is determined from $q$ using
    the Gauss sum formula).
\end{enumerate}
\end{proposition}
\begin{proof}
Let us begin by proving the first statement.  Consider the example
$\grp=\lbrace 0,1/4,1/2,3/4\rbrace=\mathbb{Z}_4$ equipped with
the quadratic form
\begin{align}
  q(0)&=0\text{ mod }1\\
  q(1/4)&=\frac{7}{8}\text{ mod }1\\
  q(1/2)&=0\text{ mod }1\\
  q(3/4)&=\frac{3}{8}\text{ mod }1\\
\end{align}
A straightforward verification shows that this is a generalized
quadratic form, i.e. $q(x+y)-q(x)-q(y)+q(0)=b(x,y)$ is bilinear, and
the associated bilinear form on the generator is just
\begin{equation}
  b(1/4,1/4)=\frac{1}{4}
\end{equation}
It is \textit{not pure} (i.e. $q(nx)\neq n^2 q(x)$ for every $x\in\grp$),
but $q(0)=0$.

Now to show the second claim.  Consider the
same group and quadratic form $(\grp,q)$.  Let us calculate the Gauss sum
\begin{equation}
  \frac{1}{\sqrt{\lvert\grp\rvert}}\sum_{x\in\grp}{\exp{(2\pi i
    q(x))}}=\text{exp}(2\pi i C/8)
\end{equation}
The LHS is easily computed to equal $1$.  So we conclude
that $C\equiv 0\text{ mod }8$.
Suppose for a contradiction that $(\grp,q)$ lifts to
an \textbf{even} lattice $(\Lambda, B)$.  By a lift we mean that there is
an even lattice $(\Lambda, B)$ such that the signature of $B$ satisfies
\begin{equation}
 \text{sign} (B)\equiv C\equiv 0\text{ mod }8
\end{equation}
and the bilinear form $B$ descends to the bilinear form $b(1/4,1/4)=\frac{1}{4}$.

On the other hand it is straightforward to compute all possible \textit{pure}
quadratic forms on $\mathbb{Z}_4$ with bilinear form $b(1/4,1/4)=\frac{1}{4}$ 
by simply enforcing the purity condition
\begin{equation}
  q(nx)=n^2 q(x)
\end{equation}
There are two pure quadratic refinements of this $b$.  The first is
\begin{align}
  q_1(0)&=0\text{ mod }1\\
  q_1(1/4)&=\frac{1}{8}\text{ mod }1\\
  q_1(1/2)&=\frac{1}{2}\text{ mod }1\\
  q_1(3/4)&=\frac{1}{8}\text{ mod }1\\
\end{align}
Computing the Gauss sum implies that $C=1\text{ mod }8$.
The second is:
\begin{align}
  q_2(0)&=0\text{ mod }1\\
  q_2(1/4)&=\frac{5}{8}\text{ mod }1\\
  q_2(1/2)&=\frac{1}{2}\text{ mod }1\\
  q_2(3/4)&=\frac{5}{8}\text{ mod }1\\
\end{align}
Computing the Gauss sum implies that $C\equiv 5\text{ mod }8$.

Now we appeal to theorem~(\ref{thm:gausssum}).  Consider again the (supposed for contradiction)
lift of the original quadratic form $q$ - this is an \textit{even} lattice $(\Lambda,B)$ with
discriminant group $\grp=\mathbb{Z}_4$, induced bilinear form $b(1/4,1/4)=\frac{1}{4}$,
and the signature is $\text{sign}(B)\equiv 0\text{ mod }8$.
Since the lattice is even there is an induced quadratic refinement
$Q$ which descends to a \textit{pure} quadratic refinement $q$.  We
already calculated all possible pure quadratic refinements for this $b$ 
($q_1$ and $q_2$ above).  Applying the theorem we see that the signature
for the lattice must be either
\begin{equation}
  \text{sign}(B)\equiv C\equiv 1\text{ mod }8\quad\text{or }5\text{ mod 8}
\end{equation}
which contradicts the fact that we assumed above that the signature must be
\begin{equation}
    0\text{ mod }8
\end{equation}
Hence the original quadratic form $q$ does \textit{not} lift to an
even lattice.
\end{proof}

\bibliographystyle{alpha}
\bibliography{thesis}

%\begin{thesisauthorvita}
%Spencer David Stirling graduated Valedictorian from
%Taylorsville High School in Taylorsville, Utah, in 1996.
%That same year he
%entered the University of Utah in Salt Lake City, Utah, in both the Mathematics and
%Physics departments. During the
%summer of 2000 he researched in the Physics Department at 
%The Ohio State University in Columbus, Ohio.  He graduated Magna Cum Laude with
%a Bachelor of Science in Mathematics
%and a Bachelor of Science in Physics from the University of Utah in 2001.  Spencer
%entered the Graduate School at the University of Texas Department of Mathematics in 2001.  
%During the
%academic year 2002-2003 he studied in the Department of Theoretical Physics at the 
%Universiteit
%Utrecht in Utrecht, the Netherlands.  He returned to the University of Texas in 2003.
%\end{thesisauthorvita}

\end{document}